\documentclass[a4paper,onecolumn,accepted=2024-07-12]{quantumarticle}
\pdfoutput=1
\usepackage[utf8]{inputenc}
\usepackage[english]{babel}
\usepackage[T1]{fontenc}
\usepackage{amsmath}
\usepackage{amssymb}
\usepackage{hyperref}
\usepackage{abraces} 
\usepackage{comment}
\usepackage[numbers,sort&compress]{natbib}

\usepackage{tikz}
\usepackage{lipsum}

\newcommand{\abs}[1]{\lvert #1 \rvert}
\newcommand{\ket}[1]{\lvert #1 \rangle}
\newcommand{\bra}[1]{\langle #1 \rvert}
\newcommand{\rket}[1]{\lvert #1 )}

\newcommand{\braket}[2]{\langle #1 \vert #2 \rangle}
\newcommand{\rbraket}[2]{( #1 \vert #2 )}

\newcommand{\rmd}{\mathrm{d}}
\newcommand{\tr}{\operatorname{tr}}

\begin{document}

\title{Hayden-Preskill recovery in chaotic and integrable unitary circuit dynamics}

\author{Michael A. Rampp}
\affiliation{Max Planck Institute for the Physics of Complex Systems, 01187 Dresden, Germany}
\orcid{0000-0002-8455-8460}
\author{Pieter W. Claeys}
\orcid{0000-0001-7150-8459}
\affiliation{Max Planck Institute for the Physics of Complex Systems, 01187 Dresden, Germany}

\maketitle

\begin{abstract}
The Hayden-Preskill protocol probes the capability of information recovery from local subsystems after unitary dynamics. As such it resolves the capability of quantum many-body systems to dynamically implement a quantum error-correcting code. 
The transition to coding behavior has been mostly discussed using effective approaches, such as entanglement membrane theory.
Here, we present exact results on the use of Hayden-Preskill recovery as a dynamical probe of scrambling in local quantum many-body systems. We investigate certain classes of unitary circuit models, both structured Floquet (dual-unitary) and Haar-random circuits. We discuss different dynamical signatures corresponding to information transport or scrambling, respectively, that go beyond effective approaches. Surprisingly, certain chaotic circuits transport information with perfect fidelity. In integrable dual-unitary circuits, we relate the information transmission to the propagation and scattering of quasiparticles. Using numerical and analytical insights, we argue that the qualitative features of information recovery extend away from these solvable points.
Our results suggest that information recovery protocols can serve to distinguish chaotic and integrable behavior, and that they are sensitive to characteristic dynamical features, such as long-lived quasiparticles or dual-unitarity.
 
\end{abstract}

\section{Introduction}

Historically, quantum many-body systems have mostly been characterized through asymptotics of correlation functions, thermodynamic probes, or transport properties~\citep{Ashcroft1976,Anderson1984,Chaikin1995}. Such a characterization is related to the low-energy properties of quantum matter. Moreover, these probes are directly accessible in experiments. In recent years a complementary point of view has emerged, with probes inspired by quantum information theory~\citep{Zeng}. This development has been driven in large part by the advent of quantum simulation platforms that enable the access to properties of individual quantum states~\citep{Georgescu2014,Preskill2018}. The quantum information theoretic perspective also offers valuable insight into questions related to temporal dynamics, such as thermalization, the emergence of hydrodynamics, and quantum chaos~\citep{Deutsch1991,Srednicki1994,Rigol2008,Khemani2018}. In particular, ergodic quantum many-body systems are dynamical error-correcting codes that hide information from their environment by encoding it non-locally~\citep{Hosur2016,Roberts2017}. This insight has been used to elucidate the stability of the volume-law phase in monitored dynamics~\citep{Gullans2020,Choi2020,Li2021,Fan2021,Yoshida2021}. Going further back, the information theoretic characterization of many-body systems has been particularly fruitful in the discussion of the black-hole information paradox~\citep{Mathur2009}. Page first modeled black holes as random unitary operators discussing the problem of extracting information from a black hole~\citep{Page1993a}. Hayden and Preskill later proposed a protocol showing that given sufficient entanglement with the initial state of the black hole, information can rapidly be recovered~\citep{Hayden2007}, leading to a program of relating black holes and quantum chaos~\citep{Shenker2014,Cotler2017,Almheiri2020}. In the mean time, closely related quantum teleportation protocols~\citep{Gao2017,Maldacena2017,Brown2023} have also been realized with quantum simulators~\citep{Landsman2019,Jafferis2022,Shapoval2023}.

While the transition to coding behavior has mostly been studied using effective approaches such as entanglement membrane theory (EMT)~\citep{Li2021,Fan2021,Gullans2021,Lovas2023,Blake2023}, numerical techniques~\citep{Sahu2024,Nakata2023}, and random models~\citep{Brown2015,Liu2021,Piroli2020a,Turkeshi2023,Gribben2024}, exact solutions for non-random models are missing. The construction of decoding protocols has also been studied~\cite{Yoshida2017,Bao2021,Leone2022,Oliviero2022,Leone2022a,Agarwal2023}. Furthermore, the presence of integrability~\citep{Lin2018,Dora2017,Gopalakrishnan2018} or localization~\citep{Chen2016,Huang2016,Smith2019,McGinley2019} may significantly impact the way information is processed. In particular, it has been demonstrated numerically that the Hayden-Preskill (HP) protocol can resolve an integrable-to-chaotic transition in a class of Sachdev-Ye-Kitaev models~\citep{Nakata2023}. Moreover, local quantum systems possess internal time and length scales that may -- together with a finite system size -- lead to rich physical behavior. In this work we discuss the Hayden-Preskill information recovery protocol as a probe of local quantum many-body dynamics. We ask if the information recovery protocol can be used to detect the presence of integrability, and how internal time scales, e.g., those related to butterfly and entanglement velocities, manifest in the recovery. On the way we develop several techniques to treat unitary circuits with finite spatial extent. In particular, we discuss how to directly simulate the dynamics of the entanglement membrane in Haar random circuits through a mapping to a non-Hermitian hopping problem, and we give general strategies to solve the dynamics of both chaotic and interacting integrable dual-unitary circuits.

Our main contribution is the determination of exact solutions of the decoding error as a function of time in several classes of unitary circuits. Along the way we develop various analytical techniques to treat the dynamics of finite unitary circuits that we expect to be valuable in other settings as well. In section~\ref{sec:background} we review the HP protocol and its relation to scrambling as well as EMT for the benefit of the reader. We also discuss our setup and fix some conventions. In Sec.~\ref{sec:chaotic} we present exact solutions of the decoding error for certain non-integrable circuits: maximally chaotic dual-unitary circuits, and Haar-random unitary circuits. We discuss the relation of the saturation at late times to scrambling and the way different time scales manifest in the signal. Then, we investigate the stability to perturbations and argue that they are representative of general chaotic dynamics. Then, in Sec.~\ref{sec:perfect} we discuss that perfect decoding is possible whenever the space-time dual of the time-evolution operator is an isometry. This directly implies the dynamical appearance of perfect decoding in dual-unitary circuits and enables the explicit construction of the decoding map. Finally, in Sec.~\ref{sec:integrable} we investigate integrable dynamics. We show that the combination of dual-unitarity and integrability leads to persistent revivals of perfect decoding. We explain this result by drawing a relation to the presence of non-dispersive quasiparticles, whose scattering determines the information transmission. We conjecture an extension of this quasiparticle picture beyond the dual-unitary point and discuss numerical results on non-dual-unitary integrable models.

\section{Background}
\label{sec:background}

\subsection{The Hayden-Preskill Protocol}

\begin{figure}[t]
  \centering
  \includegraphics[width = 0.9\textwidth]{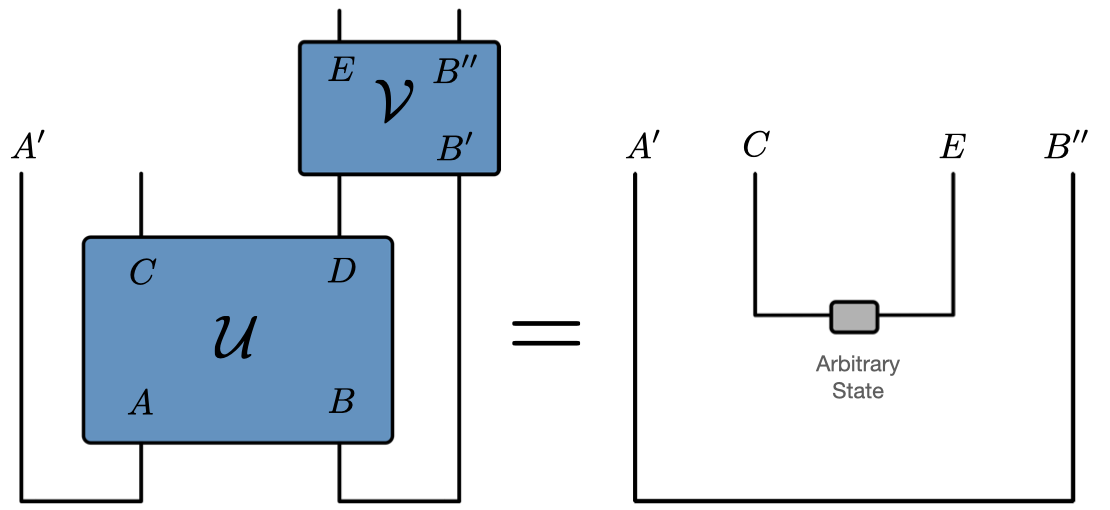}
  \caption{Circuit representation of the Hayden-Preskill protocol.}
  \label{fig:hp_decoding}
\end{figure}

Hayden and Preskill~\citep{Hayden2007} introduced the following variant of the information recovery problem [Fig.~\ref{fig:hp_decoding}]: Alice aims to destroy her diary by throwing it into a black hole. Bob wants to recover it from the outgoing Hawking radiation. Here, instead of a black hole a global unitary operator randomly drawn from the Haar ensemble is considered. Alice's diary is stored in a register $A$ of Hilbert space dimension $d_A$ that is maximally entangled with a memory register $A'$. Bob is in possession of a register $B'$ of  dimension $d_B$ that is maximally entangled with the subsystem $B$, the black hole's initial state. Moreover, the unitary operator is known to Bob. After the unitary operator has acted on the composite system $AB$ (of total dimension $d=d_A d_B$) Bob has access to subsystem $D$ (dimension $d_D$) of the final state, in addition to $B'$. He attempts to decode the information by applying a unitary recovery map on $DB'$. We define decoding to be possible if a map $V$ exists such that following its application Bob is in possession of a subsystem that is maximally entangled with the memory $A'$. We define approximate decoding with fidelity $1-\varepsilon$ if the fidelity of the output state with a maximally entangled state is $\geq1-\varepsilon$. 

Using arguments from decoupling theory~\citep{Hayden2008}, Hayden and Preskill showed that on average Bob is able to decode the information with high fidelity if the subsystem $D$ is larger than Alice's diary $A$,
\begin{equation}
    1-F \leq \frac{d_A}{d_D}.
\end{equation}

By monogamy of entanglement, if a subsystem of $DB'$ is maximally entangled with $A'$, the subsystem $C$ must be completely decoupled from $A'$ and the reduced density matrix $\rho_{A'C}$ has to factorize, $\rho_{A'C}=\rho_{A'}\otimes \rho_C$. Since this condition is hard to treat, usually the special case of a maximally mixed density matrix on $A'C$, $\rho_{A'C}=\mathbbm{1}_{A'}/d_A \otimes \mathbbm{1}_C/d_C$, is considered. The distance to the maximally mixed state can be quantified using the purity $\operatorname{tr}[\rho_{A'C}^2]$. Defining
\begin{equation}
    \delta = d_A d_C \operatorname{tr}[\rho_{A'C}^2] - 1,
\end{equation}
which is small when $\rho_{A'C}$ is close to the maximally mixed state, the decoding fidelity is lower bounded as $F\geq1-\mathcal{O}(\sqrt{\delta})$~\citep{Yoshida2017}.

Yoshida and Kitaev related the decoding fidelity to the concept of scrambling~\citep{Yoshida2017}, as expressed via the decay of out-of-time-ordered correlators (OTOCs)~\citep{Hosur2016,Roberts2017}. The scrambling condition is expressed as the vanishing of OTOCs of the form
\begin{equation}
    \langle \mathcal{O}_A(t)\mathcal{O}_D\mathcal{O}_A(t)\mathcal{O}_D\rangle,
\end{equation}
where $\mathcal{O}_{A,D}$ are traceless operators on $A,D$. Yoshida and Kitaev demonstrated that this implies the bound
\begin{equation}
    \delta \leq \frac{d_A^2}{d_D^2}.
\end{equation}
In chaotic local quantum many-body systems we expect the scrambling condition to be approximately fulfilled at late times and for small enough subsystems $A,D$~\citep{Xu2022}.

\subsection{Setup and conventions}

\begin{figure}[t]
  \centering
  \includegraphics[width = 0.8\textwidth]{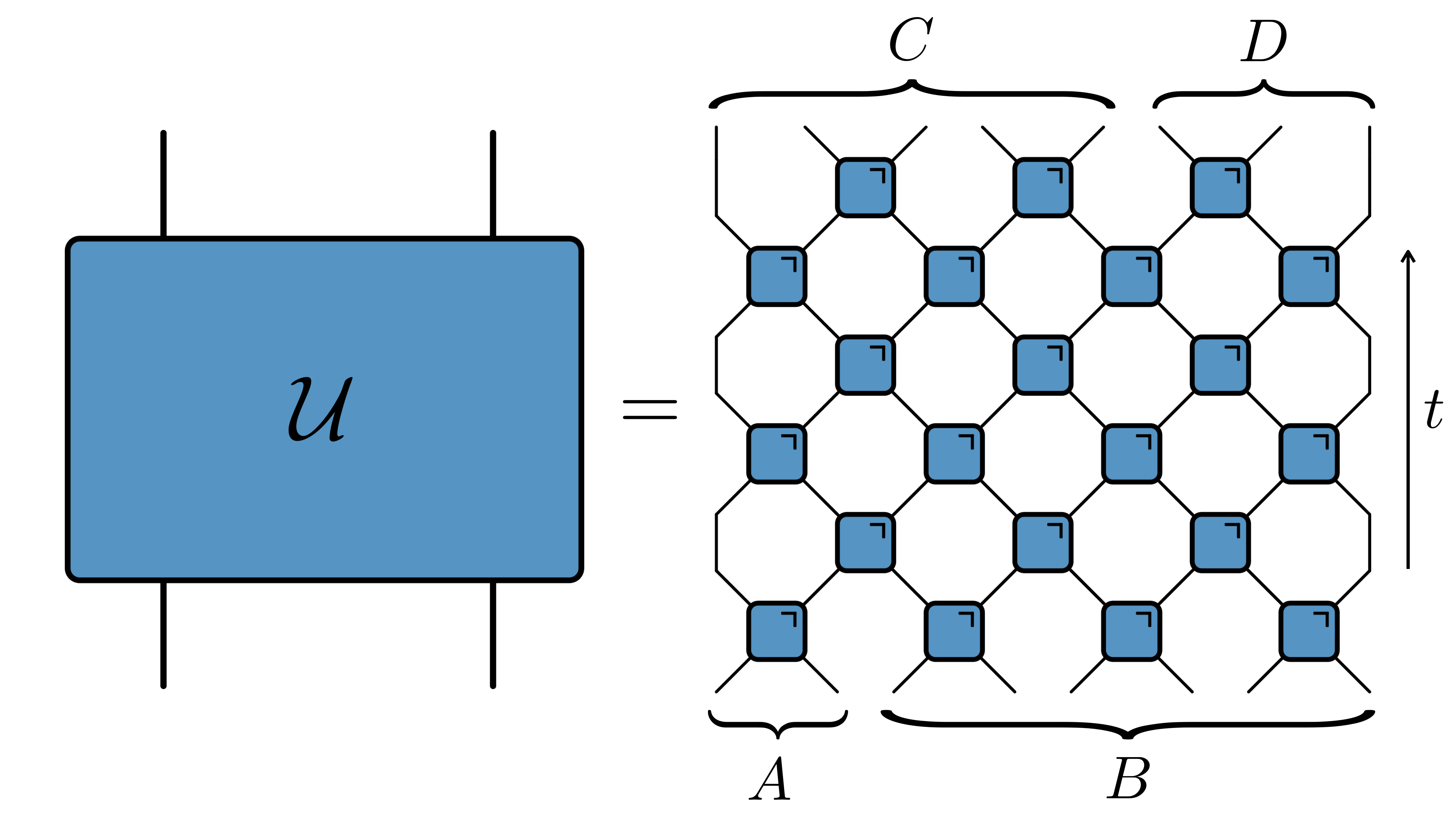}
  \caption{Unitary time-evolution operator as brickwork circuits of unitary two-site gates.}
  \label{fig:te_op_circuit}
\end{figure}

We consider a one-dimensional chain of $L$ qudits (with $q$ internal levels) with open boundary conditions. Alice's subsystem $A$ is composed of the $L_A$ leftmost qudits of the chain, such that $d_A=q^{L_A}$. The remaining qudits form subsystem $B$ ($d_B=q^{L_B}=q^{L-L_A}$). Similarly, subsystem $D$ is composed of the $L_D$ rightmost qudits. We study the decoding error, where the unitary transformation is given by a local brickwork quantum circuit~\citep{Fisher2023} constructed from individual two-site unitary gates [Fig.~\ref{fig:te_op_circuit}]. The number of layers corresponds to time.

We will frequently make use of the \emph{scaling limit} $t\rightarrow\infty,\,L\rightarrow\infty,\,\tau:=t/L=\mathrm{const.}$

\subsection{Entanglement Membrane Theory}

\begin{figure}[t]
  \centering
  \includegraphics[width = \textwidth]{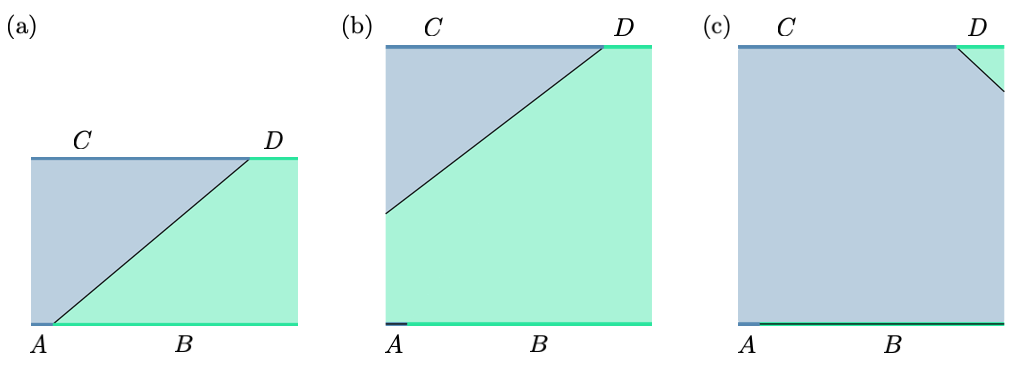}
  \caption{Schematic illustration of the coding transition in the framework of EMT. (a) At early times the minimal membrane connects the top and bottom boundary (corresponding to initial and final state). (b) At late times the minimal membrane connects the top to a spatial boundary. The bottom boundary condition in region $A$ is violated. (This latter scenario is quantitatively equivalent to a second domain wall connecting the left and bottom boundary, which is kinematically forbidden [Sec.~\ref{sec:ruc}].)  (c) Subleading contribution at late times connecting to the opposite boundary.}
  \label{fig:emt}
\end{figure}

Entanglement membrane theory (EMT) is an effective description of entanglement and operator growth in chaotic systems~\citep{Nahum2017,Jonay2018,Zhou2019}. The problem of computing the bipartite entanglement of a tensor network representing a time-evolved quantum state can be recast as the minimization of an entanglement membrane emanating from the boundary between the regions. The central object containing the microscopic information about the evolution is the entanglement line tension $\mathcal{E}(v)$, determining the entanglement cost of a given curve in space time. The entanglement line tension is a convex function of the velocity $v$ and has the properties
\begin{equation}
    \mathcal{E}(0) = v_E, \quad \mathcal{E}(v) = v, \, v\geq v_B,
\end{equation}
where $v_E$ and $v_B$ are the entanglement and butterfly velocity. These two velocities arise naturally in the context of entanglement growth and operator scrambling respectively~\cite{Keyserlingk2018,Nahum2018,Bertini2020b}.

When applied to a Hayden-Preskill-like scenario, EMT predicts a coding transition coinciding with a change in topology of the minimal membrane~\citep{Li2021,Gullans2021,Sahu2024}. For early times, there is a single membrane passing through the circuit, implying a non-vanishing mutual information between $C$ and $A$. At late times, there are two disconnected pieces implying a small mutual information [Fig.~\ref{fig:emt}]. Namely, from EMT we find that at late times the configuration in Fig.~\ref{fig:emt}(b) costs entropy
\begin{equation}
    \frac{S_2}{\log q} = v_B*\left(\frac{L_C}{v_B}\right) + L_A = L_C + L_A = L + L_A - L_D.
\end{equation}
We could also imagine the flipped configuration where the domain wall attaches to the right spatial boundary [Fig.~\ref{fig:emt}(c)]. This configuration has an entropic cost
\begin{equation}
    \frac{S_2}{\log q} = v_B*\left(\frac{L_D}{v_B}\right) + L_B = L - L_A + L_D.
\end{equation}
For $L_D>L_A$, the case we are interested in, the first choice is always favored. Taking only the dominant contributions into account EMT predicts
\begin{equation}
    \operatorname{tr}[\rho_{A'C}^2] = e^{-S_2} = \frac{1}{d_A d_C},
\end{equation}
corresponding to a maximally mixed state, i.e. perfect decoding. However, adding the subleading contributions the correct scaling with subsystem size is recovered,~\footnote{We are grateful to Tianci Zhou and the anonymous referee for pointing this out.} similar to the entropic correction for the entanglement of a bipartite state~\cite{Zhou2019}.
\begin{equation}
    \operatorname{tr}[\rho_{A'C}^2] = \frac{1}{d_A d_C} + \frac{1}{d_B d_D} \quad \implies \quad \delta = \frac{d_A^2}{d_D^2}.
\end{equation}
We will show later that there is a correction to this value when $L_A$ is small, which indeed appears in the exact solutions. This can be understood as arising from a modification of the cost of domain walls along the lower boundary.

At early times on the other hand, the entropy cost of the single domain wall configuration is
\begin{equation}
    \frac{S_2}{\log q} = t\mathcal{E}\left(\frac{L_C-L_A}{t}\right).
\end{equation}
This becomes unfavorable at large enough $t$. The transition happens roughly when $(L_C-L_A)/t\approx v_B$, as $\mathcal{E}(v_B)=v_B$.

\section{Chaotic dynamics}
\label{sec:chaotic}

In this section we discuss the dynamical behavior of the decoding error in several paradigmatic models of non-integrable dynamics, namely Floquet dual-unitary circuits~\citep{Akila2016,Bertini2019,Gopalakrishnan2019} and random unitary circuits~\citep{Oliveira2007,Znidaric2008,Keyserlingk2018,Nahum2018}. Based on these solvable points, we argue about the decoding error in generic chaotic circuits. Finally, we provide a connection between the late-time value of $\delta$ and scrambling in terms of operator strings.

The purity $\tr[\rho_{A'C}^2]$ can be represented as a tensor network where each line carries dimension $q^4$ and each gate is four times replicated, $U\otimes U^*\otimes U\otimes U^*$. For convenience we use the same graphical symbol for the gate itself and its replicated version, leading to
\begin{equation}
    \tr[\rho_{A'C}^2] = \frac{1}{q^{2L}} \,\, \vcenter{\hbox{\includegraphics[width=0.4\columnwidth]{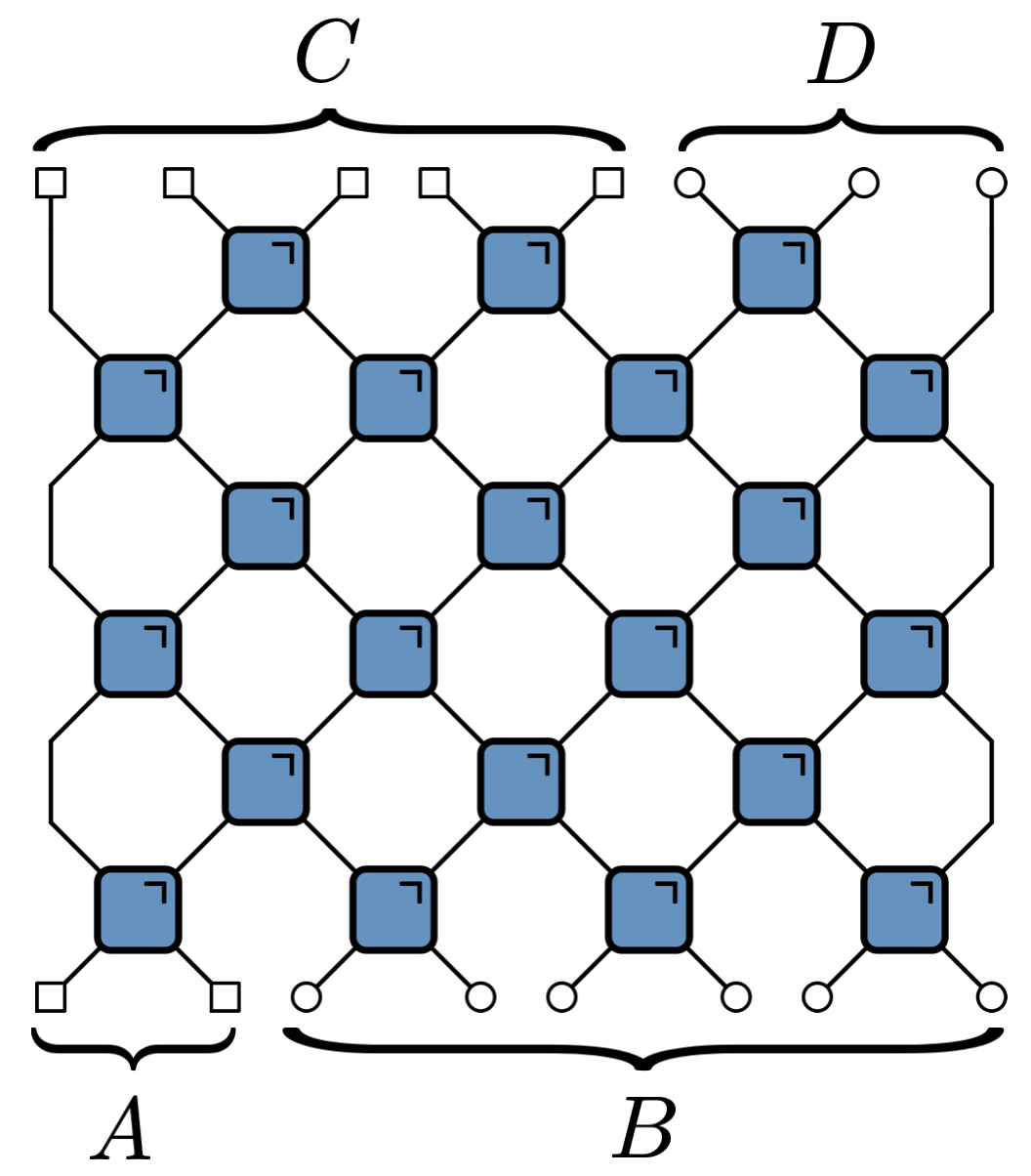}}}, \label{eq:operator_purity}
\end{equation}
where
\begin{align}
\vcenter{\hbox{\includegraphics[height=0.025\textheight]{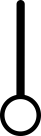}}} \,= \, \vcenter{\hbox{\includegraphics[height=0.025\textheight]{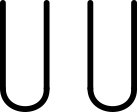}}}\,, \,\, \vcenter{\hbox{\includegraphics[height=0.025\textheight]{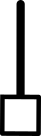}}}\, = \,\vcenter{\hbox{\includegraphics[height=0.025\textheight]{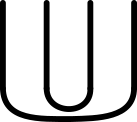}}}\,. \,\,
\end{align}
It is observed that $\tr[\rho_{A'C}^2]$ is equivalent to the purity of the time-evolution operator for a particular partition of initial and final Hilbert spaces. Evaluating this tensor network for different choices of unitary gates is the primary task of the following sections.

\subsection{Dual-unitary circuits}
\label{sec:duc}

Dual-unitary circuits are paradigmatic models of many-body quantum chaos. Despite being generically chaotic, many quantities of interest in non-equilibrium physics can be computed exactly~\citep{Akila2016,Bertini2019,Gopalakrishnan2019,Bertini2018,Piroli2020,Claeys2021,Fritzsch2021,Aravinda2021,Stephen2022,Claeys2022,Claeys2022a,Suzuki2022,Brahmachari2022,Borsi2022,Suchsland2023,Logaric2023}. Dual-unitary circuits are defined by requiring that the two-site gates satisfy unitarity not only in the time but also in the space direction~\cite{Bertini2019,Gopalakrishnan2019}. 

To compute the decoding error, we have to contract the four-layer tensor network Eq.~\eqref{eq:operator_purity}. In the time range $L<t<3L$ a light-cone transfer matrix (LCTM) can be identified that enables the evaluation of the tensor network in the scaling limit by constructing the leading eigenspace of the LCTM. For simplicity, we focus on the case $L_A=L_D=1$ and $t$ even in the main text, and treat the other cases in an Appendix [App.~\ref{app:duc}]. The results remain unchanged when one-site unitary gates are placed on the boundary sites during even timesteps. Using unitarity we simplify the tensor network to
\begin{align}
    \tr[\rho_{A'C}^2] &= \frac{1}{q^{2L}}\,\, \vcenter{\hbox{\includegraphics[width=0.6\columnwidth]{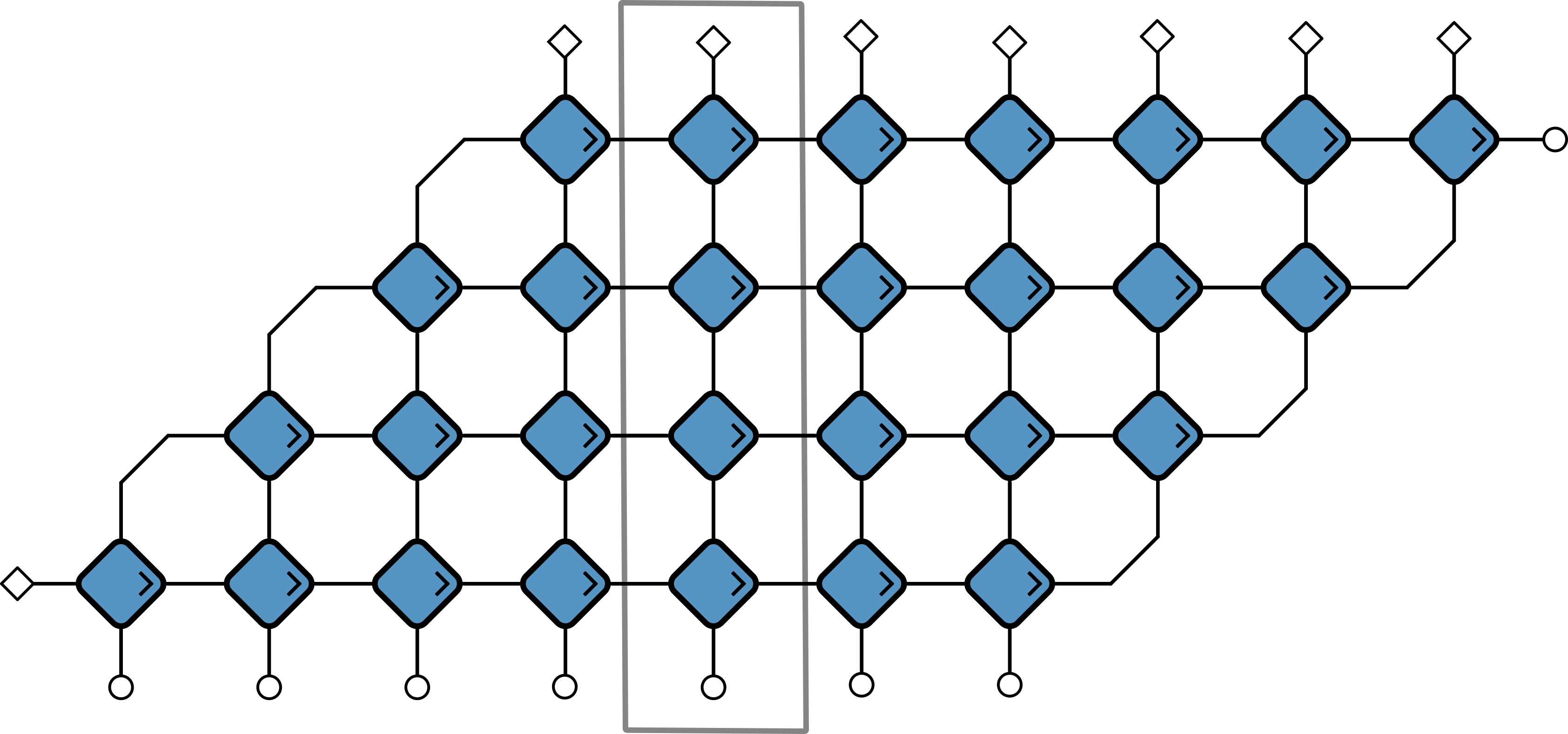}}}\nonumber\\ &= \frac{1}{q^L q^n}\left( \triangleleft_{n} \right| T_{n}^{L-n} \left| \triangleright_{n}\right) \label{eq:circuit-inter},
\end{align}
where $n=(t-L+2)/2$. The tensor network consists of three pieces: a rectangular bulk region that can be understood as the repeated multiplication of the LCTM, and two triangular boundary regions. In the limit $L-n\rightarrow\infty$ the bulk of the tensor network becomes a projector on the leading eigenspace of the LCTM. In fact, the LCTM appearing here is well-known from the study of local operator entanglement and OTOCs~\citep{Bertini2020,Claeys2020,Rampp2023,Huang2023}. For so-called maximally chaotic dual-unitary circuits the leading eigenspace of this transfer matrix can be constructed analytically. The notion of a maximally chaotic dual-unitary gate was introduced in Ref.~\cite{Bertini2020}, and a randomly drawn dual-unitary gate is generically expected to fall into this class. For these maximally chaotic gates an exhaustive set of leading eigenstates can be constructed as simple linear combinations of product states of the vectorized identity and swap operators, graphically represented as
\begin{align}
    \rket{n,k} = \frac{1}{q^n}\,\overbrace{\vcenter{\hbox{\includegraphics[width=0.012\textheight]{Figs/circle.png}}}\,\,\dots\,\,\vcenter{\hbox{\includegraphics[width=0.012\textheight]{Figs/circle.png}}}}^{k}\,\,\overbrace{\vcenter{\hbox{\includegraphics[width=0.012\textheight]{Figs/square.png}}}\,\,\dots\,\,\vcenter{\hbox{\includegraphics[width=0.012\textheight]{Figs/square.png}}}}^{n-k}.
\end{align}
The vectors are orthonormalized as
\begin{equation}
    \rket{\overline{n,0}} := \rket{n,0}, \qquad \rket{\overline{n,k}} := \frac{q\rket{n,k}-\rket{n,k-1}}{\sqrt{q^2-1}},\quad k\geq1.
\end{equation}
Hence, we can write
\begin{equation}
    \delta = q^L\tr[\rho_{RC}^2] - 1 = \frac{1}{q^n}\sum_{k=0}^n \left( \triangleleft_{n} \Big| \overline{n,k} \right)  \left( \overline{n,k} \Big| \triangleright_{n}\right) - 1.
\end{equation}

It remains to compute the overlaps with the boundary conditions. In the orthonormal basis we find (see App.~\ref{app:duc})
\begin{equation}
    \left( \triangleleft_{n} \Big| \overline{n,k} \right) = \begin{cases} q & k=0, \\ 0 & k=1, \\ \frac{q^{k}}{\sqrt{q^2-1}}\left(1-\frac{1}{q^2}\right) & k\geq2, \end{cases} \quad \left( \overline{n,k} \Big| \triangleright_{n}\right) = \begin{cases} q^{n-1} & k=0, \\  0 & 1\leq k\leq n-1, \\ \sqrt{q^2-1} & k=n. \end{cases} \label{eq:triang_overlaps}
\end{equation}
Crucially, the simplification of the boundary conditions is only possible for dual-unitary gates.
Combining these results, we obtain
\begin{equation}
    \delta = \frac{1}{q^n}\left( q^n + q^n \left(1-\frac{1}{q^2}\right)\right) - 1 = 1 - \frac{1}{q^2}.
\end{equation}
The case of general $L_A,L_D$ is treated by a slight modification of the boundary conditions (see App.~\ref{app:duc}). We find
\begin{equation}
    \delta(t) = \left(1-\frac{1}{q^{2L_A}}\right) \frac{1}{q^{2(L_D-L_A)}} = \left(1-d_A^{-2}\right)\frac{d_A^2}{d_D^2}.
\end{equation}

In maximally chaotic dual-unitary circuits, after an initial "dip" corresponding to perfect information transport (see Sec.~\ref{sec:perfect} for more details), the information is immediately scrambled, expressed by the decoding error being below the scrambling bound. Indeed, we will show in Sec.~\ref{sec:scrambling} that the scrambling assumption implies that the decoding error is close to the universal value $\left(1-d_A^{-2}\right)\frac{d_A^2}{d_D^2}$. In finite size systems, relaxation will not be immediate, instead there will be deviations associated with the gap of the light-cone transfer matrix.

\begin{figure}[t]
  \centering
  \includegraphics[width = 0.9\textwidth]{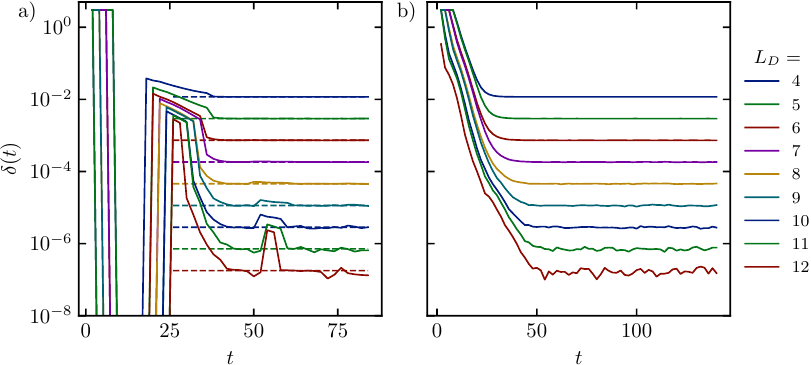}
  \caption{Decoding error $\delta$ as a function of time in non-integrable Floquet circuits for different values of $L_D$ (system size $L=14$). (a) \emph{Dual-unitary circuit.} At early times a "dip" is visible, where $\delta$ is several orders of magnitude smaller than its eventual saturation value (see Sec.~\ref{sec:perfect}). (b) \emph{Non-dual-unitary circuit.} The dip is absent. In both cases the eventual saturation values of the two models are identical and coincide with the theoretical predictions of the main text. The gates are of the form $U=(u_+\otimes u_-)\exp\left(-i(J_{xy}(XX+YY) +J_z ZZ)\right)(v_+\otimes v_-)$. For (a) $J_{xy}=\pi/4$ and $J_z/J_{xy}=-3/10$, and for (b) $J_{xy}=\pi/5$ and $J_z/J_{xy}=-3/10$. The single site gates are fixed, but randomly selected.}
  \label{fig:du_erg}
\end{figure}

While times $t>3L$ are not accessible analytically, we do not expect any deviation from the saturation value to appear. As we show in Sec.~\ref{sec:scrambling}, saturation to $\delta_\infty$ signals that information about the initial state has been scrambled non-locally. It is unlikely that this information can be unscrambled again, except for isolated times where approximate revivals occur. We expect such revivals to occur on timescales $\sim\exp(\exp(L))$. Moreover, this prediction is in very good agreement with numerical results for small system sizes obtained by exact computation of the entanglement of the time-evolution operator. These data are presented in Fig.~\ref{fig:du_erg}(a) and show excellent agreement with our predictions. The deviations for large $L_D$ at intermediate times are likely finite size effects that occur because $L_D\approx L$.

\subsection{Perturbed dual-unitary circuits}

Fig.~\ref{fig:du_erg}(b) shows that the saturation value $\delta_\infty$ coincides with the value computed in Sec.~\ref{sec:duc} also for non-dual-unitary circuits. To better understand this agreement, we investigate the effect of weak perturbations to dual unitarity and
demonstrate that these do not change the saturation value. This result is consistent with general arguments presented in section~\ref{sec:scrambling} that this value is generic in non-integrable circuits and with our result from Haar random circuits [Sec.~\ref{sec:ruc}]. We apply the method of projection of the LCTM to the maximally chaotic subspace and evaluate the resulting path-integral formula in the large-$q$ limit. This approach has been shown to reproduce the generic hydrodynamics of information spreading in chaotic systems~\cite{Rampp2023}, and important spectral features of the so-obtained projected LCTM coincide with the expected behavior in generic chaotic systems~\cite{Huang2023}.

To simplify the calculation we consider only the gates in the bulk of the TN~\eqref{eq:circuit-inter} to have broken dual unitarity. The gates in the boundary regions (correspoding to the initial and final $L$ layers) are taken to be dual unitary. We find
\begin{equation}
    \delta(t)- \delta_\infty = q^2 F_{z_1}(n,L-n) + \mathcal{O}(1/q), \quad F_{z_1}(x,y) = x \binom{y}{x}B_{z_1}(x,y-x+1), \label{eq:perturbed}
\end{equation}
where $B_{z_1}$ denotes the incomplete beta-function
\begin{equation}
    B_{z_1}(p,q) = \int_0^{z_1} \mathrm{d}t\, t^{p-1}(1-t)^{q-1}.
\end{equation}
$F_{z_1}$ can be identified with the profile of the OTOC of the homogeneously perturbed circuit in the thermodynamic limit, i.e. agnostic of the boundaries, in so-called light-cone coordinates. In the regime of validity of this perturbative approach $F_{z_1}$ is exponentially small, signalling scrambling. Hence, the saturation value is stable to weak breaking of dual unitarity.

This finding is corroborated by numerical data for generic non-dual-unitary Floquet circuits displaying relaxation to the same saturation value, while the early-time dip is no longer present (c.f. the discussion in Sec.~\ref{sec:perfect}).

\subsection{Random unitary circuits}
\label{sec:ruc}

We investigated the dynamics of HP recovery in dual-unitary circuits, which display many aspects of generic behavior, but are nevertheless fine tuned to some degree. In order to isolate the aspects common to generic non-integrable dynamics, we turn to another paradigmatic model of non-integrable many-body dynamics, Haar random circuits~\citep{Khemani2018,Keyserlingk2018,Nahum2018}. We demonstrate that $\delta_\infty$ coincides with the value in DUCs, and discuss aspects of relaxation. Mapping the problem to a directed random walk model for the entanglement membrane enables us to perform numerical simulations for large system sizes and long times.

We consider random unitary circuits with gates drawn independently from the Haar measure over $U(q^2)$ at each space-time point. The problem of computing the averaged purity
\begin{equation}
    \mathbbm{E}\left[\operatorname{tr}[\rho_{A'C}^2]\right]
\end{equation}
can be mapped to a classical statistical mechanics model. Using standard procedures we represent the purity as the partition function of a statistical mechanics model of a 2-state system on a triangular lattice with interactions on the downward pointing triangles~\citep{Nahum2017,HunterJones2019}. This triangular lattice is given by 
\begin{equation}
    \vcenter{\hbox{\includegraphics[width=0.45\textheight]{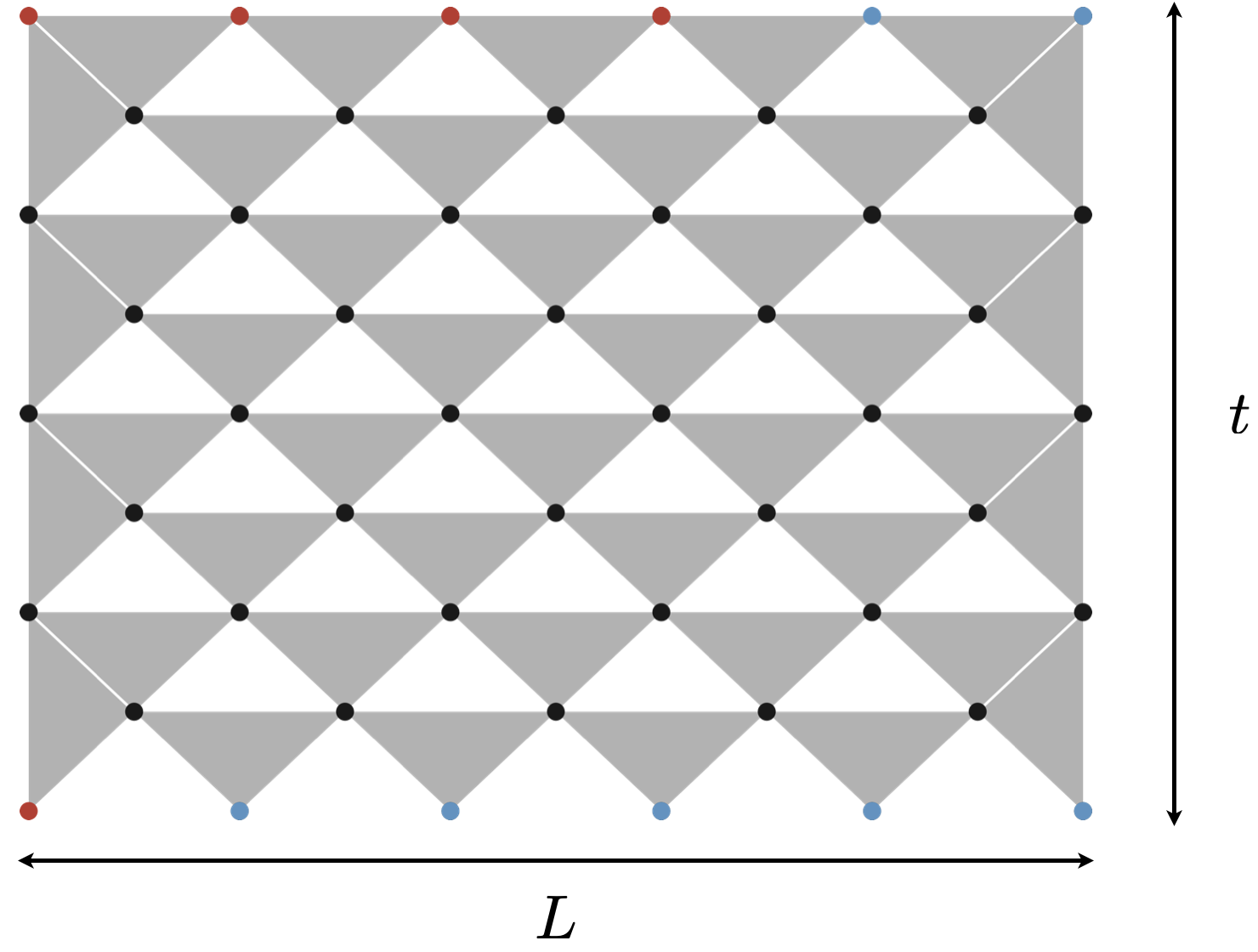}}}.
\end{equation}
where every vertex can take the values $\sigma = \pm 1$. The colored vertices are subject to boundary conditions.
The bulk weights read
\begin{align}
    J^{\sigma_1}_{\sigma_2\sigma_3} = \vcenter{\hbox{\includegraphics[width=0.15\textheight]{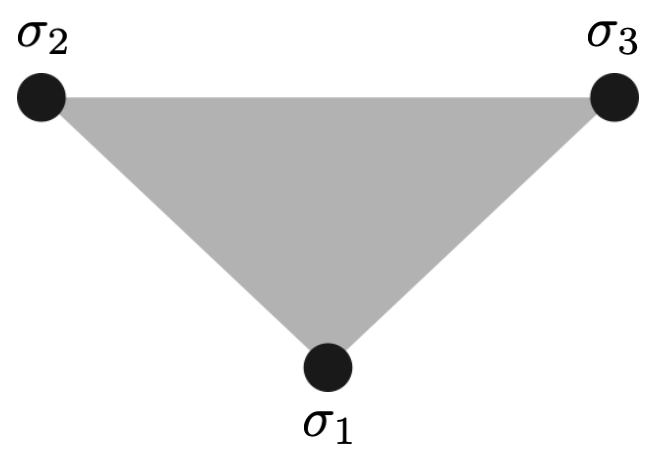}}} = \begin{cases}
        1, & \sigma_1=\sigma_2=\sigma_3, \\
        0, & \sigma_1\neq\sigma_2=\sigma_3, \\
        \frac{q}{q^2+1}, & \sigma_1=\sigma_2\neq\sigma_3 \,\,\mathrm{or}\,\,\sigma_1=\sigma_3\neq\sigma_2.
    \end{cases}
\end{align}
These weights correspond to a ferromagnet. Spin up ($\sigma=+1$) corresponds to $\vcenter{\hbox{\includegraphics[width=0.006\textheight]{Figs/square.png}}}$, while spin down ($\sigma=-1$) corresponds to $\vcenter{\hbox{\includegraphics[width=0.006\textheight]{Figs/circle.png}}}$. The model has a global $\mathbbm{Z}_2$ symmetry which is broken by the top and bottom boundary conditions implementing the particular operator space entropy. The bulk weights allow the propagation of domain walls, but due to unitarity domain walls cannot end or nucleate in the bulk, as expressed by the conditions $J^+_{--}=J^-_{++}=0$. The spatial open boundary conditions, however, lead to triangles of the following geometrical structure
\begin{equation}
    \mathrm{Left}\,\mathrm{boundary:}\,\, J^{\sigma_1}_{\sigma_2\sigma_3} = \vcenter{\hbox{\includegraphics[width=0.1\textheight]{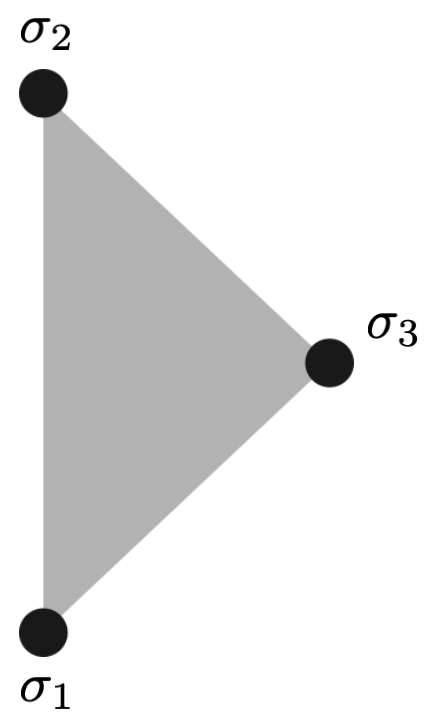}}}, \quad \mathrm{Right}\,\mathrm{boundary:}\,\, J^{\sigma_1}_{\sigma_2\sigma_3} = \vcenter{\hbox{\includegraphics[width=0.1\textheight]{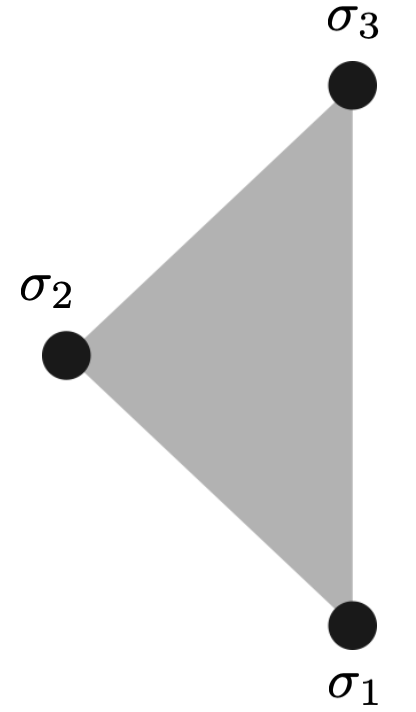}}}.
\end{equation}
These weights restrict the domain walls (coming from the top of the diagram) in such a way that they can end on the boundaries, but they cannot be nucleated there.
Our approach is to compute the partition function of the domain wall exactly in certain limits. The problem is that of a directed polymer~\cite{Kardar2007}, but with absorbing boundary conditions.

\begin{figure}[t]
  \centering
  \includegraphics[width = 0.75\textwidth]{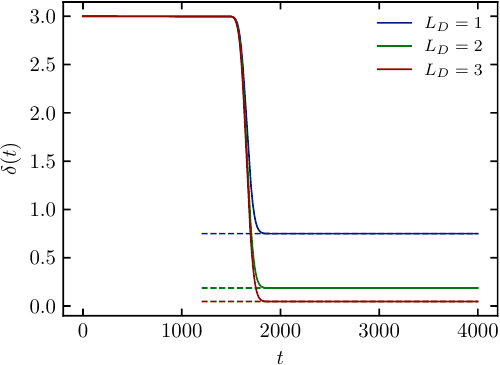}
  \caption{Averaged decoding error as a function of time in Haar random brickwork circuits for $L=1000,\,L_A=1$ and $L_D=1,2,3$. The data are generated by an exact mapping of the statistical mechanics model to the random walk of the entanglement membrane, enabling efficient simulations. The dashed lines denote the theoretical predictions for $\delta_\infty$.}
  \label{fig:ruc}
\end{figure}

\begin{figure}[t]
  \centering
  \includegraphics[width = 0.75\textwidth]{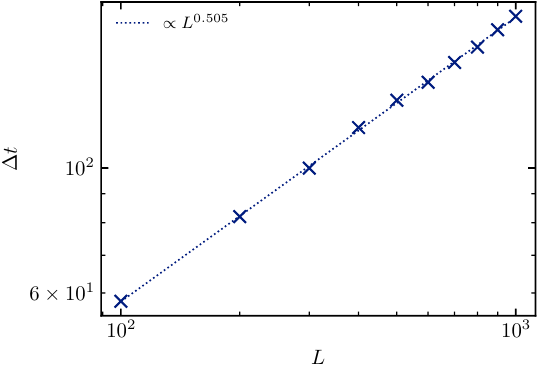}
  \caption{Width of the front region  as a function of system size in random unitary circuits (averaged). The width increases as a power law $\Delta t\sim L^\alpha$ with an exponent of $\alpha\approx0.505$, consistent with diffusion.}
  \label{fig:broadening}
\end{figure}

To that end, we focus on the transfer matrix of the statistical mechanics model in the temporal direction:
\begin{equation}
    T_{\sigma_1\dots\sigma_N,\sigma'_1\dots\sigma'_N} = \sum_{\{\tau_i\}} \vcenter{\hbox{\includegraphics[width=0.45\textheight]{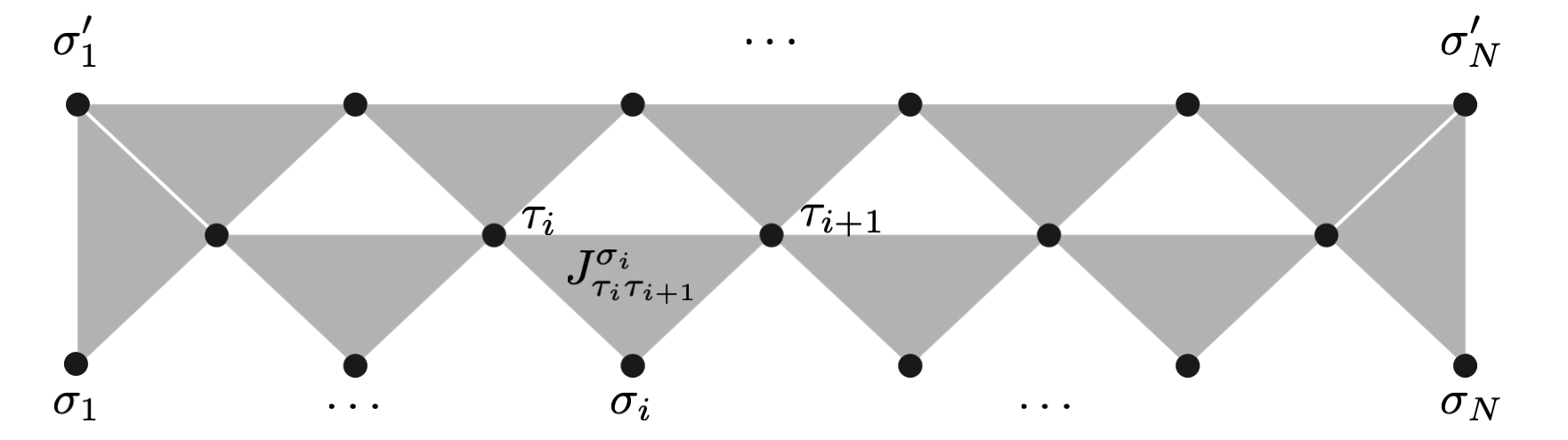}}}.
\end{equation}
The spins interact via the weights $J^{\sigma}_{\tau\tau'}$ on the shaded triangles. Then we can represent the averaged purity (for odd $t$) as
\begin{align}
    \mathbbm{E}\left[\operatorname{tr}[\rho_{A'C}^2]\right] &= \left(q^3(\bra{\uparrow}+\bra{\downarrow})_1 \prod_{i>1}(q^4\bra{\uparrow}+q^2\bra{\downarrow})_i \right)T^{\frac{t-1}{2}} \ket{\uparrow_1\dots\uparrow_{\frac{L_C}{2}}\downarrow_{\frac{L_C}{2}+1}\dots\downarrow_N} \nonumber\\
    &\equiv \bra{L_N}T^{\frac{t-1}{2}}\ket{R_N}.
\end{align}
Here, the top boundary condition $\ket{R_N}$ fixes the spin configuration at the top to be a domain wall with spins up in region $C$ and spins down in region $D$. Crucially, the transfer matrix leaves the subspace of domain wall states invariant which is a subspace of linear size in $L$, thus drastically reducing the computational complexity and allowing for numerical evaluation of large systems. We present numerical data in Fig.~\ref{fig:ruc} showing the averaged decoding error $\delta$ as a function of time. We observe a crossover from an $\mathcal{O}(1)$ value at early times to the same saturation value $\delta_\infty$ as in dual-unitary circuits at times $t\approx L/v_B$. The front regions broadens diffusively $\sim\sqrt{L}$, as shown in Fig.~\ref{fig:broadening}. This broadening results from the diffusive broadening of the operator front~\cite{Keyserlingk2018,Nahum2018}, or equivalently, using an entanglement membrane description, from a fluctuation effect of the membrane~\cite{Sahu2024}.

We can now describe the full dynamics by restricting the transfer matrix to a domain wall subspace without loss of generality. 
This domain wall subspace is spanned by the vectors
\begin{equation}
    \ket{m}:=\ket{\uparrow_1\dots\uparrow_m\downarrow_{m+1}\dots\downarrow_N},\,\,m=1,\dots,N \quad\ket{0}:=\ket{\downarrow_1\dots\downarrow_N}.
\end{equation}
As the domain wall can only propagate to the left or right through a triangular plaquette, upon one application of the transfer matrix (comprising two time steps) it can only either move one step to the left or right or not at all. As such, only the matrix elements $\bra{n}T\ket{n\pm0,1}$ can be non-zero. Explicit construction gives the transfer matrix in the domain wall subspace
\begin{equation}
    T = \begin{pmatrix}
        1 & z & & & & & & \\
        0 & 2z & z & & & & & \\
          & z & 2z & z & & & & \\
          &  & \ddots  & \ddots & \ddots  & & & \\
          & &  & \ddots  & \ddots & \ddots & &  \\
          & & & & z & 2z & z & \\
          & & &  &  & z & 2z & 0 \\
          & & &  &  &  & z & 1
    \end{pmatrix}, \quad z = \left(J^+_{+-}\right)^2 = \left(\frac{q}{q^2+1} \right)^2. \label{eq:dw_tm}
\end{equation}
The vanishing of the matrix elements $\bra{1}T\ket{0}$ and $\bra{N-1}T\ket{N}$ encodes the open boundary conditions of the qudit chain. The domain wall can end on the spatial boundaries. These conditions also make the transfer matrix non-Hermitian. The transfer matrix can be understood as the Hamiltonian of a non-Hermitian hopping problem on a line, where the non-Hermiticity is due to the absorbing boundaries.

All left- and right-eigenvectors of $T$ can be found analytically. The leading eigenvectors determine the saturation of the decoding error for $t\rightarrow\infty$, while the remaining eigenvectors contribute to the dynamics. There are two degenerate leading right- (left-) eigenvectors that are localized on the spatial boundaries. This gives a further justification for the EMT result, where the saturation value is determined by membranes terminating on the boundary. There is a gap of $\Delta=1-4z$ to the subleading eigenvectors that take the form of standing waves. Therefore, the asymptotic relaxation is given by $\sim(4z)^{t/2}=q^{-v_E t}$, where $v_E= \log\left(\frac{q^2+1}{2q}\right)/\log(q)$ is the entanglement velocity~\citep{Keyserlingk2018,Nahum2018}. A careful calculation reveals power-law corrections to this relaxation, as demonstrated further below.

Let us now derive the saturation value. From the inspection of Eq.~\eqref{eq:dw_tm} it follows immediately that $\ket{r_1}=\ket{0}$ and $\ket{r_2}=\ket{N}$ are right eigenvectors of $T$ with eigenvalue one. These states are exactly localized on the boundaries. To find a leading left eigenvector, $\bra{\ell}=\sum_i a_i\bra{i}$, we consider the recurrence 
\begin{subequations}
\label{eq:leadingleftrec}
    \begin{align}
        a_0 &= a_0, \\
        z(a_{j-2} + 2a_{j-1} + a_j) &= a_{j-1}, \,\, j=2,\dots,N, \\
        a_N &= a_N.
    \end{align}
\end{subequations}
We find
\begin{subequations}
\label{eq:leadingleft}
\begin{align}
    a^{(1)}_j &= \frac{q^{2(N-j)}-q^{-2(N-j)}}{q^{2N} - q^{-2N}}\sim q^{-2j}, \\
    a^{(2)}_j &= \frac{q^{2j}-q^{-2j}}{q^{2N} - q^{-2N}}\sim q^{2(j-N)}.  
\end{align}
\end{subequations}
The left leading eigenvectors are exponentially localized on a length scale $\sim \log(q)^{-1}$.
We obtain the saturation value by replacing $T^{\frac{t-1}{2}}$ with the projector on the leading eigenspace of $T$. In the TDL this results in
\begin{equation}
    \delta_\infty = \braket{L_N}{r_1}\braket{\ell_1}{R_N} + \braket{L_N}{r_2}\braket{\ell_2}{R_N} - 1 = \frac{q^{2L_A}-1}{q^{2L_D}}. \label{eq:delta_inf}
\end{equation}
The exponentially small finite-size corrections are discussed in App.~\ref{app:ruc}. At this point it is worth reflecting on the appearance of the $\mathcal{O}(d_A^{-2})$ correction to the EMT result. It stems from the modified boundary condition on the lower boundary, which implies that having a domain wall running along the lower boundary is slightly more costly than it would be otherwise. This modification appears because the operator purity differs slightly from the partition function used to define the entanglement line tension~\cite{Zhou2020}.

The remaining eigenvalues and -vectors can be found in a similar manner. For a left eigenvector with eigenvalue $\lambda$ the recurrence relation becomes
\begin{equation}
    a_j = (\lambda z^{-1}-2)a_{j-1} - a_{j-2},
\end{equation}
with boundary conditions $a_0=a_N=0$. The case of right eigenvectors is treated in the Appendix. The solutions are standing waves $a_j=a\sin(\pi nj/N)$ with the spectrum
\begin{equation}
    \lambda_n = 2z\left( 1 + \cos\left(\frac{\pi n}{N}\right) \right), \quad n=1,\dots, N-1. \label{eq:spectrum}
\end{equation}
The gap is thus $\Delta = 1-4z$ in the TDL. This gap defines a velocity scale $v=\log(\sqrt{4z})/\log(q)$ that coincides exactly with the entanglement velocity for the second R\'{e}nyi entropy $v_E$. We also note that there are no Jordan blocks.

These results enable us to write down an exact expression for the average decoding error in random unitary circuits as
\begin{equation}
    \delta(t) 
    = 1-\frac{1}{q^{2L_A}} + q^{L+L_A-L_D}\sum_{n=1}^{N-1} \braket{L_N}{r_n}\braket{\ell_n}{R_N} \lambda_n^{\frac{t-1}{2}}.
\end{equation}
In the thermodynamic limit the contribution of the subleading parts can be written in integral form as
\begin{align}
    &\sum_{n=1}^{N-1}\lambda_n^{\frac{t-1}{2}} \rbraket{L}{r_n}\rbraket{\ell_n}{R}, \nonumber\\
    =& \int_0^\pi \rmd k\lambda(k)^{\frac{t-1}{2}}\sin(k N_C)\left( \frac{2z\sin(k)}{\lambda(k)-1}\left(q^{-L_A}+q^{L_A-L}\right) + iq^{L_A+2}\frac{e^{2i k}-1}{(e^{i k}-q^2)(e^{ik }q^2-1)} \right). \label{eq:ruc_subleading}
\end{align}
This integral can be evaluated asymptotically for $t/N\rightarrow\infty$ [App.~\ref{app:ruc}], leading to
\begin{equation}
    \delta(t)-\delta_\infty \sim q^{L-v_E t} \frac{L}{t^{3/2}}. \label{eq:relax} 
\end{equation}
The decay at late times is exponential with power law corrections. The corrections appear because the integrand has a zero at the saddle point.

We have observed from the exact solution that in random circuits the relaxation rate to the saturation value is given by the entanglement velocity. This result suggests that in EMT subleading contributions from nearly vertical membranes have to be taken into account. When considering generic non-integrable dynamics, there is also a possible scenario in which the late-time saturation is governed by a "magnon" mode~\cite{Jonay2023}, a bound state of two membranes. These two scenarios have to be carefully distinguished.

\subsection{Relation of the saturation to scrambling}
\label{sec:scrambling}

The fact that the saturation value shows up in two independent models and is also robust to perturbations suggests that it has a common, universal origin. In the following we show that this saturation value can indeed be understood as a generic consequence of scrambling. Our result gives an interpretation in terms of operator strings~\citep{Keyserlingk2018,Nahum2018} of the bound by Yoshida and Kitaev~\citep{Yoshida2017}. It implies that in a scrambling system $\delta$ is not only smaller than $d_A^2/d_D^2$, it is in fact close to $\delta_\infty=\left(1-d_A^{-2}\right)d_A^2/d_D^2$.

First, we relate the decoding error to the OTOC averaged over operators on $A$ and $D$. We write $\delta=d_A^2\Delta-1$, where $\Delta$ is the average OTOC
\begin{equation}
    \Delta(t) = \mathbb{E}_{\mathcal{O}_A,\mathcal{O}_D}\left[\langle\mathcal{O}_A(t)\mathcal{O}_D\mathcal{O}_A(t)\mathcal{O}_D\rangle\right].
\end{equation}
The average over operators on $D$ can be rewritten as the probability that the evolved operator $\mathcal{O}_A(t)$ acts on $D$ as the identity~\citep{Xu2022}
\begin{equation}
    \Delta = \mathbb{E}_{\mathcal{O}_A}\left[\sum_{\mathcal{S},\mathcal{S}_D=\mathbbm{1}}\abs{\alpha_A(\mathcal{S},t)}^2\right],
\end{equation}
where we have decomposed $\mathcal{O}_A(t)$ in terms of Pauli strings
\begin{equation}
    \mathcal{O}_A(t) = \sum_\mathcal{S} \alpha_A(\mathcal{S},t)\mathcal{S}.
\end{equation}
Now, we make the assumption that in a chaotic system at late times all operator strings appear in the expansion of $\mathcal{O}_A(t)$ with equal probability. This is the assumption that the initial information is uniformly "scrambled" over the local subsystem $D$~\citep{Xu2022}. In particular this assumption also implies a vanishing average OTOC on $D$. The only exception is when the initial operator is the identity $\mathcal{O}_A(t=0)=\mathbbm{1}_A$, which is preserved due to unitarity. With this assumption, we can write
\begin{equation}
    \sum_{\mathcal{S},\mathcal{S}_D=\mathbbm{1}}\abs{\alpha_A(\mathcal{S},t)}^2 = \frac{1}{q^{2L_D}}\left(1-\delta_{\mathcal{O}_A,\mathbbm{1}}\right) + \delta_{\mathcal{O}_A,\mathbbm{1}}.
\end{equation}
There are $q^{2L_A}-1$ non-identity operators on $A$, hence,
\begin{equation}
    \Delta_\infty = \frac{1}{q^{2L_A}} \left(1 + \frac{q^{2L_A}-1}{q^{2L_D}}\right) \quad\implies\quad \delta_\infty = \frac{q^{2L_A}-1}{q^{2L_D}} = \left(1-d_A^{-2}\right)\frac{d_A^2}{d_D^2}.
\end{equation}

It is also possible to relate the decoding error $\delta(t)$ to the OTOC averaged over non-identity operators
\begin{equation}
    \delta(t) = \delta_\infty + (d_A^2-1)(1-d_D^{-2})F(t), \quad F(t) = \mathbb{E}_{\mathcal{O}_A,\mathcal{O}_D\neq\mathbbm{1}}\left[\langle\mathcal{O}_A(t)\mathcal{O}_D\mathcal{O}_A(t)\mathcal{O}_D\rangle\right].
\end{equation}
This result shows again that in systems that scramble, i.e. where $F(t)\ll 1$ for $t\rightarrow\infty$, the decoding error relaxes to $\delta_\infty$.

\section{Perfect decoding}
\label{sec:perfect}

Scrambling of quantum information guarantees a decoding error that is exponentially small in the size of $D$. In the following we discuss the conditions under which the information can be recovered \emph{perfectly}. The same condition was discussed in a different context in Ref.~\citep{Lie2022}. Moreover, the condition also enables the explicit construction of the decoding map as a quantum circuit.

The key insight is that the decoding map in the HP protocol [Fig.~\ref{fig:hp_decoding}] acts on the \emph{space-time dual} of the time-evolution operator. Hence, decoding is associated to finding an approximate (unidirectional) inverse to the space-time dual of $\mathcal{U}$. In particular, perfect decoding is possible if and only if such an inverse exists. This is the case if the time-evolution operator itself is dual unitary, a condition that is fulfilled in dual-unitary circuits (if $L_A=L_D$) for $t=L$. This implies that any dual-unitary circuit, chaotic or integrable, can be used to transport information with perfect fidelity. In the dynamical signal $\delta(t)$ for chaotic dual-unitary circuits this leads to a characteristic dip-plateau structure. In generic chaotic circuits, this dip is washed out in the scaling limit and a finite length scale emerges, over which information can be transmitted with high fidelity.

More precisely, the decoding fidelity, i.e., the overlap with a maximally entangled state on $A'B''$, $\ket{\psi_{\mathrm{out}}}$,
\begin{equation}
    F = \bra{\psi_{\mathrm{out}}} P_{A'B''} \ket{\psi_{\mathrm{out}}}
\end{equation}
can be written diagrammatically as~\citep{Yoshida2017}
\begin{align}
F = \frac{1}{d_A^2 d_B}\,\, \vcenter{\hbox{\includegraphics[width=0.4\columnwidth]{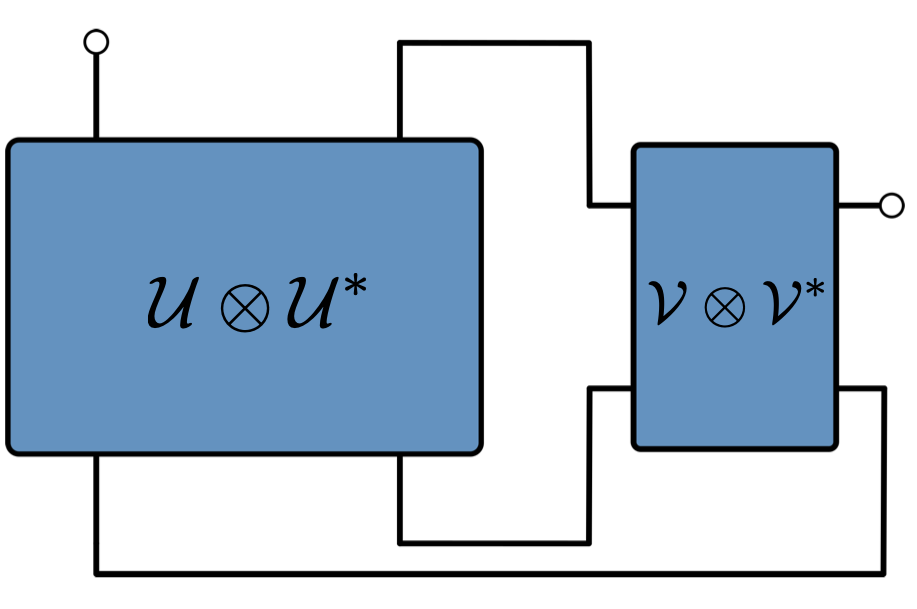}}} \label{eq:perfect},
\end{align}
hence, the condition for perfect decoding is that the space-time dual of $\mathcal{U}$ is an \emph{isometry}, that is
\begin{equation}
    \Tilde{\mathcal{U}}\Tilde{\mathcal{U}}^\dagger = \sqrt{\frac{d_B}{d_C}}\, \Xi_{CE}\otimes\mathbbm{1}_{A} = \sqrt{\frac{d_B}{d_C}} \,\, \vcenter{\hbox{\includegraphics[width=0.34\columnwidth]{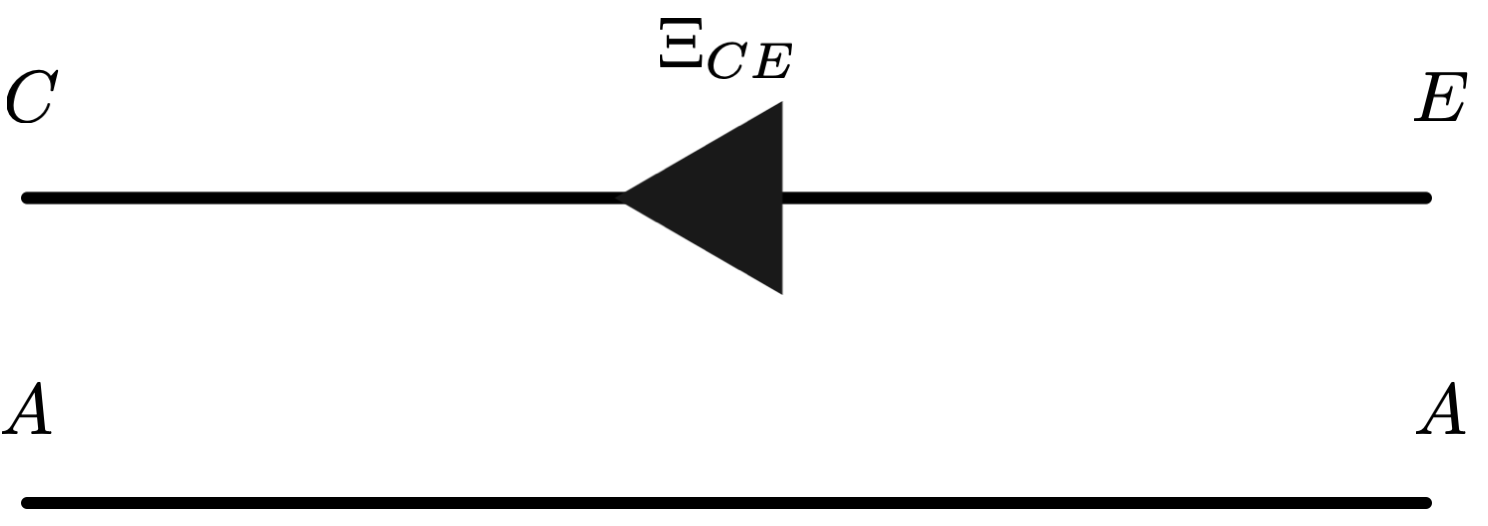}}},
\end{equation}
where $\Xi_{CE}$ is an embedding of C in E. If this is the case, the optimal decoding strategy is choosing $\mathcal{V}=\Tilde{\mathcal{U}}^\dagger$, i.e. the (unidirectional) inverse of $\Tilde{\mathcal{U}}$, which leads to
\begin{align}
    F = \frac{\tr_{A,C,E}\left[\Tilde{\mathcal{U}}\Tilde{\mathcal{U}}^\dagger \otimes \Tilde{\mathcal{U}}^*\Tilde{\mathcal{U}}^T\right]}{d_A^2 d_B} = \frac{d_B}{d_C}\frac{\tr\left[\Xi_{CE}\,\Xi_{CE}^\dagger\right]}{d_B} = 1,
\end{align}
as $\tr\left[\Xi_{CE}\Xi_{CE}^\dagger\right]=\operatorname{min}(d_C,d_E)$, and for the cases of our interest $d_C<d_E$.

What distinguishes the special case $t=L-1$ presented here from the generic, scrambling, case is that the information is not encoded non-locally. Instead the local subspace encoding the information is perfectly preserved and at time $t=L-1$ it is part of Bob's subsystem $D$.

In the following we discuss what happens to the dip as dual-unitarity is broken. We focus on the case $L_A=L_D=1$ and consider a finite system size $L$. This is equivalent to asking the question, how well the information stored in a single site is transported along the light-cone over a distance $(L-1)$. It turns out that the answer is directly related to a class of generalized operator entanglement entropies that appeared previously in the context of quantifying the deviation from dual unitarity~\citep{Rampp2023}.

Diagrammatically, we have [c.f. Eq.~\eqref{eq:operator_purity}]
\begin{equation}
    \operatorname{tr}\left[\rho_{A'C}^2\right] = \frac{1}{q^{2L}}\,\underbrace{\vcenter{\hbox{\includegraphics[width=0.4\columnwidth]{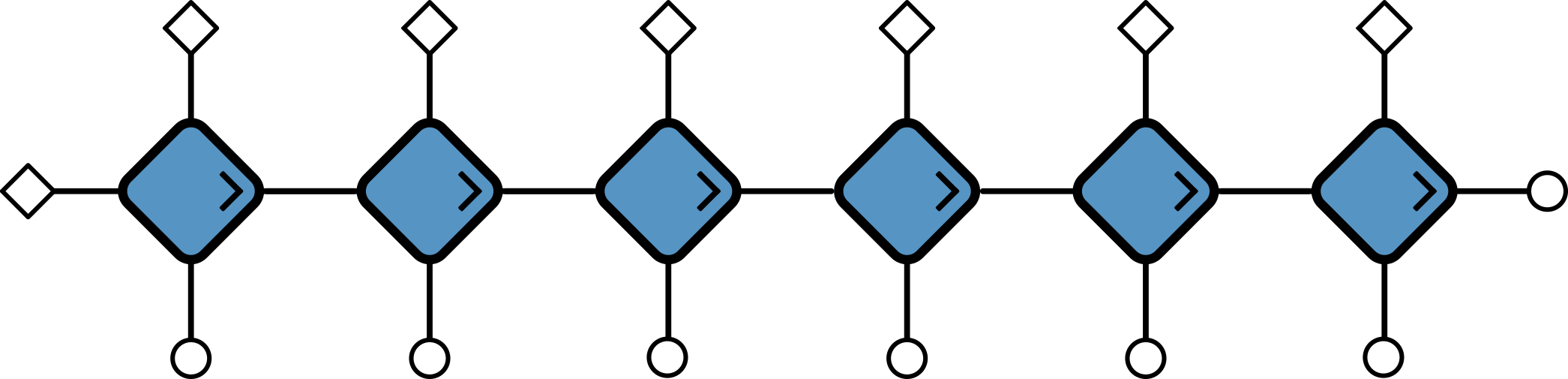}}}}_{L-1} 
\end{equation}
Crucially, this is nothing but the operator purity of a $(L-2)$-fold diagonally composed time evolution operator
\begin{equation}
    \operatorname{tr}\left[\rho_{A'C}^2\right] = \frac{1}{q^{2L}}\,\vcenter{\hbox{\includegraphics[width=0.3\columnwidth]{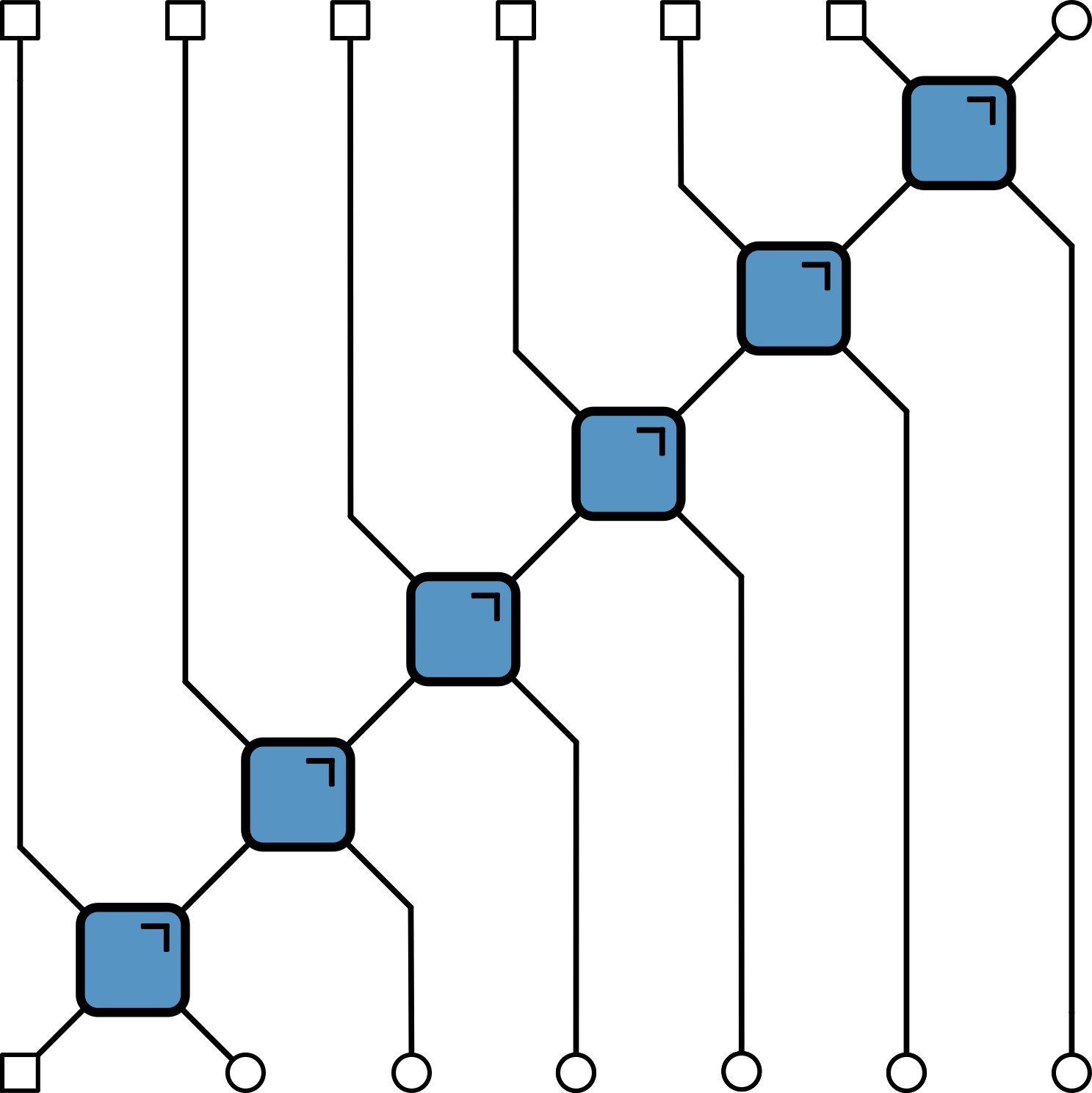}}} \equiv \frac{1}{q^L}B_{L-1},
\end{equation}
which is itself directly related to the quantity $B_{L-1}$ introduced in Ref.~\cite{Rampp2023} to quantify the deviation from dual-unitarity.

We can relate $B_{L-1}$ directly to the decoding error as
\begin{equation}
    \delta(t=L-1) = B_{L-1}-1.
\end{equation}
We conclude that $B_{L-1}$ quantifies how well information can be transmitted over a distance $(L-1)$ along the light cone. For dual-unitary gates, $B_{L-1}=1$ for all $L$, and thus information transmission is perfect. For non-dual-unitary gates it was shown that $B_{L-1}\rightarrow q^2$ for $L\rightarrow\infty$. Thus, no information is transmitted and the dip is absent. For weakly broken dual unitarity perturbative arguments [App.~\ref{app:perturbed}] show that information is preserved on the light cone up to a distance
\begin{equation}
    \frac{1}{\ell} \sim -\log\left(\frac{q^2 E(U)}{q^2-1}\right),
\end{equation}
with  $E(U) = (q^2-B_1)/q^2$ the operator entanglement of the gate, which is maximized for dual-unitary gates~\citep{Rather2020}. Beyond this scale the breaking of dual-unitarity becomes apparent and the information is situated inside the geometric light cone.

\section{Integrable dynamics}
\label{sec:integrable}

\begin{figure}[t]
  \centering
  \includegraphics[width = 0.7\textwidth]{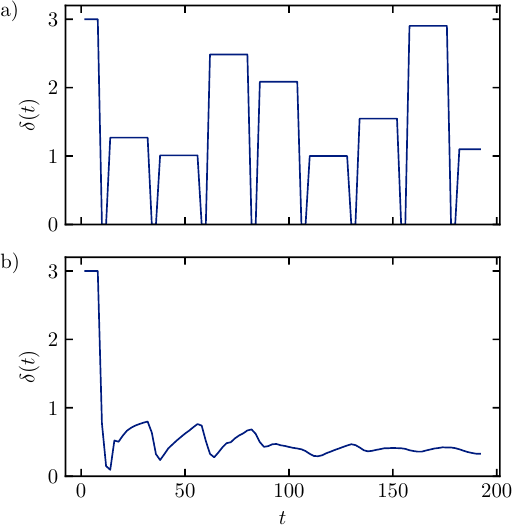}
  \caption{Decoding error as function of time in interacting integrable models (system size $L=12$, subsystem $L_D=2$, open boundary conditions). (a) \emph{Dual-unitary circuit.} The DUC displays persistent revivals of perfect decoding at odd multiples of system size owing to the non-dispersiveness of quasiparticles. (b) \emph{Non-dual-unitary circuit.} At early times the quasiparticles are still sufficiently localized that pronounced dips are visible. For long times these are washed away and saturation sets in. The gates are of the form $U=\exp\left(-i(J_{xy}(XX+YY) +J_z ZZ)\right)$. For (a) $J_{xy}=\pi/4$ and $J_z/J_{xy}=\sqrt{2}/2$, and for (b) $J_{xy}=\pi/5$ and $J_z/J_{xy}=\sqrt{2}/2$.}
  \label{fig:du_iint}
\end{figure}

\begin{figure}[t]
  \centering
  \includegraphics[width = 0.3\textwidth]{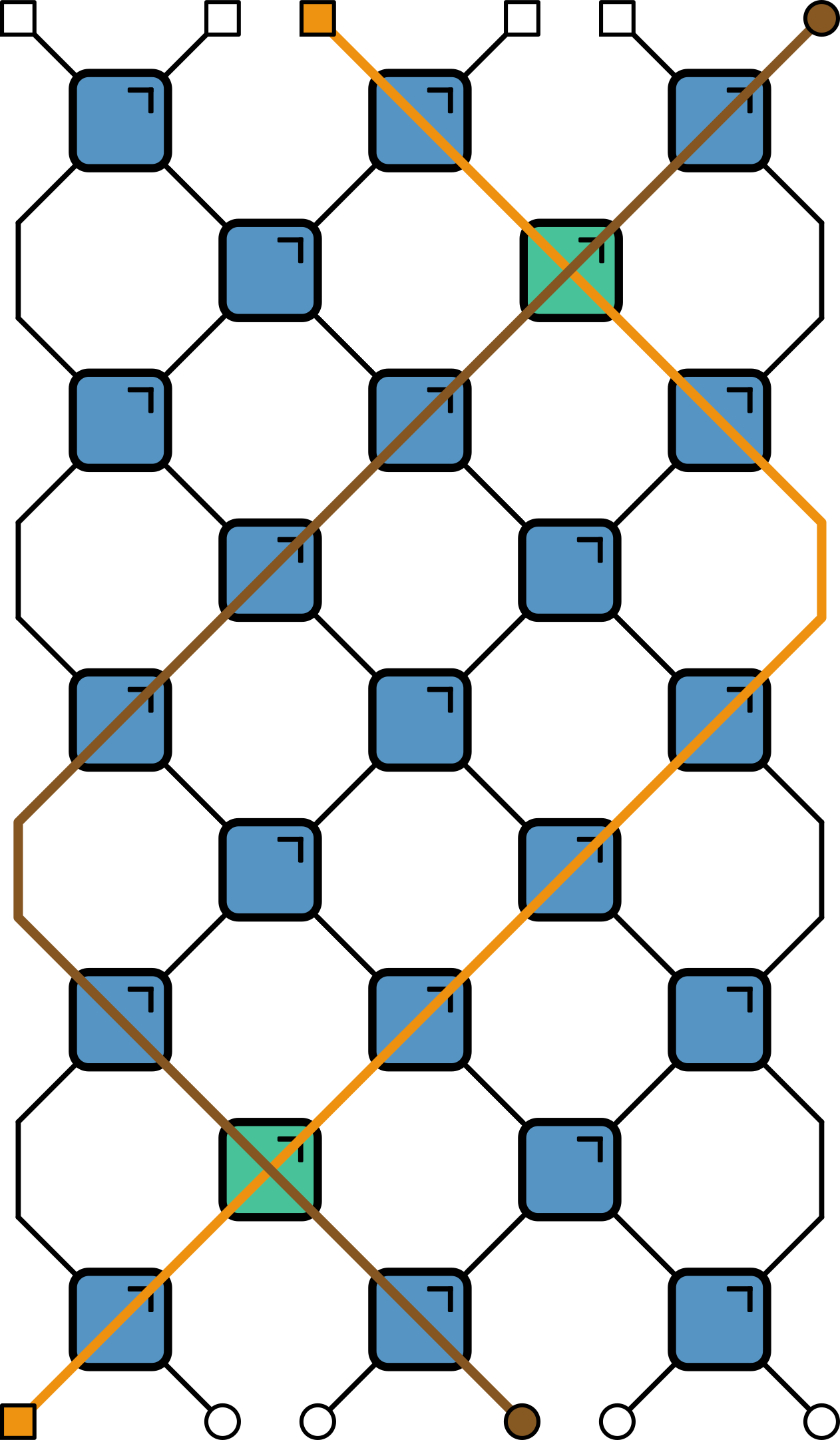}
  \caption{Quasiparticle picture of Hayden-Preskill recovery in interacting integrable DUCs. The information transmission is governed by the scattering processes (green) between the quasiparticle ray outgoing from $A$ (orange) and the quasiparticle ray incoming into $D$ (brown).}
  \label{fig:qp_scattering}
\end{figure}

As a contrast to the models of chaotic (non-integrable) dynamics studied above, in the following we investigate models of integrable dynamics. While a precise definition of quantum integrability does not exist, integrability is commonly associated with the presence of an extensive number of (quasi-)local conservation laws~\citep{Caux2011}. These conservation laws heavily constrain the dynamics and lead to non-ergodic behavior and anomalous hydrodynamics~\citep{Rigol2007,Essler2016,CastroAlvaredo2016,Bertini2021}. A precise understanding of the capacity of integrable systems to scramble information is, however, lacking so far. (For some investigations, see Refs.~\citep{Lin2018,Dora2017,Gopalakrishnan2018,Dubail2017,Alba2019,Bertini2020a,LopezPiqueres2021,Medenjak2022}.)
In the following we focus on a class of unitary circuit models that is integrable by Bethe Ansatz techniques, the Trotterized XXZ chain~\citep{Vanicat2018,Ljubotina2019,Claeys2022b}. We give exact solutions for the decoding error when the Trotter step is tuned to the dual-unitary point, and develop a quasiparticle picture to interpret these. Away from this special point we turn to numerical investigation and find qualitative differences to non-integrable dynamics for finite system sizes. In particular, integrable circuits appear to scramble information less efficiently. Finally, we comment on a possible generalization of the quasiparticle picture of scrambling to non-dual-unitary integrable circuits. Since the effect of scrambling is expected to be most pronounced in interacting systems we focus on the Trotterized XXZ chain away from the points where the dynamics can be mapped to that of noninteracting (free) fermions.

\subsection{Integrable dual-unitary circuits}

The combination of integrability and dual-unitarity is special, as it constrains quasiparticles to move with maximal velocity~\citep{Bertini2020a}. Quasiparticles in integrable DUCs are thus non-dispersive, even in the interacting case. This restriction has interesting physical consequences. As all quasiparticles occupy the same position at a given point in time, perfect recovery is possible, whenever this point lies in subsystem $D$, leading to perfect revivals of perfect recovery at arbitrarily long times. Moreover, the non-dispersiveness allows to analyze the scattering of all quasiparticles collectively, implying a simple expression for the information transmission in terms of the quasiparticle scattering matrix.

\begin{figure}[h]
  \centering
  \includegraphics[width = 0.75\textwidth]{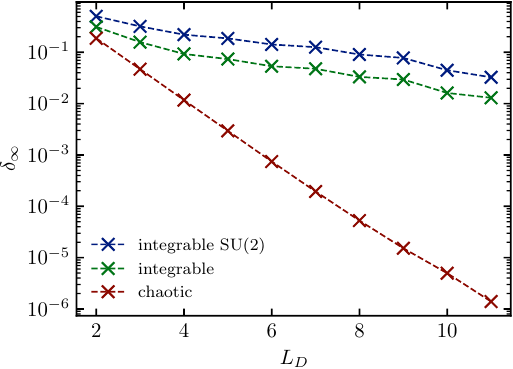}
  \caption{Scaling of $\delta_\infty$ with subsystem size $L_D$ in integrable (with and without $SU(2)$ symmetry) and chaotic models (system size $L=12$, periodic boundary conditions). The saturation value in the integrable models is consistently larger than in the chaotic model by several orders of magnitude. In the presence of $SU(2)$ symmetry we observe an increased $\delta_\infty$. The form of the scaling with $L_D$ is consistent with an exponential. The integrable gates both have $J_{xy}=\pi/5$. $J_z/J_{xy}=1$ for the $SU(2)$-symmetric gate and $J_z/J_{xy}=\sqrt{2}/2$ for the non-symmetric gate. The chaotic gate coincides with the gate used in Fig.~\ref{fig:du_erg}(b).}
  \label{fig:gen_iint}
\end{figure}

In the following we show explicitly that this picture holds in a large class of interacting integrable DUCs, including the Trotterized XXZ chain at the dual-unitary point. We assume that scattering processes at adjacent bonds commute, i.e., defining the scattering operator $V:=US$, that $V_{12}V_{23}=V_{23}V_{12}$ holds. The independence of the scattering processes leads to a significant simplification of the problem. For simplicity we also assume that the gate is non-chiral, i.e., commutes with the swap. Both these conditions are met on the dual-unitary XXZ line, where $V=\,\vcenter{\hbox{\includegraphics[height=0.03\textheight]{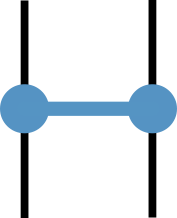}}}\,=e^{-i\frac{\pi}{4}\left(1 + (J-1)ZZ\right)}$, ($J=J_z/J_{xy}$ with $J_{xy}=\pi/4$). The factorization of the total gate reads 
\begin{equation}
   U_{\mathrm{XXZ}}\left(J_{xy}=\frac{\pi}{4}\right) = \,\vcenter{\hbox{\includegraphics[height=0.05\textheight]{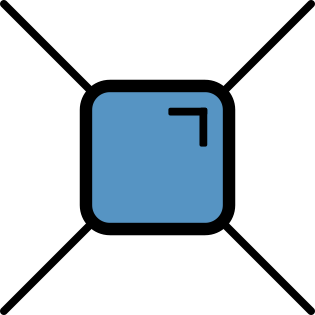}}}\, = \,\vcenter{\hbox{\includegraphics[height=0.05\textheight]{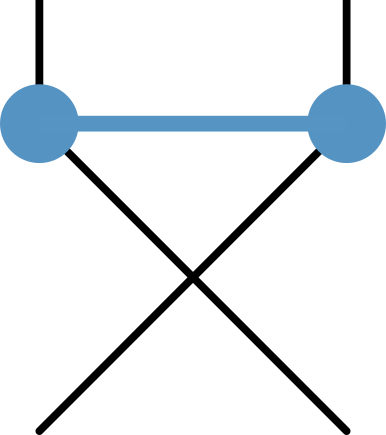}}}\,. \label{eq:du_factorization}
\end{equation}
This factorization enables the rewriting of the circuit as
\begin{equation}
    \,\vcenter{\hbox{\includegraphics[height=0.2\textheight]{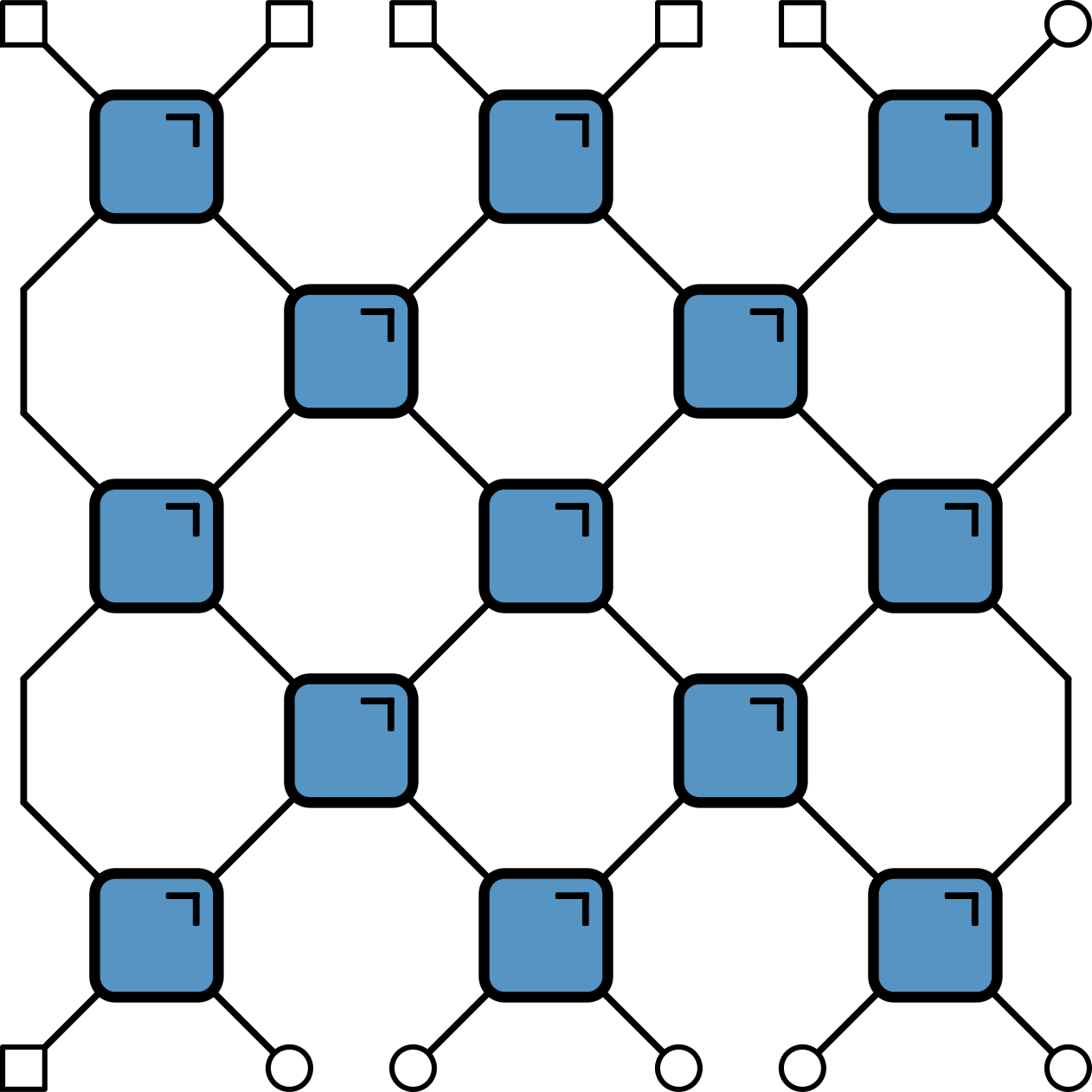}}}\, =  \,\vcenter{\hbox{\includegraphics[height=0.2\textheight]{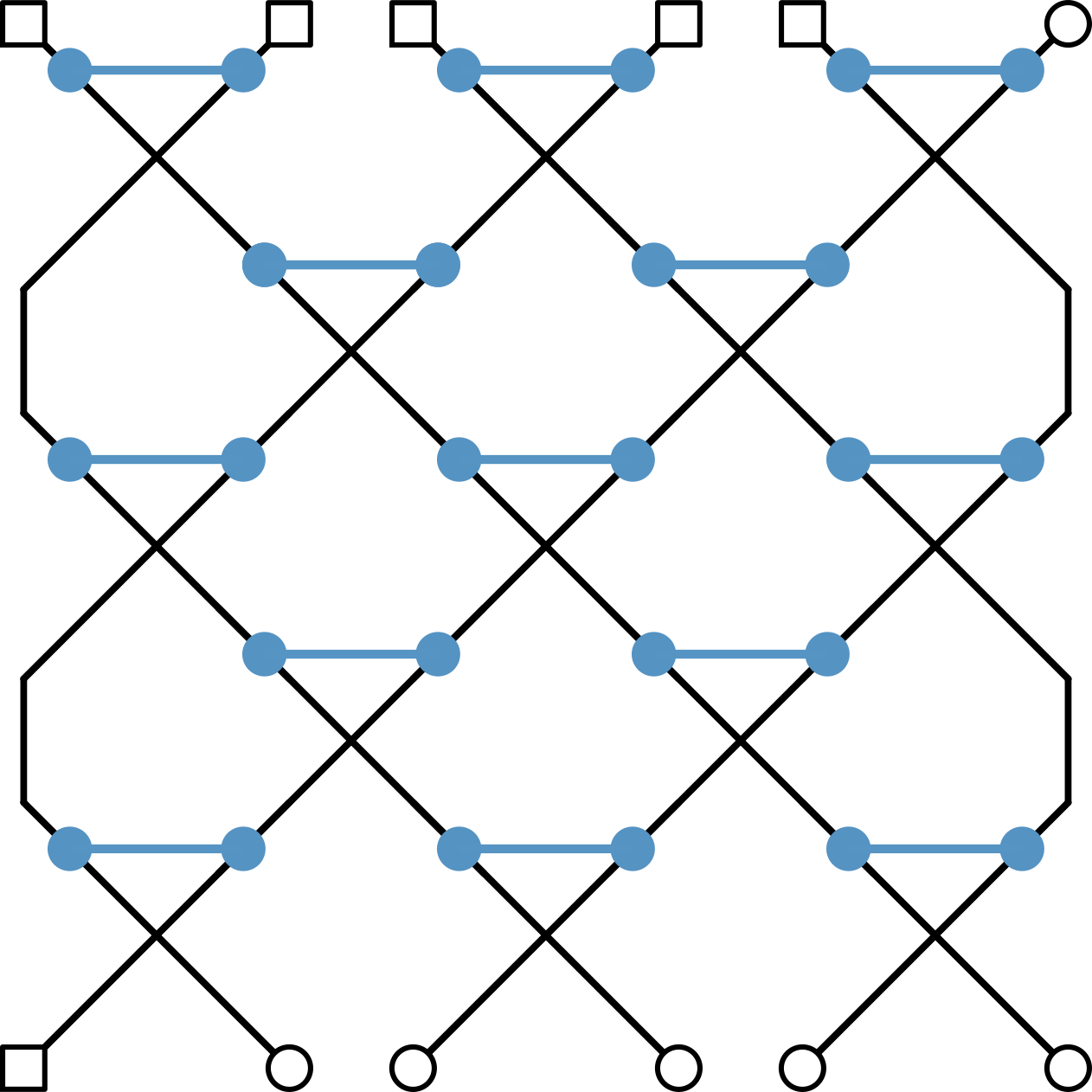}}}\,.
\end{equation}
Since the scattering gates $V$ all commute, it suffices to keep track of which pairs of gates a scattering operator has acted on. Consequently, the circuit can be factorized into a part due to SWAP gates and a part due to (now non-local) two-body interactions (we concentrate on the simplest case $L_A=L_D=1$, however, the arguments can be readily extended to different subsystem sizes)
\begin{align}
    \operatorname{tr}[\rho_{A'C}^2] = q^{-2L} \, \vcenter{\hbox{\includegraphics[height=0.1\textheight]{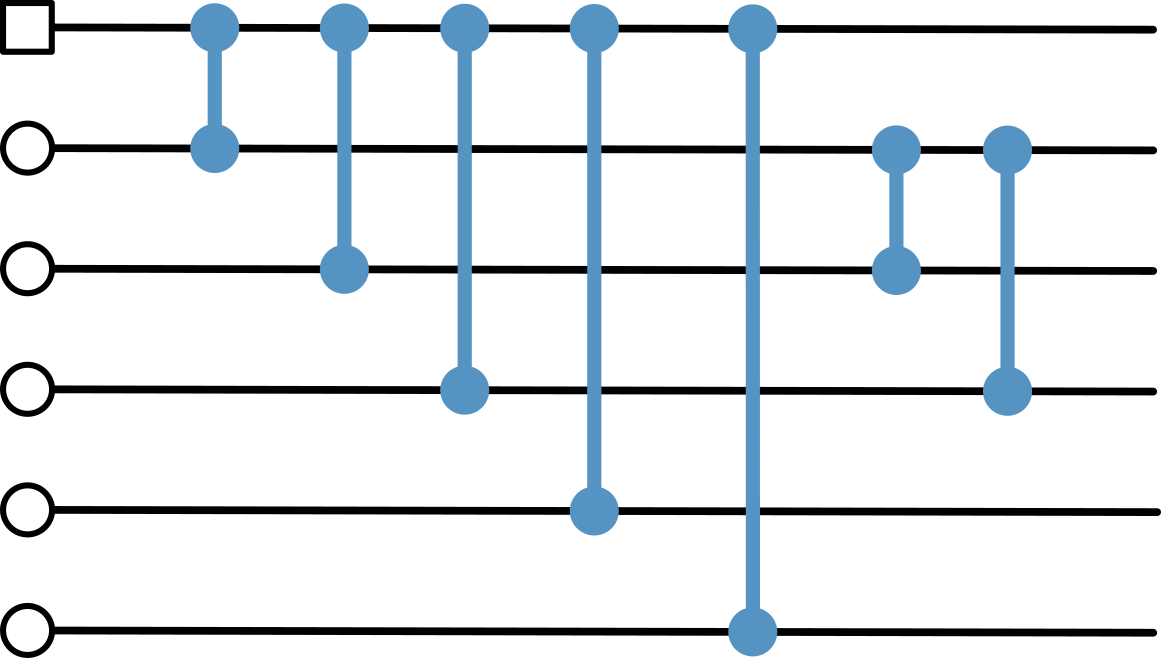}}}\,\dots\,\vcenter{\hbox{\includegraphics[height=0.1\textheight]{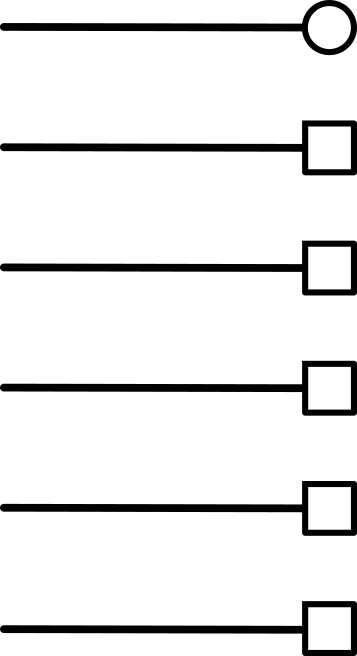}}}.
\end{align}
The SWAP gates now merely lead to a relative shift of the final state (in the replica picture) with respect to the initial state. For $t=(2k+1)L$ the initial and final permutations are perfectly anti-aligned. All but two scattering gates can be canceled due to unitarity ($\vcenter{\hbox{\includegraphics[height=0.02\textheight]{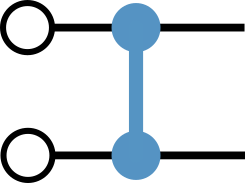}}}=\vcenter{\hbox{\includegraphics[height=0.02\textheight]{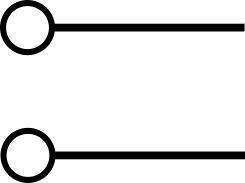}}},\,\vcenter{\hbox{\includegraphics[height=0.02\textheight]{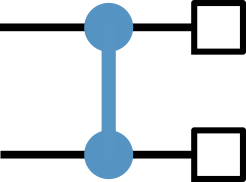}}}=\vcenter{\hbox{\includegraphics[height=0.02\textheight]{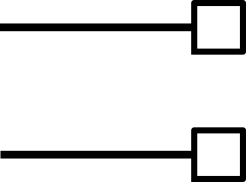}}}$), and the final two gates can be canceled due to T-duality of $V$ ($\vcenter{\hbox{\includegraphics[height=0.02\textheight]{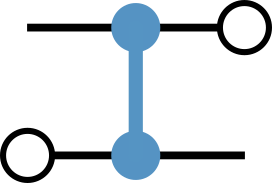}}}=\vcenter{\hbox{\includegraphics[height=0.02\textheight]{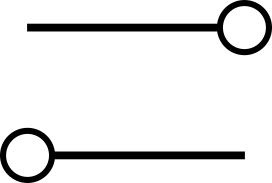}}}$). It follows that $\operatorname{tr}[\rho_{A'C}^2]=q^{-L}\,\implies\,\delta=0$, implying perfect recovery.

For the intermediate time steps, the permutations in the initial and final state are not perfectly anti-aligned,
\begin{align}
    \operatorname{tr}[\rho_{A'C}^2] = q^{-2L} \, \vcenter{\hbox{\includegraphics[height=0.1\textheight]{Figs/iint_unraveled_1.png}}}\,\dots\,\vcenter{\hbox{\includegraphics[height=0.1\textheight]{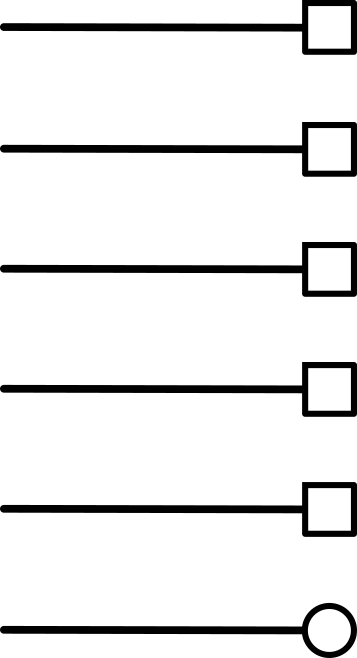}}} = q^{-L-2} \, \vcenter{\hbox{\includegraphics[height=0.0265\textheight]{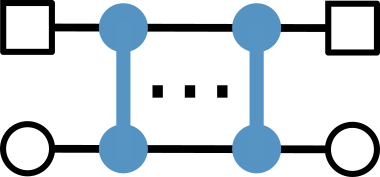}}}.
    \label{eq:inter_bubble}
\end{align}
There remain two anti-aligned states, namely the qudit initially in $A$ and the qudit in $D$ at the final time. Unitarity allows to cancel all gates that do not act on the pair $AD$, i.e., that do not directly couple the quasiparticles emanating from $A$ with those reaching $D$ at the final time. After passing through the system $2k-1$ times with $k\in\mathbb{N}$, there are $2k$ such scattering processes. Therefore the total scattering operator between $A$ and $D$ is $V^{2k}=(US)^{2k}$. The diagram appearing on the right-hand side of Eq.~\eqref{eq:inter_bubble} can be related to the linear operator entanglement of $V^{2k}$, $E(V^{2k})$, as
\begin{equation}
    \delta(t) = q^2\left(E_{\mathrm{max}}-E((US)^{2k})\right), \quad (2k-1)L < t < (2k+1)L, \quad E_{\mathrm{max}} = 1 - \frac{1}{q^2}.
    \label{eq:intermediate}
\end{equation}
This equation directly relates the value of the decoding error between the perfect revivals to the total scattering operator after $2k$ scattering processes. It can be interpreted as information being transmitted \emph{via} scattering of quasiparticles from Alice's light ray to Bob's. For the dual-unitary XXZ model, the sequence of values $E((US)^{2k})$ is periodic if $J=J_z/(\pi/4)$ is rational, and aperiodic if it is not. Note that there are three special points where the evolution is free: the XX point where $J_z=0$ and the Heisenberg points where $J_z=\pm \pi/4$. For these gates $\delta(t)=q^2-1$ between revivals, signalling the absence of interaction effects. 

In Fig.~\ref{fig:du_iint}(a) we show numerical results that are in perfect agreement with this picture.

\subsection{General integrable circuits}

In general interacting integrable circuits the situation is more complicated. While an exact solution is no longer available, we observe qualitative differences to the non-integrable case in finite-size systems. In particular, for open boundary conditions at early times there are pronounced dips of the decoding error close to odd multiples of the system size, Fig.~\ref{fig:du_iint}(b). We interpret these as being due to quasiparticles connecting $A$ and $D$.

In Fig.~\ref{fig:gen_iint} the late-time saturation of $\delta$ in integrable interacting circuits with and without $SU(2)$ symmetry is compared with results from a generic non-integrable circuit. Here, we use periodic boundary conditions in a chain of $L=12$ qubits to allow the decoding error to relax more quickly. While the non-integrable circuit clearly shows exponential decay as a function of subsystem size $L_D$, for the integrable circuits the data are inconclusive. One feature that can be observed, however, is that for finite-size chains $\delta_\infty$ in integrable circuits is consistently larger (for sufficiently large $L_D$ the discrepancy is several orders of magnitude) than in non-integrable circuits, meaning that finite-size integrable circuits scramble information less effectively than non-integrable circuits. We also observe that the $SU(2)$-symmetric circuit scrambles information less effectively than the non-symmetric circuit. For relatively small $L_D$ the decay of $\delta_\infty$ with $L_D$ is consistent with a power law $\delta_\infty\sim L_D^{-\kappa}$, where $\kappa\approx 1.2\, (1.5)$ for $SU(2)$-symmetric (non-symmetric) circuits (see App.~\ref{app:numerics})). However, we stress that we can only access small system sizes, such that our data are inconclusive.

In the following we attempt to generalize the quasiparticle picture from the exact dual-unitary solution to the general case. In the original formulation of the HP protocol the initial state is a collection of Bell states. Such a state corresponds to a collection of pairs of quasiparticles of the integrable model~\citep{Calabrese2020}. However, in the HP setup only one partner of the pair is subjected to non-trivial unitary dynamics. The relevant entanglement is thus carried by quasiparticles in the ancillae ($A'$, $B'$) that are entangled with quasiparticles that are in the appropriate subsystem ($D$, $C$) at time $t$. As every quasiparticle undergoing non-trivial dynamics has an entangled partner in the ancilla spaces we only have to consider the dynamics of single quasiparticles, instead of pairs as in the usual investigations of quench dynamics.

How would a formula for the decoding error based on this quasiparticle picture look like? In order to avoid complications arising from the R\'{e}nyi entropy~\citep{Calabrese2020}, we focus directly on the von Neumann mutual information which also bounds the fidelity of error correction~\citep{Schumacher2002}
\begin{equation}
    I(A:C) = S_A + S_C - S_{AC}.
\end{equation}
Keep in mind that the entropies are \emph{operator space entanglement entropies}. Due to unitarity we can simplify $S_A=L_A\log(q)$ and $S_C=L_C\log(q)$. It remains to compute $S_{AC}$. Based on the arguments above we conjecture the expression
\begin{align}
    S_{AC} =& \sum_{\eta=\pm} \int_{x_A\in A}\rmd x_A\int_{x_D\in D}\rmd x_D\int\rmd p s(p) \delta(x_D-x_A-\eta v(p)t) +\ldots \nonumber\\  &\ldots+ \sum_{\eta=\pm} \int_{x_B\in B}\rmd x_B\int_{x_C\in C}\rmd x_C\int\rmd p s(p) \delta(x_C-x_B-\eta v(p)t).
\end{align}
Here $s(p)$ is the entropy carried by quasiparticles of momentum $p$. Note the presence of only one delta-function per integration, reflecting that one partner of the entangled pair is stored in ancilla degrees of freedom. Verifying this conjecture in more general integrable dynamics will be the subject of future work.

\section{Conclusion and outlook}

We have investigated the capacity of the Hayden-Preskill protocol to differentiate chaotic and integrable dynamics, and presented several exact results for paradigmatic models both with and without randomness. In chaotic models we found universal relaxation of the decoding error to a universal value. We derived this value, which can be understood as a consequence of random-matrix like behavior, in chaotic dual-unitary circuits without any randomness. In dual-unitary chaotic models we additionally identified a pronounced dip in the decoding error at early time, corresponding to perfect decoding. Dual-unitary circuits can thus transport information perfectly over arbitrary distances, despite their chaoticity. In interacting integrable models, our numerical results suggest that the decoding error relaxes to a larger value, expressing the fact that integrable circuits are less effective at scrambling information. For interacting integrable dual-unitary circuits we derived the appearance of revivals of perfect decoding at evenly spaced intervals in time, indicating that these models keep information intact, even after scattering with boundaries. We linked this property to the presence of dispersionless quasiparticles, suggesting that conformally invariant models might display similar dynamics~\citep{Bernard2016}.

Beyond dual-unitary circuits however, a precise understanding of scrambling in interacting integrable models is still missing. The knowledge of exact eigenstates through the Bethe equations does not provide any simplification, and generalized hydrodynamics~\citep{CastroAlvaredo2016} is not straightforwardly applicable in the regime of interest (for OTOCs this corresponds to the region deep inside the light cone). The development of an analogous form of generalized hydrodynamics for operator space entanglement growth would remedy this problem.
Furthermore, the influence of symmetries~\citep{Nakata2023a}, Hilbert-space fragmentation, or localization on Hayden-Preskill recovery pose interesting questions.

Finally, we hope that the analytical techniques to treat dynamics of finite size systems presented here will contribute to shed light on other questions of quantum dynamics. In particular, the technique introduced for chaotic dual-unitary circuits can be straightforwardly adapted to treat out-of-time-order correlators and local operator entanglement entropies. For integrable dual-unitary circuits satisfying a factorization like Eq.~\eqref{eq:du_factorization} with a commuting scattering part the technique is even more powerful and enables in principle the computation of any observable of interest that can be represented as a tensor-network diagram.

\acknowledgements

We acknowledge useful discussions with Youenn Le Gal, Cathy Li, Roderich Moessner, Suhail A. Rather, Subhayan Sahu, Zack Weinstein, and Tianci Zhou.

\appendix

\section{Details on the calculation of the decoding error}

\subsection{Maximally chaotic dual-unitary circuits}
\label{app:duc}

In this section we collect some details on the calculation of $\delta$ in maximally chaotic DUCs. Namely, we discuss the necessary overlaps with the triangular boundary conditions in Eq.~\eqref{eq:circuit-inter} and the generalization of Eq.~\eqref{eq:circuit-inter} to general values of $L_A,L_D$.

We begin with $L_A=L_D=1$, where we compute the overlaps in the basis of product states. Graphically,
\begin{align}
    \left( \triangleleft_{n} \big| n,k \right) = \frac{1}{q^n} \, \vcenter{\hbox{\includegraphics[height=0.2\textheight]{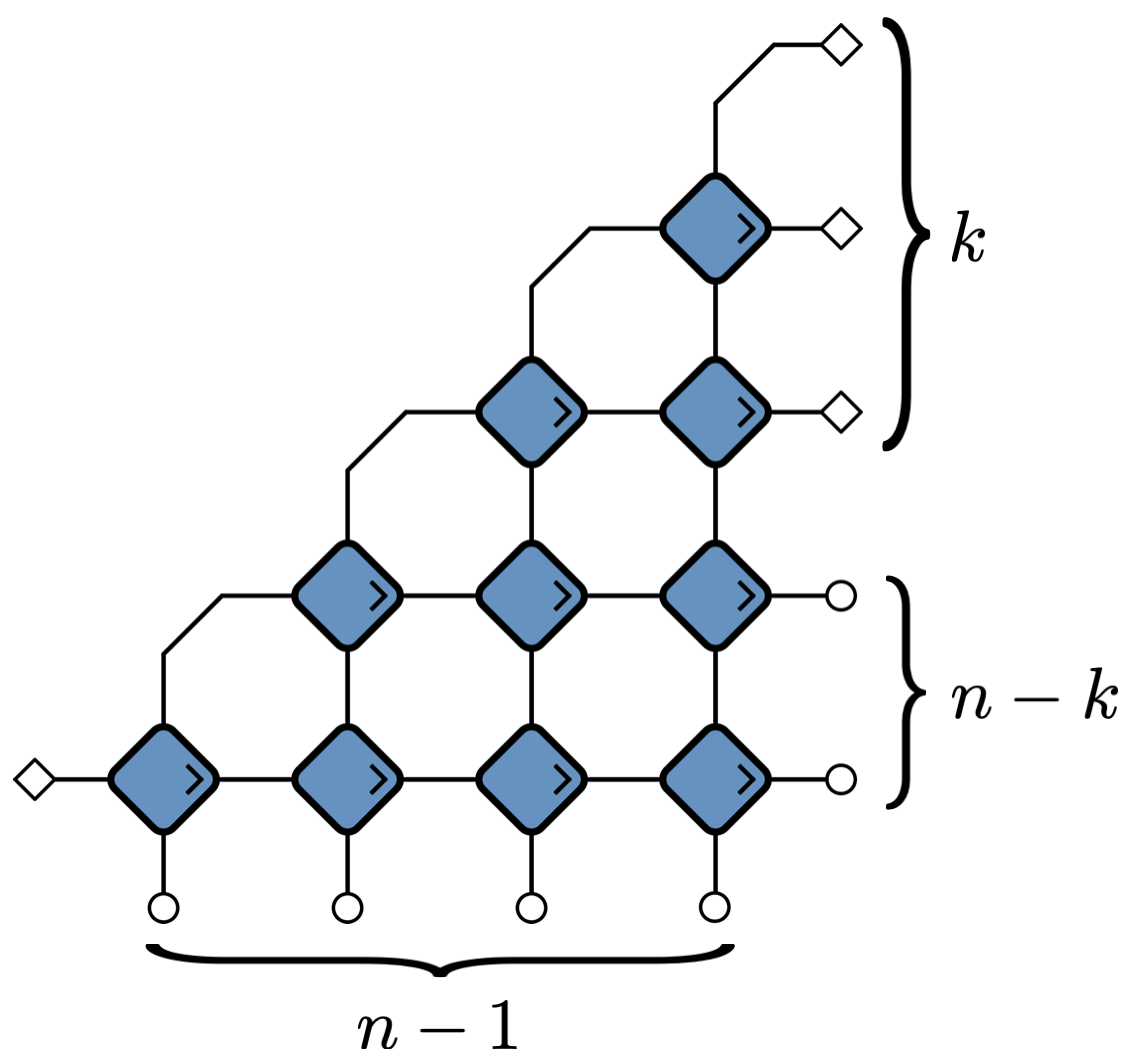}}},\quad \left( n,k | \triangleright_{n}\right)= \frac{1}{q^n} \, \vcenter{\hbox{\includegraphics[height=0.2\textheight]{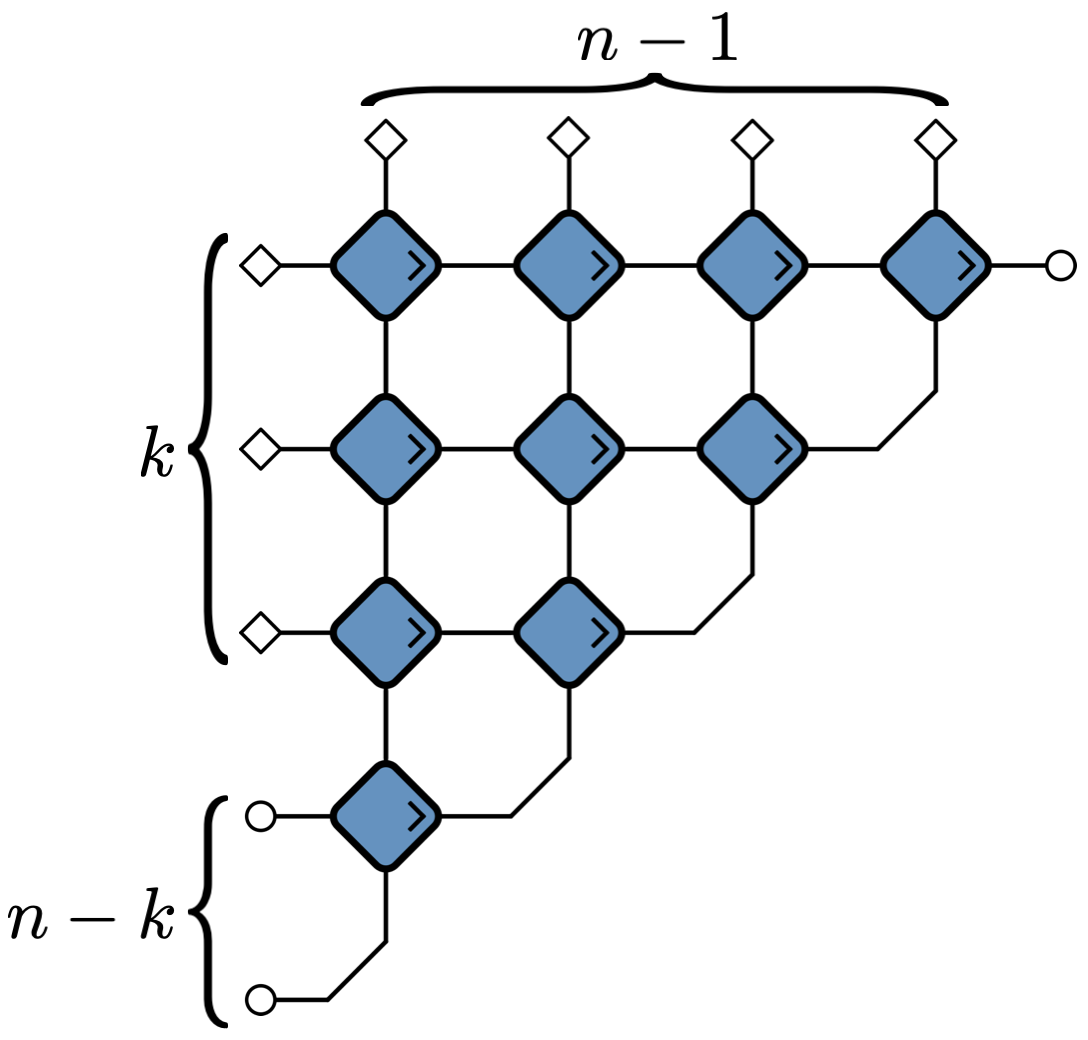}}}.
\end{align}
Using unitarity and dual-unitarity yields 
\begin{equation}
    \left( \triangleleft_{n} \big| n,k \right) = \begin{cases} q & k=0, \\ q^{k-1} & k\geq1, \end{cases} \qquad \left( n,k | \triangleright_{n}\right) = \begin{cases} q^{n-k-1} & k\leq n-1, \\ q & k=n. \end{cases}
\end{equation}
Notice the $k\longleftrightarrow n-k$ symmetry between the two expressions that is also apparent graphically. Eq.~\eqref{eq:triang_overlaps} follows directly from the above.

Finally, we consider the general case where both $L_A$ and $L_D$ are arbitrary. In this case we have
\begin{equation}
    \delta + 1 = \frac{1}{q^{n+L_D-L_A}} \left( \triangleleft_{n,L_A} \right| T_{n}^{L-n} \left| \triangleright_{n,L_D}\right).
\end{equation} 
The boundary conditions are modified as follows
\begin{align}
     ( \triangleleft_{n,L_A} | =  \, \vcenter{\hbox{\includegraphics[height=0.2\textheight]{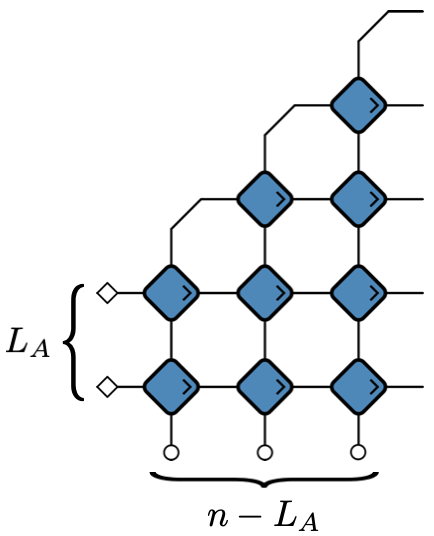}}}\,,\quad | \triangleright_{n,L_D})=  \, \vcenter{\hbox{\includegraphics[height=0.2\textheight]{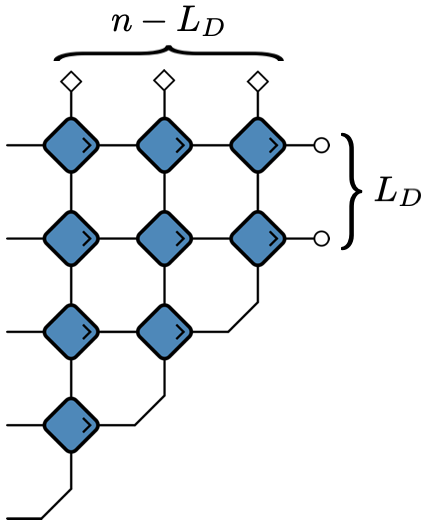}}}.
\end{align}
We compute the overlaps with the modified left boundary condition
\begin{equation}
    \left( \triangleleft_{n,L_A} | n,k \right) = q^{\abs{k-L_A}} \,\implies\, \left( \triangleleft_{n,L_A} \Big| \overline{n,k} \right) = \begin{cases} q^{L_A} & k=0, \\ 0 & 1\leq k\leq L_A, \\ \frac{q^{k+1-L_A}}{\sqrt{q^2-1}}\left(1-\frac{1}{q^2}\right) & k>L_A. \end{cases}
\end{equation}
To compute the overlaps with the right boundary condtion we assume $L_D<2n$ (which is the relevant regime for saturation if $D$ remains finite in the TDL) and find 
\begin{align}
    &\left( n,k | \triangleright_{n,L_D}\right) = \begin{cases} q^{n-k-L_D} & k\leq n-L_D, \\ q^{L_D+k-n} & k>n-L_D, \end{cases}\\
    &\qquad\,\implies\, \left( \overline{n,k} \Big| \triangleright_{n,L_D}\right) = \begin{cases} q^{n-L_D} & k=0, \\ 0 & 1\leq k\leq n-L_D, \\ q^{L_D-1+k-n}\sqrt{q^2-1} & k>n-L_D. \end{cases}
\end{align}
Altogether,
\begin{align}
    \delta+1 &= \frac{1}{q^{n+L_D-L_A}} \left(q^{n+L_A-L_D} + \left(1-\frac{1}{q^2}\right)\sum_{k=\max(n+1-L_D,L_A+1)}^n q^{2k+L_D-L_A-n} \right), \nonumber\\
    & = \frac{1}{q^{L_D-L_A}} \left(q^{L_A-L_D} + \left(1-\frac{1}{q^2}\right) \frac{q^{1-L_A}}{q^{L_D-1}} \sum_{k=n+1-L_D}^n q^{2(k-n-(L_D-1))} \right), \nonumber\\
    & = 1 + \left(1-\frac{1}{q^{2L_A}}\right)\frac{1}{q^{2(L_D-L_A)}}.
\end{align}
For large enough $n$ the summation starts from $k=n+1-L_D$. This result shows that the prefactor of the exponential depends on the size of the input register. A smaller input can be decoded with higher fidelity. For large inputs the decoding error approaches the Yoshida-Kitaev bound. We can also state the result as
\begin{equation}
    \delta = \left(1-d_A^{-2}\right)\frac{d_A^2}{d_D^2}.
\end{equation}

\subsection{Perturbed dual-unitary circuits}
\label{app:perturbed}

We evaluate the TN in Eq.~\eqref{eq:circuit-inter}, where we replace the bulk gates by gates with weakly perturbed gates, such that they are no longer dual-unitary. The deviation from dual unitarity can be quantified with the operator entanglement $E(U)$ of the gate, which is maximized only for dual-unitary gates~\citep{Rather2020}. It is convenient to use $z_1 = 1- q^2 E(U)/(q^2-1)$ to quantify the deviation from maximal operator entanglement.

We follow Ref.~\citep{Rampp2023} and project the LCTM to the MCS. We work in the one-step approximation, which is exact in the large-$q$ limit. The LCTM is then represented as the $(n+1)\times(n+1)$-matrix
\begin{equation}
    T_n|_{\mathrm{MCS}} \approx \begin{pmatrix}
        1 & \sqrt{q^2-1}z_1 &  & & \\
         & 1-z_1 & qz_1 & & \\
         & & \ddots & \ddots & \\
         & & & 1-z_1 & qz_1  \\
         & & & & 1-z_1 
    \end{pmatrix}.
\end{equation}
We denote the $j$-th column of this matrix by $\rket{u_j}$, $j=0,\ldots,n$. We find
\begin{equation}
    \delta(t) -\delta_\infty +1 \approx \frac{1}{q^n}\sum_{\nu=0}^{L-n}\binom{L-n}{\nu}\sum_{j_1,j_\nu} \binom{\nu-1}{j_\nu-j_1} (-1)^{j_\nu-j_1+\nu-1} q^{j_\nu-j_1} z_1^{\nu-1}\rbraket{\triangleleft_{n}}{u_{j_1}}\rbraket{j_\nu}{\triangleright_{n}}.
\end{equation}
The overlaps with the boundary conditions can be readily evaluated as
\begin{subequations}
    \begin{align}
        \rbraket{\triangleleft_{n}}{u_j} = q^2 z_1 \delta_{j,1}, \qquad
        \rbraket{j}{\triangleright_{n}} = \sqrt{q^2-1}\delta_{j,n}.
    \end{align}
\end{subequations}
Using
\begin{equation}
    F_{z}(n,m) = (-1)^n \sum_{\nu=n}^m \binom{m}{\nu} \binom{\nu-1}{n-1} (-z)^\nu,
\end{equation}
we obtain the result Eq.~\eqref{eq:perturbed} from the main text.

\subsection{Random unitary circuits}
\label{app:ruc}

\subsubsection{Eigenvectors of transfer matrix}

The leading left eigenvectors of the transfer matrix are found by solving the recurrence Eq.~\eqref{eq:leadingleftrec}. 
We first determine the roots of its characteristic polynomial
\begin{equation}
    p(\theta) = \theta^2 - \left(z^{-1}-2\right)\theta + 1,
\end{equation}
which has roots 
\begin{equation}
    \theta_\pm  = q^{\pm2}.
\end{equation}
Leading left eigenvectors hence have coefficients of the form $a_j = c_+\theta_+^j + c_-\theta_-^j$. Requiring biorthogonality with $\{\rket{r_1},\rket{r_2}\}$ yields the conditions 
\begin{align}
    \rket{\ell_1}: \quad a_N = 0 \,\, \mathrm{and} \,\, a_0 = 1, \nonumber\\
    \rket{\ell_2}: \quad a_0 = 0 \,\, \mathrm{and} \,\, a_N = 1. \nonumber
\end{align}
The solution Eq.~\eqref{eq:leadingleft} follows.

It is in fact possible to solve for all eigenvectors. The solutions to the left-eigenvector equation at $\abs{\lambda}<1$ are standing waves of the form
\begin{equation}
    a_j = a\sin(k j ).
\end{equation}
The condition $a_N=0$ leads to the quantization condition
\begin{equation}
    k = \frac{\pi n}{N}, \quad n\in\mathbb{N}. \label{eq:quantization}
\end{equation}
To obtain the spectrum we consider the characteristic polynomial
\begin{equation}
    p(\theta) = \theta^2 - \left(\lambda z^{-1}-2\right)\theta + 1.
\end{equation}
Defining $\Omega:=(\lambda z^{-1}-2)/2$ the roots read
\begin{equation}
    \theta_\pm = \Omega \pm \sqrt{\Omega^2-1}.
\end{equation}
Oscillating solutions exist for $\abs{\Omega}<1$, where
\begin{equation}
    \theta_\pm = \Omega \pm i \sqrt{1-\Omega^2} \equiv e^{ik}.
\end{equation}
Inserting the quantization condition for $k$ yields Eq.~\eqref{eq:spectrum}.

The corresponding right eigenvectors can similarly be constructed. While the recurrence relation in the bulk stays the same, the following relations hold on the boundary
    \begin{align}
        a_1 = \frac{\lambda-1}{z}a_0, \qquad
        a_2 = \frac{\lambda-2z}{z}a_1, \qquad
        a_{N-2} = \frac{\lambda-2z}{z}a_{N-1}, \qquad
        a_{N-1} = \frac{\lambda-1}{z}a_N.
    \end{align}
Consider the Ansatz $a_j=a\sin(k_n j+\phi),\,1\leq l\leq N-1$ with $k_n = \pi n/N$ as above. This Ansatz has to be consistent with the boundary conditions on both sides. Using that $a_{N-j}=-a_{-j}$ we consider the ratios $a_{N-2}/a_{N-1}$ and $a_2/a_1$ that should both equal $2\Omega_n=\lambda_n z^{-1} -2$. We write
\begin{equation}
    \frac{a_{2}}{a_{1}} = \frac{\sin(2k_n+\phi)}{\sin(k_n+\phi)} = 2\Omega_n
\end{equation}
We recall that $\Omega_n=\cos k_n$ and hence the equation is solved for $\phi=0$. (Recall that $\sin(2\phi)/\sin(\phi)=2\cos\phi$.) Then, the other equation
\begin{equation}
    \frac{a_{N-2}}{a_{N-1}} = \frac{\sin(2k_n-\phi)}{\sin(k_n-\phi)} = 2\Omega_n
\end{equation}
is also fulfilled. The components of the right eigenvectors are thus given by
\begin{equation}
    \rbraket{j}{r_n} = \begin{cases}
        \frac{z}{\lambda_n-1} a \sin\left(\frac{\pi n}{N}\right), & j=0,N, \\
        a \sin\left(\frac{\pi n}{N}j\right), &\mathrm{otherwise}.
    \end{cases}
\end{equation}
It remains to biorthonormalize the eigenvectors. We recall that in the continuum the functions $\sqrt{2}\sin(k_nx)$ form an ONB on the interval $x\in[0,N]$. Hence, for large $N$ we choose the prefactor $a=\sqrt{2}$ for both left- and right-eigenvectors.

\subsubsection{Finite-size corrections}

To compute the saturation value we need the overlaps
\begin{align}
    \rbraket{\ell_i}{R} = a_{L_C/2}^{(i)} &= \begin{cases} \frac{q^{L_D}-q^{-L_D}}{q^L-q^{-L}}, & i=1,\\ \frac{q^{L-L_D}-q^{L_D-L}}{q^L-q^{-L}}, & i=2,\end{cases}.
\end{align}
Superficially it seems as if the overlaps might be different for $L_C$ odd, but the prefactors indeed cancel and the result is the same.
Moreover,
\begin{align}
        \rbraket{L}{r_1} = q^{-L_A}, \quad \rbraket{L}{r_2} = q^{L_A-L}.        
\end{align}
This yields
\begin{equation}
    \delta + 1 = q^{L-L_D}\frac{q^{L_D}-q^{-L_D}}{q^L-q^{-L}} +  q^{2L_A-L_D}\frac{q^{L-L_D}-q^{L_D-L}}{q^L-q^{-L}}.
\end{equation}
In the thermodynamic limit this reduces to the result Eq.~\eqref{eq:delta_inf}.

\subsubsection{Asymptotic relaxation}

In the following Eq.~\eqref{eq:relax} is derived. The integral Eq.~\eqref{eq:ruc_subleading} is evaluated asymptotically for $t/L\rightarrow\infty$, i.e. the situation corresponds to infinite time in a finite (but large, such that the continuum approximation is valid) system.

The exponent $\log\lambda( k)$ is maximal at $ k=0$ and the Taylor expansion reads
\begin{equation}
    \log\lambda( k) = \log(2z+2z\cos k) \approx \log4z - \frac{ k^2}{4}.
\end{equation}
The exponent is maximal at the boundary of the integration range. Introducing $\tau:=(t-1)/2$
\begin{align}
    \mathcal{I}_1 &:= \int_0^\pi\rmd k \lambda( k)^\tau \frac{\sin( k N_c)\sin(k)}{\lambda( k)-1} \approx (4z)^\tau \int_0^\infty\rmd k e^{-\tau\frac{ k^2}{4}}\left(\frac{N_C  k^2}{4z-1}+\mathcal{O}( k^4)\right) \nonumber\\ 
    &\approx  - \frac{\sqrt{\pi}}{1-4z} (4z)^\tau \frac{L}{\tau^{\frac{3}{2}}}, \\
    \mathcal{I}_2 &:= \int_0^\pi\rmd k \lambda( k)^\tau \frac{i(e^{2i k}-1)\sin( k N_c)}{(e^{i k }-q^2)(e^{i k }q^2-1)} \approx (4z)^\tau \int_0^\infty\rmd k e^{-\tau\frac{ k^2}{4}} \left(\frac{2N_C  k^2}{(q^2-1)^2}+\mathcal{O}( k^4)\right) \nonumber\\ 
    &\approx \frac{(4z)^\tau}{\tau^{3/2}} \frac{2\sqrt{\pi}L}{(q^2-1)^2}.
\end{align}
We have dropped terms suppressed in $L$. The individual contributions to the decoding error are
\begin{subequations}
    \begin{align}
        \delta_1(t) &= -\frac{\sqrt{8\pi z}}{(1-4z)q^{L_D}} \frac{L}{t^{\frac{3}{2}}} q^{L-v_E t}, \\
        \delta_2(t) &= -\frac{\sqrt{8\pi}q^{4+L_A}}{(q^2-1)^2 q^{L_D}\sqrt{z}} \frac{L}{t^{\frac{3}{2}}} q^{L-v_E t}.
    \end{align}
\end{subequations}
Altogether
\begin{equation}
    \delta(t)-\delta_\infty = \frac{4\sqrt{2\pi}D}{v_B^2} \frac{q^{2(L_A+1)}-1}{q^{L_D}} \frac{L}{t^{\frac{3}{2}}} q^{L-v_E t},
\end{equation}
where we have introduced the butterfly velocity $v_B=(q^2-1)/(q^2+1)$ and diffusion constant $D=\sqrt{z}/2$ of random circuits~\citep{Keyserlingk2018,Nahum2018}.

\clearpage

\section{Additional numerical results}
\label{app:numerics}

\begin{figure}[h!]
  \centering
  \includegraphics[width = 0.75\textwidth]{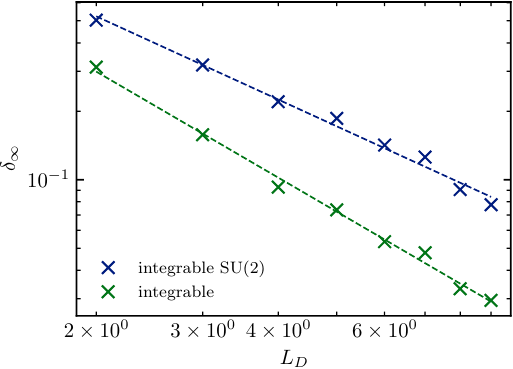}
  \caption{The decay of $\delta_\infty$ with $L_D$ in interacting integrable circuits is consistent with a power law, with the exponent differing between the $SU(2)$ symmetric and non-symmetric cases. The fitted powers are $\kappa\approx 1.2 (1.5)$ for $SU(2)$-symmetric (non-symmetric) circuits. We note that due to the small system sizes the data are inconclusive. The presented data are the same as the one used in Fig.~\ref{fig:gen_iint}.}
  \label{fig:power_law}
\end{figure}

\bibliographystyle{apsrev4-2}
\setcitestyle{numbers,open={[},close={]}}
\bibliography{haydenpreskill}

\begin{thebibliography}{106}%
\makeatletter
\providecommand \@ifxundefined [1]{%
 \@ifx{#1\undefined}
}%
\providecommand \@ifnum [1]{%
 \ifnum #1\expandafter \@firstoftwo
 \else \expandafter \@secondoftwo
 \fi
}%
\providecommand \@ifx [1]{%
 \ifx #1\expandafter \@firstoftwo
 \else \expandafter \@secondoftwo
 \fi
}%
\providecommand \natexlab [1]{#1}%
\providecommand \enquote  [1]{``#1''}%
\providecommand \bibnamefont  [1]{#1}%
\providecommand \bibfnamefont [1]{#1}%
\providecommand \citenamefont [1]{#1}%
\providecommand \href@noop [0]{\@secondoftwo}%
\providecommand \href [0]{\begingroup \@sanitize@url \@href}%
\providecommand \@href[1]{\@@startlink{#1}\@@href}%
\providecommand \@@href[1]{\endgroup#1\@@endlink}%
\providecommand \@sanitize@url [0]{\catcode `\\12\catcode `\$12\catcode
  `\&12\catcode `\#12\catcode `\^12\catcode `\_12\catcode `\%12\relax}%
\providecommand \@@startlink[1]{}%
\providecommand \@@endlink[0]{}%
\providecommand \url  [0]{\begingroup\@sanitize@url \@url }%
\providecommand \@url [1]{\endgroup\@href {#1}{\urlprefix }}%
\providecommand \urlprefix  [0]{URL }%
\providecommand \Eprint [0]{\href }%
\providecommand \doibase [0]{https://doi.org/}%
\providecommand \selectlanguage [0]{\@gobble}%
\providecommand \bibinfo  [0]{\@secondoftwo}%
\providecommand \bibfield  [0]{\@secondoftwo}%
\providecommand \translation [1]{[#1]}%
\providecommand \BibitemOpen [0]{}%
\providecommand \bibitemStop [0]{}%
\providecommand \bibitemNoStop [0]{.\EOS\space}%
\providecommand \EOS [0]{\spacefactor3000\relax}%
\providecommand \BibitemShut  [1]{\csname bibitem#1\endcsname}%
\let\auto@bib@innerbib\@empty
\bibitem [{\citenamefont {Ashcroft}\ and\ \citenamefont
  {Mermin}(1976)}]{Ashcroft1976}%
  \BibitemOpen
  \bibfield  {author} {\bibinfo {author} {\bibfnamefont {N.~W.}\ \bibnamefont
  {Ashcroft}}\ and\ \bibinfo {author} {\bibfnamefont {N.~D.}\ \bibnamefont
  {Mermin}},\ }\href@noop {} {\emph {\bibinfo {title} {Solid state physics}}}\
  (\bibinfo  {publisher} {Holt, Rinehart and Winston},\ \bibinfo {year}
  {1976})\ p.\ \bibinfo {pages} {826}\BibitemShut {NoStop}%
\bibitem [{\citenamefont {Anderson}(1984)}]{Anderson1984}%
  \BibitemOpen
  \bibfield  {author} {\bibinfo {author} {\bibfnamefont {P.~W.}\ \bibnamefont
  {Anderson}},\ }\href@noop {} {\emph {\bibinfo {title} {Basic notions of
  condensed matter physics}}}\ (\bibinfo  {publisher} {Benjamin/Cummings Pub.
  Co., Advanced Book Program},\ \bibinfo {year} {1984})\ p.\ \bibinfo {pages}
  {549}\BibitemShut {NoStop}%
\bibitem [{\citenamefont {Chaikin}\ and\ \citenamefont
  {Lubensky}(1995)}]{Chaikin1995}%
  \BibitemOpen
  \bibfield  {author} {\bibinfo {author} {\bibfnamefont {P.~M.}\ \bibnamefont
  {Chaikin}}\ and\ \bibinfo {author} {\bibfnamefont {T.~C.}\ \bibnamefont
  {Lubensky}},\ }\href@noop {} {\emph {\bibinfo {title} {Principles of
  condensed matter physics}}}\ (\bibinfo  {publisher} {Cambridge University
  Press},\ \bibinfo {year} {1995})\ p.\ \bibinfo {pages} {699}\BibitemShut
  {NoStop}%
\bibitem [{\citenamefont {Zeng}\ \emph {et~al.}(2019)\citenamefont {Zeng},
  \citenamefont {Chen}, \citenamefont {Zhou},\ and\ \citenamefont
  {Wen}}]{Zeng}%
  \BibitemOpen
  \bibfield  {author} {\bibinfo {author} {\bibfnamefont {B.}~\bibnamefont
  {Zeng}}, \bibinfo {author} {\bibfnamefont {X.}~\bibnamefont {Chen}}, \bibinfo
  {author} {\bibfnamefont {D.-L.}\ \bibnamefont {Zhou}},\ and\ \bibinfo
  {author} {\bibfnamefont {X.-G.}\ \bibnamefont {Wen}},\ }\href@noop {} {\emph
  {\bibinfo {title} {Quantum Information Meets Quantum Matter : From Quantum
  Entanglement to Topological Phases of Many-Body Systems}}}\ (\bibinfo
  {publisher} {Springer},\ \bibinfo {year} {2019})\ p.\ \bibinfo {pages}
  {364}\BibitemShut {NoStop}%
\bibitem [{\citenamefont {Georgescu}\ \emph {et~al.}(2014)\citenamefont
  {Georgescu}, \citenamefont {Ashhab},\ and\ \citenamefont
  {Nori}}]{Georgescu2014}%
  \BibitemOpen
  \bibfield  {author} {\bibinfo {author} {\bibfnamefont {I.}~\bibnamefont
  {Georgescu}}, \bibinfo {author} {\bibfnamefont {S.}~\bibnamefont {Ashhab}},\
  and\ \bibinfo {author} {\bibfnamefont {F.}~\bibnamefont {Nori}},\ }\href
  {https://doi.org/10.1103/revmodphys.86.153} {\bibfield  {journal} {\bibinfo
  {journal} {Reviews of Modern Physics}\ }\textbf {\bibinfo {volume} {86}},\
  \bibinfo {pages} {153} (\bibinfo {year} {2014})}\BibitemShut {NoStop}%
\bibitem [{\citenamefont {Preskill}(2018)}]{Preskill2018}%
  \BibitemOpen
  \bibfield  {author} {\bibinfo {author} {\bibfnamefont {J.}~\bibnamefont
  {Preskill}},\ }\href {https://doi.org/10.22331/q-2018-08-06-79} {\bibfield
  {journal} {\bibinfo  {journal} {Quantum}\ }\textbf {\bibinfo {volume} {2}},\
  \bibinfo {pages} {79} (\bibinfo {year} {2018})}\BibitemShut {NoStop}%
\bibitem [{\citenamefont {Deutsch}(1991)}]{Deutsch1991}%
  \BibitemOpen
  \bibfield  {author} {\bibinfo {author} {\bibfnamefont {J.~M.}\ \bibnamefont
  {Deutsch}},\ }\href {https://doi.org/10.1103/physreva.43.2046} {\bibfield
  {journal} {\bibinfo  {journal} {Physical Review A}\ }\textbf {\bibinfo
  {volume} {43}},\ \bibinfo {pages} {2046} (\bibinfo {year}
  {1991})}\BibitemShut {NoStop}%
\bibitem [{\citenamefont {Srednicki}(1994)}]{Srednicki1994}%
  \BibitemOpen
  \bibfield  {author} {\bibinfo {author} {\bibfnamefont {M.}~\bibnamefont
  {Srednicki}},\ }\href {https://doi.org/10.1103/physreve.50.888} {\bibfield
  {journal} {\bibinfo  {journal} {Physical Review E}\ }\textbf {\bibinfo
  {volume} {50}},\ \bibinfo {pages} {888} (\bibinfo {year} {1994})}\BibitemShut
  {NoStop}%
\bibitem [{\citenamefont {Rigol}\ \emph {et~al.}(2008)\citenamefont {Rigol},
  \citenamefont {Dunjko},\ and\ \citenamefont {Olshanii}}]{Rigol2008}%
  \BibitemOpen
  \bibfield  {author} {\bibinfo {author} {\bibfnamefont {M.}~\bibnamefont
  {Rigol}}, \bibinfo {author} {\bibfnamefont {V.}~\bibnamefont {Dunjko}},\ and\
  \bibinfo {author} {\bibfnamefont {M.}~\bibnamefont {Olshanii}},\ }\href
  {https://doi.org/10.1038/nature06838} {\bibfield  {journal} {\bibinfo
  {journal} {Nature}\ }\textbf {\bibinfo {volume} {452}},\ \bibinfo {pages}
  {854} (\bibinfo {year} {2008})}\BibitemShut {NoStop}%
\bibitem [{\citenamefont {Khemani}\ \emph {et~al.}(2018)\citenamefont
  {Khemani}, \citenamefont {Vishwanath},\ and\ \citenamefont
  {Huse}}]{Khemani2018}%
  \BibitemOpen
  \bibfield  {author} {\bibinfo {author} {\bibfnamefont {V.}~\bibnamefont
  {Khemani}}, \bibinfo {author} {\bibfnamefont {A.}~\bibnamefont
  {Vishwanath}},\ and\ \bibinfo {author} {\bibfnamefont {D.~A.}\ \bibnamefont
  {Huse}},\ }\href {https://doi.org/10.1103/physrevx.8.031057} {\bibfield
  {journal} {\bibinfo  {journal} {Physical Review X}\ }\textbf {\bibinfo
  {volume} {8}},\ \bibinfo {pages} {031057} (\bibinfo {year}
  {2018})}\BibitemShut {NoStop}%
\bibitem [{\citenamefont {Hosur}\ \emph {et~al.}(2016)\citenamefont {Hosur},
  \citenamefont {Qi}, \citenamefont {Roberts},\ and\ \citenamefont
  {Yoshida}}]{Hosur2016}%
  \BibitemOpen
  \bibfield  {author} {\bibinfo {author} {\bibfnamefont {P.}~\bibnamefont
  {Hosur}}, \bibinfo {author} {\bibfnamefont {X.-L.}\ \bibnamefont {Qi}},
  \bibinfo {author} {\bibfnamefont {D.~A.}\ \bibnamefont {Roberts}},\ and\
  \bibinfo {author} {\bibfnamefont {B.}~\bibnamefont {Yoshida}},\ }\href
  {https://doi.org/10.1007/JHEP02(2016)004} {\bibfield  {journal} {\bibinfo
  {journal} {Journal of High Energy Physics}\ }\textbf {\bibinfo {volume}
  {2016}} (\bibinfo {year} {2016})}\BibitemShut {NoStop}%
\bibitem [{\citenamefont {Roberts}\ and\ \citenamefont
  {Yoshida}(2017)}]{Roberts2017}%
  \BibitemOpen
  \bibfield  {author} {\bibinfo {author} {\bibfnamefont {D.~A.}\ \bibnamefont
  {Roberts}}\ and\ \bibinfo {author} {\bibfnamefont {B.}~\bibnamefont
  {Yoshida}},\ }\href {https://doi.org/10.1007/JHEP04(2017)121} {\bibfield
  {journal} {\bibinfo  {journal} {Journal of High Energy Physics}\ }\textbf
  {\bibinfo {volume} {2017}} (\bibinfo {year} {2017})}\BibitemShut {NoStop}%
\bibitem [{\citenamefont {Gullans}\ and\ \citenamefont
  {Huse}(2020)}]{Gullans2020}%
  \BibitemOpen
  \bibfield  {author} {\bibinfo {author} {\bibfnamefont {M.~J.}\ \bibnamefont
  {Gullans}}\ and\ \bibinfo {author} {\bibfnamefont {D.~A.}\ \bibnamefont
  {Huse}},\ }\href {https://doi.org/10.1103/physrevx.10.041020} {\bibfield
  {journal} {\bibinfo  {journal} {Physical Review X}\ }\textbf {\bibinfo
  {volume} {10}},\ \bibinfo {pages} {041020} (\bibinfo {year}
  {2020})}\BibitemShut {NoStop}%
\bibitem [{\citenamefont {Choi}\ \emph {et~al.}(2020)\citenamefont {Choi},
  \citenamefont {Bao}, \citenamefont {Qi},\ and\ \citenamefont
  {Altman}}]{Choi2020}%
  \BibitemOpen
  \bibfield  {author} {\bibinfo {author} {\bibfnamefont {S.}~\bibnamefont
  {Choi}}, \bibinfo {author} {\bibfnamefont {Y.}~\bibnamefont {Bao}}, \bibinfo
  {author} {\bibfnamefont {X.-L.}\ \bibnamefont {Qi}},\ and\ \bibinfo {author}
  {\bibfnamefont {E.}~\bibnamefont {Altman}},\ }\href
  {https://doi.org/10.1103/physrevlett.125.030505} {\bibfield  {journal}
  {\bibinfo  {journal} {Physical Review Letters}\ }\textbf {\bibinfo {volume}
  {125}},\ \bibinfo {pages} {030505} (\bibinfo {year} {2020})}\BibitemShut
  {NoStop}%
\bibitem [{\citenamefont {Li}\ and\ \citenamefont {Fisher}(2021)}]{Li2021}%
  \BibitemOpen
  \bibfield  {author} {\bibinfo {author} {\bibfnamefont {Y.}~\bibnamefont
  {Li}}\ and\ \bibinfo {author} {\bibfnamefont {M.~P.~A.}\ \bibnamefont
  {Fisher}},\ }\href {https://doi.org/10.1103/physrevb.103.104306} {\bibfield
  {journal} {\bibinfo  {journal} {Physical Review B}\ }\textbf {\bibinfo
  {volume} {103}},\ \bibinfo {pages} {104306} (\bibinfo {year}
  {2021})}\BibitemShut {NoStop}%
\bibitem [{\citenamefont {Fan}\ \emph {et~al.}(2021)\citenamefont {Fan},
  \citenamefont {Vijay}, \citenamefont {Vishwanath},\ and\ \citenamefont
  {You}}]{Fan2021}%
  \BibitemOpen
  \bibfield  {author} {\bibinfo {author} {\bibfnamefont {R.}~\bibnamefont
  {Fan}}, \bibinfo {author} {\bibfnamefont {S.}~\bibnamefont {Vijay}}, \bibinfo
  {author} {\bibfnamefont {A.}~\bibnamefont {Vishwanath}},\ and\ \bibinfo
  {author} {\bibfnamefont {Y.-Z.}\ \bibnamefont {You}},\ }\href
  {https://doi.org/10.1103/physrevb.103.174309} {\bibfield  {journal} {\bibinfo
   {journal} {Physical Review B}\ }\textbf {\bibinfo {volume} {103}},\ \bibinfo
  {pages} {174309} (\bibinfo {year} {2021})}\BibitemShut {NoStop}%
\bibitem [{\citenamefont {Yoshida}(2021)}]{Yoshida2021}%
  \BibitemOpen
  \bibfield  {author} {\bibinfo {author} {\bibfnamefont {B.}~\bibnamefont
  {Yoshida}},\ }\href {https://doi.org/10.48550/arXiv.2109.08691} {\bibfield
  {journal} {\bibinfo  {journal} {arXiv:2109.08691}\ } (\bibinfo {year}
  {2021})}\BibitemShut {NoStop}%
\bibitem [{\citenamefont {Mathur}(2009)}]{Mathur2009}%
  \BibitemOpen
  \bibfield  {author} {\bibinfo {author} {\bibfnamefont {S.~D.}\ \bibnamefont
  {Mathur}},\ }\href {https://doi.org/10.1088/0264-9381/26/22/224001}
  {\bibfield  {journal} {\bibinfo  {journal} {Classical and Quantum Gravity}\
  }\textbf {\bibinfo {volume} {26}},\ \bibinfo {pages} {224001} (\bibinfo
  {year} {2009})}\BibitemShut {NoStop}%
\bibitem [{\citenamefont {Page}(1993)}]{Page1993a}%
  \BibitemOpen
  \bibfield  {author} {\bibinfo {author} {\bibfnamefont {D.~N.}\ \bibnamefont
  {Page}},\ }\href {https://doi.org/10.1103/physrevlett.71.3743} {\bibfield
  {journal} {\bibinfo  {journal} {Physical Review Letters}\ }\textbf {\bibinfo
  {volume} {71}},\ \bibinfo {pages} {3743} (\bibinfo {year}
  {1993})}\BibitemShut {NoStop}%
\bibitem [{\citenamefont {Hayden}\ and\ \citenamefont
  {Preskill}(2007)}]{Hayden2007}%
  \BibitemOpen
  \bibfield  {author} {\bibinfo {author} {\bibfnamefont {P.}~\bibnamefont
  {Hayden}}\ and\ \bibinfo {author} {\bibfnamefont {J.}~\bibnamefont
  {Preskill}},\ }\href {https://doi.org/10.1088/1126-6708/2007/09/120}
  {\bibfield  {journal} {\bibinfo  {journal} {Journal of High Energy Physics}\
  }\textbf {\bibinfo {volume} {2007}},\ \bibinfo {pages} {120} (\bibinfo {year}
  {2007})}\BibitemShut {NoStop}%
\bibitem [{\citenamefont {Shenker}\ and\ \citenamefont
  {Stanford}(2014)}]{Shenker2014}%
  \BibitemOpen
  \bibfield  {author} {\bibinfo {author} {\bibfnamefont {S.~H.}\ \bibnamefont
  {Shenker}}\ and\ \bibinfo {author} {\bibfnamefont {D.}~\bibnamefont
  {Stanford}},\ }\href {https://doi.org/10.1007/jhep03(2014)067} {\bibfield
  {journal} {\bibinfo  {journal} {Journal of High Energy Physics}\ }\textbf
  {\bibinfo {volume} {2014}} (\bibinfo {year} {2014})}\BibitemShut {NoStop}%
\bibitem [{\citenamefont {Cotler}\ \emph {et~al.}(2017)\citenamefont {Cotler},
  \citenamefont {Gur-Ari}, \citenamefont {Hanada}, \citenamefont {Polchinski},
  \citenamefont {Saad}, \citenamefont {Shenker}, \citenamefont {Stanford},
  \citenamefont {Streicher},\ and\ \citenamefont {Tezuka}}]{Cotler2017}%
  \BibitemOpen
  \bibfield  {author} {\bibinfo {author} {\bibfnamefont {J.~S.}\ \bibnamefont
  {Cotler}}, \bibinfo {author} {\bibfnamefont {G.}~\bibnamefont {Gur-Ari}},
  \bibinfo {author} {\bibfnamefont {M.}~\bibnamefont {Hanada}}, \bibinfo
  {author} {\bibfnamefont {J.}~\bibnamefont {Polchinski}}, \bibinfo {author}
  {\bibfnamefont {P.}~\bibnamefont {Saad}}, \bibinfo {author} {\bibfnamefont
  {S.~H.}\ \bibnamefont {Shenker}}, \bibinfo {author} {\bibfnamefont
  {D.}~\bibnamefont {Stanford}}, \bibinfo {author} {\bibfnamefont
  {A.}~\bibnamefont {Streicher}},\ and\ \bibinfo {author} {\bibfnamefont
  {M.}~\bibnamefont {Tezuka}},\ }\href
  {https://doi.org/10.1007/JHEP05(2017)118} {\bibfield  {journal} {\bibinfo
  {journal} {Journal of High Energy Physics}\ }\textbf {\bibinfo {volume}
  {2017}} (\bibinfo {year} {2017})}\BibitemShut {NoStop}%
\bibitem [{\citenamefont {Almheiri}\ \emph {et~al.}(2020)\citenamefont
  {Almheiri}, \citenamefont {Hartman}, \citenamefont {Maldacena}, \citenamefont
  {Shaghoulian},\ and\ \citenamefont {Tajdini}}]{Almheiri2020}%
  \BibitemOpen
  \bibfield  {author} {\bibinfo {author} {\bibfnamefont {A.}~\bibnamefont
  {Almheiri}}, \bibinfo {author} {\bibfnamefont {T.}~\bibnamefont {Hartman}},
  \bibinfo {author} {\bibfnamefont {J.}~\bibnamefont {Maldacena}}, \bibinfo
  {author} {\bibfnamefont {E.}~\bibnamefont {Shaghoulian}},\ and\ \bibinfo
  {author} {\bibfnamefont {A.}~\bibnamefont {Tajdini}},\ }\href
  {https://doi.org/10.1007/JHEP05(2020)013} {\bibfield  {journal} {\bibinfo
  {journal} {Journal of High Energy Physics}\ }\textbf {\bibinfo {volume}
  {2020}} (\bibinfo {year} {2020})}\BibitemShut {NoStop}%
\bibitem [{\citenamefont {Gao}\ \emph {et~al.}(2017)\citenamefont {Gao},
  \citenamefont {Jafferis},\ and\ \citenamefont {Wall}}]{Gao2017}%
  \BibitemOpen
  \bibfield  {author} {\bibinfo {author} {\bibfnamefont {P.}~\bibnamefont
  {Gao}}, \bibinfo {author} {\bibfnamefont {D.~L.}\ \bibnamefont {Jafferis}},\
  and\ \bibinfo {author} {\bibfnamefont {A.~C.}\ \bibnamefont {Wall}},\ }\href
  {https://dx.doi.org/10.1007/jhep12(2017)151} {\bibfield  {journal} {\bibinfo
  {journal} {Journal of High Energy Physics}\ }\textbf {\bibinfo {volume}
  {2017}} (\bibinfo {year} {2017})}\BibitemShut {NoStop}%
\bibitem [{\citenamefont {Maldacena}\ \emph {et~al.}(2017)\citenamefont
  {Maldacena}, \citenamefont {Stanford},\ and\ \citenamefont
  {Yang}}]{Maldacena2017}%
  \BibitemOpen
  \bibfield  {author} {\bibinfo {author} {\bibfnamefont {J.}~\bibnamefont
  {Maldacena}}, \bibinfo {author} {\bibfnamefont {D.}~\bibnamefont
  {Stanford}},\ and\ \bibinfo {author} {\bibfnamefont {Z.}~\bibnamefont
  {Yang}},\ }\href {https://doi.org/10.1002/prop.201700034} {\bibfield
  {journal} {\bibinfo  {journal} {Fortschritte der Physik}\ }\textbf {\bibinfo
  {volume} {65}} (\bibinfo {year} {2017})}\BibitemShut {NoStop}%
\bibitem [{\citenamefont {Brown}\ \emph {et~al.}(2023)\citenamefont {Brown},
  \citenamefont {Gharibyan}, \citenamefont {Leichenauer}, \citenamefont {Lin},
  \citenamefont {Nezami}, \citenamefont {Salton}, \citenamefont {Susskind},
  \citenamefont {Swingle},\ and\ \citenamefont {Walter}}]{Brown2023}%
  \BibitemOpen
  \bibfield  {author} {\bibinfo {author} {\bibfnamefont {A.~R.}\ \bibnamefont
  {Brown}}, \bibinfo {author} {\bibfnamefont {H.}~\bibnamefont {Gharibyan}},
  \bibinfo {author} {\bibfnamefont {S.}~\bibnamefont {Leichenauer}}, \bibinfo
  {author} {\bibfnamefont {H.~W.}\ \bibnamefont {Lin}}, \bibinfo {author}
  {\bibfnamefont {S.}~\bibnamefont {Nezami}}, \bibinfo {author} {\bibfnamefont
  {G.}~\bibnamefont {Salton}}, \bibinfo {author} {\bibfnamefont
  {L.}~\bibnamefont {Susskind}}, \bibinfo {author} {\bibfnamefont
  {B.}~\bibnamefont {Swingle}},\ and\ \bibinfo {author} {\bibfnamefont
  {M.}~\bibnamefont {Walter}},\ }\href
  {https://doi.org/10.1103/prxquantum.4.010320} {\bibfield  {journal} {\bibinfo
   {journal} {{PRX} Quantum}\ }\textbf {\bibinfo {volume} {4}},\ \bibinfo
  {pages} {010320} (\bibinfo {year} {2023})}\BibitemShut {NoStop}%
\bibitem [{\citenamefont {Landsman}\ \emph {et~al.}(2019)\citenamefont
  {Landsman}, \citenamefont {Figgatt}, \citenamefont {Schuster}, \citenamefont
  {Linke}, \citenamefont {Yoshida}, \citenamefont {Yao},\ and\ \citenamefont
  {Monroe}}]{Landsman2019}%
  \BibitemOpen
  \bibfield  {author} {\bibinfo {author} {\bibfnamefont {K.~A.}\ \bibnamefont
  {Landsman}}, \bibinfo {author} {\bibfnamefont {C.}~\bibnamefont {Figgatt}},
  \bibinfo {author} {\bibfnamefont {T.}~\bibnamefont {Schuster}}, \bibinfo
  {author} {\bibfnamefont {N.~M.}\ \bibnamefont {Linke}}, \bibinfo {author}
  {\bibfnamefont {B.}~\bibnamefont {Yoshida}}, \bibinfo {author} {\bibfnamefont
  {N.~Y.}\ \bibnamefont {Yao}},\ and\ \bibinfo {author} {\bibfnamefont
  {C.}~\bibnamefont {Monroe}},\ }\href
  {https://doi.org/10.1038/s41586-019-0952-6} {\bibfield  {journal} {\bibinfo
  {journal} {Nature}\ }\textbf {\bibinfo {volume} {567}},\ \bibinfo {pages}
  {61} (\bibinfo {year} {2019})}\BibitemShut {NoStop}%
\bibitem [{\citenamefont {Jafferis}\ \emph {et~al.}(2022)\citenamefont
  {Jafferis}, \citenamefont {Zlokapa}, \citenamefont {Lykken}, \citenamefont
  {Kolchmeyer}, \citenamefont {Davis}, \citenamefont {Lauk}, \citenamefont
  {Neven},\ and\ \citenamefont {Spiropulu}}]{Jafferis2022}%
  \BibitemOpen
  \bibfield  {author} {\bibinfo {author} {\bibfnamefont {D.}~\bibnamefont
  {Jafferis}}, \bibinfo {author} {\bibfnamefont {A.}~\bibnamefont {Zlokapa}},
  \bibinfo {author} {\bibfnamefont {J.~D.}\ \bibnamefont {Lykken}}, \bibinfo
  {author} {\bibfnamefont {D.~K.}\ \bibnamefont {Kolchmeyer}}, \bibinfo
  {author} {\bibfnamefont {S.~I.}\ \bibnamefont {Davis}}, \bibinfo {author}
  {\bibfnamefont {N.}~\bibnamefont {Lauk}}, \bibinfo {author} {\bibfnamefont
  {H.}~\bibnamefont {Neven}},\ and\ \bibinfo {author} {\bibfnamefont
  {M.}~\bibnamefont {Spiropulu}},\ }\href
  {https://doi.org/10.1038/s41586-022-05424-3} {\bibfield  {journal} {\bibinfo
  {journal} {Nature}\ }\textbf {\bibinfo {volume} {612}},\ \bibinfo {pages}
  {51} (\bibinfo {year} {2022})}\BibitemShut {NoStop}%
\bibitem [{\citenamefont {Shapoval}\ \emph {et~al.}(2023)\citenamefont
  {Shapoval}, \citenamefont {Su}, \citenamefont {de~Jong}, \citenamefont
  {Urbanek},\ and\ \citenamefont {Swingle}}]{Shapoval2023}%
  \BibitemOpen
  \bibfield  {author} {\bibinfo {author} {\bibfnamefont {I.}~\bibnamefont
  {Shapoval}}, \bibinfo {author} {\bibfnamefont {V.~P.}\ \bibnamefont {Su}},
  \bibinfo {author} {\bibfnamefont {W.}~\bibnamefont {de~Jong}}, \bibinfo
  {author} {\bibfnamefont {M.}~\bibnamefont {Urbanek}},\ and\ \bibinfo {author}
  {\bibfnamefont {B.}~\bibnamefont {Swingle}},\ }\href
  {https://doi.org/10.22331/q-2023-10-12-1138} {\bibfield  {journal} {\bibinfo
  {journal} {Quantum}\ }\textbf {\bibinfo {volume} {7}},\ \bibinfo {pages}
  {1138} (\bibinfo {year} {2023})}\BibitemShut {NoStop}%
\bibitem [{\citenamefont {Gullans}\ \emph {et~al.}(2021)\citenamefont
  {Gullans}, \citenamefont {Krastanov}, \citenamefont {Huse}, \citenamefont
  {Jiang},\ and\ \citenamefont {Flammia}}]{Gullans2021}%
  \BibitemOpen
  \bibfield  {author} {\bibinfo {author} {\bibfnamefont {M.~J.}\ \bibnamefont
  {Gullans}}, \bibinfo {author} {\bibfnamefont {S.}~\bibnamefont {Krastanov}},
  \bibinfo {author} {\bibfnamefont {D.~A.}\ \bibnamefont {Huse}}, \bibinfo
  {author} {\bibfnamefont {L.}~\bibnamefont {Jiang}},\ and\ \bibinfo {author}
  {\bibfnamefont {S.~T.}\ \bibnamefont {Flammia}},\ }\href
  {https://doi.org/10.1103/physrevx.11.031066} {\bibfield  {journal} {\bibinfo
  {journal} {Physical Review X}\ }\textbf {\bibinfo {volume} {11}},\ \bibinfo
  {pages} {031066} (\bibinfo {year} {2021})}\BibitemShut {NoStop}%
\bibitem [{\citenamefont {Lovas}\ \emph {et~al.}(2023)\citenamefont {Lovas},
  \citenamefont {Agrawal},\ and\ \citenamefont {Vijay}}]{Lovas2023}%
  \BibitemOpen
  \bibfield  {author} {\bibinfo {author} {\bibfnamefont {I.}~\bibnamefont
  {Lovas}}, \bibinfo {author} {\bibfnamefont {U.}~\bibnamefont {Agrawal}},\
  and\ \bibinfo {author} {\bibfnamefont {S.}~\bibnamefont {Vijay}},\ }\href
  {https://doi.org/10.48550/arXiv.2304.02664} {\bibfield  {journal} {\bibinfo
  {journal} {arXiv:2304.02664}\ } (\bibinfo {year} {2023})}\BibitemShut
  {NoStop}%
\bibitem [{\citenamefont {Blake}\ and\ \citenamefont
  {Thompson}(2023)}]{Blake2023}%
  \BibitemOpen
  \bibfield  {author} {\bibinfo {author} {\bibfnamefont {M.}~\bibnamefont
  {Blake}}\ and\ \bibinfo {author} {\bibfnamefont {A.~P.}\ \bibnamefont
  {Thompson}},\ }\bibfield  {journal} {\bibinfo  {journal} {Journal of High
  Energy Physics}\ }\textbf {\bibinfo {volume} {2023}},\ \href
  {https://doi.org/10.1007/jhep11(2023)016} {10.1007/jhep11(2023)016} (\bibinfo
  {year} {2023})\BibitemShut {NoStop}%
\bibitem [{\citenamefont {Sahu}\ and\ \citenamefont {Jian}(2024)}]{Sahu2024}%
  \BibitemOpen
  \bibfield  {author} {\bibinfo {author} {\bibfnamefont {S.}~\bibnamefont
  {Sahu}}\ and\ \bibinfo {author} {\bibfnamefont {S.-K.}\ \bibnamefont
  {Jian}},\ }\href {https://doi.org/10.1103/physreva.109.042414} {\bibfield
  {journal} {\bibinfo  {journal} {Physical Review A}\ }\textbf {\bibinfo
  {volume} {109}},\ \bibinfo {pages} {042414} (\bibinfo {year}
  {2024})}\BibitemShut {NoStop}%
\bibitem [{\citenamefont {Nakata}\ and\ \citenamefont
  {Tezuka}(2024)}]{Nakata2023}%
  \BibitemOpen
  \bibfield  {author} {\bibinfo {author} {\bibfnamefont {Y.}~\bibnamefont
  {Nakata}}\ and\ \bibinfo {author} {\bibfnamefont {M.}~\bibnamefont
  {Tezuka}},\ }\href {https://doi.org/10.1103/physrevresearch.6.l022021}
  {\bibfield  {journal} {\bibinfo  {journal} {Physical Review Research}\
  }\textbf {\bibinfo {volume} {6}},\ \bibinfo {pages} {l022021} (\bibinfo
  {year} {2024})}\BibitemShut {NoStop}%
\bibitem [{\citenamefont {Brown}\ and\ \citenamefont
  {Fawzi}(2015)}]{Brown2015}%
  \BibitemOpen
  \bibfield  {author} {\bibinfo {author} {\bibfnamefont {W.}~\bibnamefont
  {Brown}}\ and\ \bibinfo {author} {\bibfnamefont {O.}~\bibnamefont {Fawzi}},\
  }\href {https://doi.org/10.1007/s00220-015-2470-1} {\bibfield  {journal}
  {\bibinfo  {journal} {Communications in Mathematical Physics}\ }\textbf
  {\bibinfo {volume} {340}},\ \bibinfo {pages} {867} (\bibinfo {year}
  {2015})}\BibitemShut {NoStop}%
\bibitem [{\citenamefont {Liu}\ and\ \citenamefont {Vardhan}(2021)}]{Liu2021}%
  \BibitemOpen
  \bibfield  {author} {\bibinfo {author} {\bibfnamefont {H.}~\bibnamefont
  {Liu}}\ and\ \bibinfo {author} {\bibfnamefont {S.}~\bibnamefont {Vardhan}},\
  }\href {https://doi.org/10.1007/JHEP03(2021)088} {\bibfield  {journal}
  {\bibinfo  {journal} {Journal of High Energy Physics}\ }\textbf {\bibinfo
  {volume} {2021}} (\bibinfo {year} {2021})}\BibitemShut {NoStop}%
\bibitem [{\citenamefont {Piroli}\ \emph
  {et~al.}(2020{\natexlab{a}})\citenamefont {Piroli}, \citenamefont
  {Sünderhauf},\ and\ \citenamefont {Qi}}]{Piroli2020a}%
  \BibitemOpen
  \bibfield  {author} {\bibinfo {author} {\bibfnamefont {L.}~\bibnamefont
  {Piroli}}, \bibinfo {author} {\bibfnamefont {C.}~\bibnamefont
  {Sünderhauf}},\ and\ \bibinfo {author} {\bibfnamefont {X.-L.}\ \bibnamefont
  {Qi}},\ }\href {https://doi.org/10.1007/JHEP04(2020)063} {\bibfield
  {journal} {\bibinfo  {journal} {Journal of High Energy Physics}\ }\textbf
  {\bibinfo {volume} {2020}} (\bibinfo {year}
  {2020}{\natexlab{a}})}\BibitemShut {NoStop}%
\bibitem [{\citenamefont {Turkeshi}\ and\ \citenamefont
  {Sierant}(2023)}]{Turkeshi2023}%
  \BibitemOpen
  \bibfield  {author} {\bibinfo {author} {\bibfnamefont {X.}~\bibnamefont
  {Turkeshi}}\ and\ \bibinfo {author} {\bibfnamefont {P.}~\bibnamefont
  {Sierant}},\ }\href {https://doi.org/10.48550/ARXIV.2308.06321} {\bibfield
  {journal} {\bibinfo  {journal} {arXiv:2308.06321}\ } (\bibinfo {year}
  {2023})}\BibitemShut {NoStop}%
\bibitem [{\citenamefont {Gribben}\ \emph {et~al.}(2024)\citenamefont
  {Gribben}, \citenamefont {Marino},\ and\ \citenamefont
  {Kelly}}]{Gribben2024}%
  \BibitemOpen
  \bibfield  {author} {\bibinfo {author} {\bibfnamefont {D.}~\bibnamefont
  {Gribben}}, \bibinfo {author} {\bibfnamefont {J.}~\bibnamefont {Marino}},\
  and\ \bibinfo {author} {\bibfnamefont {S.~P.}\ \bibnamefont {Kelly}},\ }\href
  {https://doi.org/10.48550/ARXIV.2401.17066} {\bibfield  {journal} {\bibinfo
  {journal} {arXiv:2401.17066}\ } (\bibinfo {year} {2024})}\BibitemShut
  {NoStop}%
\bibitem [{\citenamefont {Yoshida}\ and\ \citenamefont
  {Kitaev}(2017)}]{Yoshida2017}%
  \BibitemOpen
  \bibfield  {author} {\bibinfo {author} {\bibfnamefont {B.}~\bibnamefont
  {Yoshida}}\ and\ \bibinfo {author} {\bibfnamefont {A.}~\bibnamefont
  {Kitaev}},\ }\href {https://doi.org/10.48550/arXiv.1710.03363} {\bibfield
  {journal} {\bibinfo  {journal} {arXiv:1710.03363}\ } (\bibinfo {year}
  {2017})}\BibitemShut {NoStop}%
\bibitem [{\citenamefont {Bao}\ and\ \citenamefont {Kikuchi}(2021)}]{Bao2021}%
  \BibitemOpen
  \bibfield  {author} {\bibinfo {author} {\bibfnamefont {N.}~\bibnamefont
  {Bao}}\ and\ \bibinfo {author} {\bibfnamefont {Y.}~\bibnamefont {Kikuchi}},\
  }\href {https://doi.org/10.1007/jhep02(2021)017} {\bibfield  {journal}
  {\bibinfo  {journal} {Journal of High Energy Physics}\ }\textbf {\bibinfo
  {volume} {2021}} (\bibinfo {year} {2021})}\BibitemShut {NoStop}%
\bibitem [{\citenamefont {Leone}\ \emph {et~al.}(2022)\citenamefont {Leone},
  \citenamefont {Oliviero}, \citenamefont {Piemontese}, \citenamefont {True},\
  and\ \citenamefont {Hamma}}]{Leone2022}%
  \BibitemOpen
  \bibfield  {author} {\bibinfo {author} {\bibfnamefont {L.}~\bibnamefont
  {Leone}}, \bibinfo {author} {\bibfnamefont {S.~F.~E.}\ \bibnamefont
  {Oliviero}}, \bibinfo {author} {\bibfnamefont {S.}~\bibnamefont
  {Piemontese}}, \bibinfo {author} {\bibfnamefont {S.}~\bibnamefont {True}},\
  and\ \bibinfo {author} {\bibfnamefont {A.}~\bibnamefont {Hamma}},\ }\href
  {https://doi.org/10.1103/physreva.106.062434} {\bibfield  {journal} {\bibinfo
   {journal} {Physical Review A}\ }\textbf {\bibinfo {volume} {106}},\ \bibinfo
  {pages} {062434} (\bibinfo {year} {2022})}\BibitemShut {NoStop}%
\bibitem [{\citenamefont {Oliviero}\ \emph {et~al.}(2024)\citenamefont
  {Oliviero}, \citenamefont {Leone}, \citenamefont {Lloyd},\ and\ \citenamefont
  {Hamma}}]{Oliviero2022}%
  \BibitemOpen
  \bibfield  {author} {\bibinfo {author} {\bibfnamefont {S.~F.~E.}\
  \bibnamefont {Oliviero}}, \bibinfo {author} {\bibfnamefont {L.}~\bibnamefont
  {Leone}}, \bibinfo {author} {\bibfnamefont {S.}~\bibnamefont {Lloyd}},\ and\
  \bibinfo {author} {\bibfnamefont {A.}~\bibnamefont {Hamma}},\ }\href
  {https://doi.org/10.1103/physrevlett.132.080402} {\bibfield  {journal}
  {\bibinfo  {journal} {Physical Review Letters}\ }\textbf {\bibinfo {volume}
  {132}},\ \bibinfo {pages} {080402} (\bibinfo {year} {2024})}\BibitemShut
  {NoStop}%
\bibitem [{\citenamefont {Leone}\ \emph {et~al.}(2024)\citenamefont {Leone},
  \citenamefont {Oliviero}, \citenamefont {Lloyd},\ and\ \citenamefont
  {Hamma}}]{Leone2022a}%
  \BibitemOpen
  \bibfield  {author} {\bibinfo {author} {\bibfnamefont {L.}~\bibnamefont
  {Leone}}, \bibinfo {author} {\bibfnamefont {S.~F.~E.}\ \bibnamefont
  {Oliviero}}, \bibinfo {author} {\bibfnamefont {S.}~\bibnamefont {Lloyd}},\
  and\ \bibinfo {author} {\bibfnamefont {A.}~\bibnamefont {Hamma}},\ }\href
  {https://doi.org/10.1103/physreva.109.022429} {\bibfield  {journal} {\bibinfo
   {journal} {Physical Review A}\ }\textbf {\bibinfo {volume} {109}},\ \bibinfo
  {pages} {022429} (\bibinfo {year} {2024})}\BibitemShut {NoStop}%
\bibitem [{\citenamefont {Agarwal}\ \emph {et~al.}(2023)\citenamefont
  {Agarwal}, \citenamefont {Langlett},\ and\ \citenamefont {Xu}}]{Agarwal2023}%
  \BibitemOpen
  \bibfield  {author} {\bibinfo {author} {\bibfnamefont {L.}~\bibnamefont
  {Agarwal}}, \bibinfo {author} {\bibfnamefont {C.~M.}\ \bibnamefont
  {Langlett}},\ and\ \bibinfo {author} {\bibfnamefont {S.}~\bibnamefont {Xu}},\
  }\href {https://doi.org/10.1103/physrevlett.130.020801} {\bibfield  {journal}
  {\bibinfo  {journal} {Physical Review Letters}\ }\textbf {\bibinfo {volume}
  {130}},\ \bibinfo {pages} {020801} (\bibinfo {year} {2023})}\BibitemShut
  {NoStop}%
\bibitem [{\citenamefont {Lin}\ and\ \citenamefont
  {Motrunich}(2018)}]{Lin2018}%
  \BibitemOpen
  \bibfield  {author} {\bibinfo {author} {\bibfnamefont {C.-J.}\ \bibnamefont
  {Lin}}\ and\ \bibinfo {author} {\bibfnamefont {O.~I.}\ \bibnamefont
  {Motrunich}},\ }\href {https://doi.org/10.1103/physrevb.97.144304} {\bibfield
   {journal} {\bibinfo  {journal} {Physical Review B}\ }\textbf {\bibinfo
  {volume} {97}},\ \bibinfo {pages} {144304} (\bibinfo {year}
  {2018})}\BibitemShut {NoStop}%
\bibitem [{\citenamefont {D{\'{o}}ra}\ and\ \citenamefont
  {Moessner}(2017)}]{Dora2017}%
  \BibitemOpen
  \bibfield  {author} {\bibinfo {author} {\bibfnamefont {B.}~\bibnamefont
  {D{\'{o}}ra}}\ and\ \bibinfo {author} {\bibfnamefont {R.}~\bibnamefont
  {Moessner}},\ }\href {https://doi.org/10.1103/physrevlett.119.026802}
  {\bibfield  {journal} {\bibinfo  {journal} {Physical Review Letters}\
  }\textbf {\bibinfo {volume} {119}},\ \bibinfo {pages} {026802} (\bibinfo
  {year} {2017})}\BibitemShut {NoStop}%
\bibitem [{\citenamefont {Gopalakrishnan}\ \emph {et~al.}(2018)\citenamefont
  {Gopalakrishnan}, \citenamefont {Huse}, \citenamefont {Khemani},\ and\
  \citenamefont {Vasseur}}]{Gopalakrishnan2018}%
  \BibitemOpen
  \bibfield  {author} {\bibinfo {author} {\bibfnamefont {S.}~\bibnamefont
  {Gopalakrishnan}}, \bibinfo {author} {\bibfnamefont {D.~A.}\ \bibnamefont
  {Huse}}, \bibinfo {author} {\bibfnamefont {V.}~\bibnamefont {Khemani}},\ and\
  \bibinfo {author} {\bibfnamefont {R.}~\bibnamefont {Vasseur}},\ }\href
  {https://doi.org/10.1103/physrevb.98.220303} {\bibfield  {journal} {\bibinfo
  {journal} {Physical Review B}\ }\textbf {\bibinfo {volume} {98}},\ \bibinfo
  {pages} {220303} (\bibinfo {year} {2018})}\BibitemShut {NoStop}%
\bibitem [{\citenamefont {Chen}\ \emph {et~al.}(2016)\citenamefont {Chen},
  \citenamefont {Zhou}, \citenamefont {Huse},\ and\ \citenamefont
  {Fradkin}}]{Chen2016}%
  \BibitemOpen
  \bibfield  {author} {\bibinfo {author} {\bibfnamefont {X.}~\bibnamefont
  {Chen}}, \bibinfo {author} {\bibfnamefont {T.}~\bibnamefont {Zhou}}, \bibinfo
  {author} {\bibfnamefont {D.~A.}\ \bibnamefont {Huse}},\ and\ \bibinfo
  {author} {\bibfnamefont {E.}~\bibnamefont {Fradkin}},\ }\href
  {https://doi.org/10.1002/andp.201600332} {\bibfield  {journal} {\bibinfo
  {journal} {Annalen der Physik}\ }\textbf {\bibinfo {volume} {529}} (\bibinfo
  {year} {2016})}\BibitemShut {NoStop}%
\bibitem [{\citenamefont {Huang}\ \emph {et~al.}(2016)\citenamefont {Huang},
  \citenamefont {Zhang},\ and\ \citenamefont {Chen}}]{Huang2016}%
  \BibitemOpen
  \bibfield  {author} {\bibinfo {author} {\bibfnamefont {Y.}~\bibnamefont
  {Huang}}, \bibinfo {author} {\bibfnamefont {Y.-L.}\ \bibnamefont {Zhang}},\
  and\ \bibinfo {author} {\bibfnamefont {X.}~\bibnamefont {Chen}},\ }\href
  {https://doi.org/10.1002/andp.201600318} {\bibfield  {journal} {\bibinfo
  {journal} {Annalen der Physik}\ }\textbf {\bibinfo {volume} {529}} (\bibinfo
  {year} {2016})}\BibitemShut {NoStop}%
\bibitem [{\citenamefont {Smith}\ \emph {et~al.}(2019)\citenamefont {Smith},
  \citenamefont {Knolle}, \citenamefont {Moessner},\ and\ \citenamefont
  {Kovrizhin}}]{Smith2019}%
  \BibitemOpen
  \bibfield  {author} {\bibinfo {author} {\bibfnamefont {A.}~\bibnamefont
  {Smith}}, \bibinfo {author} {\bibfnamefont {J.}~\bibnamefont {Knolle}},
  \bibinfo {author} {\bibfnamefont {R.}~\bibnamefont {Moessner}},\ and\
  \bibinfo {author} {\bibfnamefont {D.~L.}\ \bibnamefont {Kovrizhin}},\ }\href
  {https://doi.org/10.1103/physrevlett.123.086602} {\bibfield  {journal}
  {\bibinfo  {journal} {Physical Review Letters}\ }\textbf {\bibinfo {volume}
  {123}},\ \bibinfo {pages} {086602} (\bibinfo {year} {2019})}\BibitemShut
  {NoStop}%
\bibitem [{\citenamefont {McGinley}\ \emph {et~al.}(2019)\citenamefont
  {McGinley}, \citenamefont {Nunnenkamp},\ and\ \citenamefont
  {Knolle}}]{McGinley2019}%
  \BibitemOpen
  \bibfield  {author} {\bibinfo {author} {\bibfnamefont {M.}~\bibnamefont
  {McGinley}}, \bibinfo {author} {\bibfnamefont {A.}~\bibnamefont
  {Nunnenkamp}},\ and\ \bibinfo {author} {\bibfnamefont {J.}~\bibnamefont
  {Knolle}},\ }\href {https://doi.org/10.1103/physrevlett.122.020603}
  {\bibfield  {journal} {\bibinfo  {journal} {Physical Review Letters}\
  }\textbf {\bibinfo {volume} {122}},\ \bibinfo {pages} {020603} (\bibinfo
  {year} {2019})}\BibitemShut {NoStop}%
\bibitem [{\citenamefont {Hayden}\ \emph {et~al.}(2008)\citenamefont {Hayden},
  \citenamefont {Horodecki}, \citenamefont {Winter},\ and\ \citenamefont
  {Yard}}]{Hayden2008}%
  \BibitemOpen
  \bibfield  {author} {\bibinfo {author} {\bibfnamefont {P.}~\bibnamefont
  {Hayden}}, \bibinfo {author} {\bibfnamefont {M.}~\bibnamefont {Horodecki}},
  \bibinfo {author} {\bibfnamefont {A.}~\bibnamefont {Winter}},\ and\ \bibinfo
  {author} {\bibfnamefont {J.}~\bibnamefont {Yard}},\ }\href
  {https://doi.org/10.1142/s1230161208000043} {\bibfield  {journal} {\bibinfo
  {journal} {Open Systems {\&} Information Dynamics}\ }\textbf {\bibinfo
  {volume} {15}},\ \bibinfo {pages} {7} (\bibinfo {year} {2008})}\BibitemShut
  {NoStop}%
\bibitem [{\citenamefont {Xu}\ and\ \citenamefont {Swingle}(2024)}]{Xu2022}%
  \BibitemOpen
  \bibfield  {author} {\bibinfo {author} {\bibfnamefont {S.}~\bibnamefont
  {Xu}}\ and\ \bibinfo {author} {\bibfnamefont {B.}~\bibnamefont {Swingle}},\
  }\href {https://doi.org/10.1103/prxquantum.5.010201} {\bibfield  {journal}
  {\bibinfo  {journal} {PRX Quantum}\ }\textbf {\bibinfo {volume} {5}},\
  \bibinfo {pages} {010201} (\bibinfo {year} {2024})}\BibitemShut {NoStop}%
\bibitem [{\citenamefont {Fisher}\ \emph {et~al.}(2023)\citenamefont {Fisher},
  \citenamefont {Khemani}, \citenamefont {Nahum},\ and\ \citenamefont
  {Vijay}}]{Fisher2023}%
  \BibitemOpen
  \bibfield  {author} {\bibinfo {author} {\bibfnamefont {M.~P.}\ \bibnamefont
  {Fisher}}, \bibinfo {author} {\bibfnamefont {V.}~\bibnamefont {Khemani}},
  \bibinfo {author} {\bibfnamefont {A.}~\bibnamefont {Nahum}},\ and\ \bibinfo
  {author} {\bibfnamefont {S.}~\bibnamefont {Vijay}},\ }\href
  {https://doi.org/10.1146/annurev-conmatphys-031720-030658} {\bibfield
  {journal} {\bibinfo  {journal} {Annual Review of Condensed Matter Physics}\
  }\textbf {\bibinfo {volume} {14}},\ \bibinfo {pages} {335} (\bibinfo {year}
  {2023})}\BibitemShut {NoStop}%
\bibitem [{\citenamefont {Nahum}\ \emph {et~al.}(2017)\citenamefont {Nahum},
  \citenamefont {Ruhman}, \citenamefont {Vijay},\ and\ \citenamefont
  {Haah}}]{Nahum2017}%
  \BibitemOpen
  \bibfield  {author} {\bibinfo {author} {\bibfnamefont {A.}~\bibnamefont
  {Nahum}}, \bibinfo {author} {\bibfnamefont {J.}~\bibnamefont {Ruhman}},
  \bibinfo {author} {\bibfnamefont {S.}~\bibnamefont {Vijay}},\ and\ \bibinfo
  {author} {\bibfnamefont {J.}~\bibnamefont {Haah}},\ }\href
  {https://doi.org/10.1103/physrevx.7.031016} {\bibfield  {journal} {\bibinfo
  {journal} {Physical Review X}\ }\textbf {\bibinfo {volume} {7}},\ \bibinfo
  {pages} {031016} (\bibinfo {year} {2017})}\BibitemShut {NoStop}%
\bibitem [{\citenamefont {Jonay}\ \emph {et~al.}(2018)\citenamefont {Jonay},
  \citenamefont {Huse},\ and\ \citenamefont {Nahum}}]{Jonay2018}%
  \BibitemOpen
  \bibfield  {author} {\bibinfo {author} {\bibfnamefont {C.}~\bibnamefont
  {Jonay}}, \bibinfo {author} {\bibfnamefont {D.~A.}\ \bibnamefont {Huse}},\
  and\ \bibinfo {author} {\bibfnamefont {A.}~\bibnamefont {Nahum}},\ }\href
  {https://doi.org/10.48550/arXiv.1803.00089} {\bibfield  {journal} {\bibinfo
  {journal} {arXiv:1803.00089}\ } (\bibinfo {year} {2018})}\BibitemShut
  {NoStop}%
\bibitem [{\citenamefont {Zhou}\ and\ \citenamefont {Nahum}(2019)}]{Zhou2019}%
  \BibitemOpen
  \bibfield  {author} {\bibinfo {author} {\bibfnamefont {T.}~\bibnamefont
  {Zhou}}\ and\ \bibinfo {author} {\bibfnamefont {A.}~\bibnamefont {Nahum}},\
  }\href {https://doi.org/10.1103/physrevb.99.174205} {\bibfield  {journal}
  {\bibinfo  {journal} {Physical Review B}\ }\textbf {\bibinfo {volume} {99}},\
  \bibinfo {pages} {174205} (\bibinfo {year} {2019})}\BibitemShut {NoStop}%
\bibitem [{\citenamefont {von Keyserlingk}\ \emph {et~al.}(2018)\citenamefont
  {von Keyserlingk}, \citenamefont {Rakovszky}, \citenamefont {Pollmann},\ and\
  \citenamefont {Sondhi}}]{Keyserlingk2018}%
  \BibitemOpen
  \bibfield  {author} {\bibinfo {author} {\bibfnamefont {C.}~\bibnamefont {von
  Keyserlingk}}, \bibinfo {author} {\bibfnamefont {T.}~\bibnamefont
  {Rakovszky}}, \bibinfo {author} {\bibfnamefont {F.}~\bibnamefont
  {Pollmann}},\ and\ \bibinfo {author} {\bibfnamefont {S.}~\bibnamefont
  {Sondhi}},\ }\href {https://doi.org/10.1103/physrevx.8.021013} {\bibfield
  {journal} {\bibinfo  {journal} {Physical Review X}\ }\textbf {\bibinfo
  {volume} {8}},\ \bibinfo {pages} {021013} (\bibinfo {year}
  {2018})}\BibitemShut {NoStop}%
\bibitem [{\citenamefont {Nahum}\ \emph {et~al.}(2018)\citenamefont {Nahum},
  \citenamefont {Vijay},\ and\ \citenamefont {Haah}}]{Nahum2018}%
  \BibitemOpen
  \bibfield  {author} {\bibinfo {author} {\bibfnamefont {A.}~\bibnamefont
  {Nahum}}, \bibinfo {author} {\bibfnamefont {S.}~\bibnamefont {Vijay}},\ and\
  \bibinfo {author} {\bibfnamefont {J.}~\bibnamefont {Haah}},\ }\href
  {https://doi.org/10.1103/physrevx.8.021014} {\bibfield  {journal} {\bibinfo
  {journal} {Physical Review X}\ }\textbf {\bibinfo {volume} {8}},\ \bibinfo
  {pages} {021014} (\bibinfo {year} {2018})}\BibitemShut {NoStop}%
\bibitem [{\citenamefont {Bertini}\ and\ \citenamefont
  {Piroli}(2020)}]{Bertini2020b}%
  \BibitemOpen
  \bibfield  {author} {\bibinfo {author} {\bibfnamefont {B.}~\bibnamefont
  {Bertini}}\ and\ \bibinfo {author} {\bibfnamefont {L.}~\bibnamefont
  {Piroli}},\ }\href {https://doi.org/10.1103/physrevb.102.064305} {\bibfield
  {journal} {\bibinfo  {journal} {Physical Review B}\ }\textbf {\bibinfo
  {volume} {102}},\ \bibinfo {pages} {064305} (\bibinfo {year}
  {2020})}\BibitemShut {NoStop}%
\bibitem [{\citenamefont {Akila}\ \emph {et~al.}(2016)\citenamefont {Akila},
  \citenamefont {Waltner}, \citenamefont {Gutkin},\ and\ \citenamefont
  {Guhr}}]{Akila2016}%
  \BibitemOpen
  \bibfield  {author} {\bibinfo {author} {\bibfnamefont {M.}~\bibnamefont
  {Akila}}, \bibinfo {author} {\bibfnamefont {D.}~\bibnamefont {Waltner}},
  \bibinfo {author} {\bibfnamefont {B.}~\bibnamefont {Gutkin}},\ and\ \bibinfo
  {author} {\bibfnamefont {T.}~\bibnamefont {Guhr}},\ }\href
  {https://doi.org/10.1088/1751-8113/49/37/375101} {\bibfield  {journal}
  {\bibinfo  {journal} {Journal of Physics A: Mathematical and Theoretical}\
  }\textbf {\bibinfo {volume} {49}},\ \bibinfo {pages} {375101} (\bibinfo
  {year} {2016})}\BibitemShut {NoStop}%
\bibitem [{\citenamefont {Bertini}\ \emph {et~al.}(2019)\citenamefont
  {Bertini}, \citenamefont {Kos},\ and\ \citenamefont {Prosen}}]{Bertini2019}%
  \BibitemOpen
  \bibfield  {author} {\bibinfo {author} {\bibfnamefont {B.}~\bibnamefont
  {Bertini}}, \bibinfo {author} {\bibfnamefont {P.}~\bibnamefont {Kos}},\ and\
  \bibinfo {author} {\bibfnamefont {T.}~\bibnamefont {Prosen}},\ }\href
  {https://doi.org/10.1103/physrevlett.123.210601} {\bibfield  {journal}
  {\bibinfo  {journal} {Physical Review Letters}\ }\textbf {\bibinfo {volume}
  {123}},\ \bibinfo {pages} {210601} (\bibinfo {year} {2019})}\BibitemShut
  {NoStop}%
\bibitem [{\citenamefont {Gopalakrishnan}\ and\ \citenamefont
  {Lamacraft}(2019)}]{Gopalakrishnan2019}%
  \BibitemOpen
  \bibfield  {author} {\bibinfo {author} {\bibfnamefont {S.}~\bibnamefont
  {Gopalakrishnan}}\ and\ \bibinfo {author} {\bibfnamefont {A.}~\bibnamefont
  {Lamacraft}},\ }\href {https://doi.org/10.1103/physrevb.100.064309}
  {\bibfield  {journal} {\bibinfo  {journal} {Physical Review B}\ }\textbf
  {\bibinfo {volume} {100}},\ \bibinfo {pages} {064309} (\bibinfo {year}
  {2019})}\BibitemShut {NoStop}%
\bibitem [{\citenamefont {Oliveira}\ \emph {et~al.}(2007)\citenamefont
  {Oliveira}, \citenamefont {Dahlsten},\ and\ \citenamefont
  {Plenio}}]{Oliveira2007}%
  \BibitemOpen
  \bibfield  {author} {\bibinfo {author} {\bibfnamefont {R.}~\bibnamefont
  {Oliveira}}, \bibinfo {author} {\bibfnamefont {O.~C.~O.}\ \bibnamefont
  {Dahlsten}},\ and\ \bibinfo {author} {\bibfnamefont {M.~B.}\ \bibnamefont
  {Plenio}},\ }\href {https://doi.org/10.1103/physrevlett.98.130502} {\bibfield
   {journal} {\bibinfo  {journal} {Physical Review Letters}\ }\textbf {\bibinfo
  {volume} {98}},\ \bibinfo {pages} {130502} (\bibinfo {year}
  {2007})}\BibitemShut {NoStop}%
\bibitem [{\citenamefont {{\v{Z}}nidari{\v{c}}}(2008)}]{Znidaric2008}%
  \BibitemOpen
  \bibfield  {author} {\bibinfo {author} {\bibfnamefont {M.}~\bibnamefont
  {{\v{Z}}nidari{\v{c}}}},\ }\href {https://doi.org/10.1103/physreva.78.032324}
  {\bibfield  {journal} {\bibinfo  {journal} {Physical Review A}\ }\textbf
  {\bibinfo {volume} {78}},\ \bibinfo {pages} {032324} (\bibinfo {year}
  {2008})}\BibitemShut {NoStop}%
\bibitem [{\citenamefont {Bertini}\ \emph {et~al.}(2018)\citenamefont
  {Bertini}, \citenamefont {Kos},\ and\ \citenamefont {Prosen}}]{Bertini2018}%
  \BibitemOpen
  \bibfield  {author} {\bibinfo {author} {\bibfnamefont {B.}~\bibnamefont
  {Bertini}}, \bibinfo {author} {\bibfnamefont {P.}~\bibnamefont {Kos}},\ and\
  \bibinfo {author} {\bibfnamefont {T.}~\bibnamefont {Prosen}},\ }\href
  {https://doi.org/10.1103/physrevlett.121.264101} {\bibfield  {journal}
  {\bibinfo  {journal} {Physical Review Letters}\ }\textbf {\bibinfo {volume}
  {121}},\ \bibinfo {pages} {264101} (\bibinfo {year} {2018})}\BibitemShut
  {NoStop}%
\bibitem [{\citenamefont {Piroli}\ \emph
  {et~al.}(2020{\natexlab{b}})\citenamefont {Piroli}, \citenamefont {Bertini},
  \citenamefont {Cirac},\ and\ \citenamefont {Prosen}}]{Piroli2020}%
  \BibitemOpen
  \bibfield  {author} {\bibinfo {author} {\bibfnamefont {L.}~\bibnamefont
  {Piroli}}, \bibinfo {author} {\bibfnamefont {B.}~\bibnamefont {Bertini}},
  \bibinfo {author} {\bibfnamefont {J.~I.}\ \bibnamefont {Cirac}},\ and\
  \bibinfo {author} {\bibfnamefont {T.}~\bibnamefont {Prosen}},\ }\href
  {https://doi.org/10.1103/physrevb.101.094304} {\bibfield  {journal} {\bibinfo
   {journal} {Physical Review B}\ }\textbf {\bibinfo {volume} {101}},\ \bibinfo
  {pages} {094304} (\bibinfo {year} {2020}{\natexlab{b}})}\BibitemShut
  {NoStop}%
\bibitem [{\citenamefont {Claeys}\ and\ \citenamefont
  {Lamacraft}(2021)}]{Claeys2021}%
  \BibitemOpen
  \bibfield  {author} {\bibinfo {author} {\bibfnamefont {P.~W.}\ \bibnamefont
  {Claeys}}\ and\ \bibinfo {author} {\bibfnamefont {A.}~\bibnamefont
  {Lamacraft}},\ }\href {https://doi.org/10.1103/physrevlett.126.100603}
  {\bibfield  {journal} {\bibinfo  {journal} {Physical Review Letters}\
  }\textbf {\bibinfo {volume} {126}},\ \bibinfo {pages} {100603} (\bibinfo
  {year} {2021})}\BibitemShut {NoStop}%
\bibitem [{\citenamefont {Fritzsch}\ and\ \citenamefont
  {Prosen}(2021)}]{Fritzsch2021}%
  \BibitemOpen
  \bibfield  {author} {\bibinfo {author} {\bibfnamefont {F.}~\bibnamefont
  {Fritzsch}}\ and\ \bibinfo {author} {\bibfnamefont {T.}~\bibnamefont
  {Prosen}},\ }\href {https://doi.org/10.1103/physreve.103.062133} {\bibfield
  {journal} {\bibinfo  {journal} {Physical Review E}\ }\textbf {\bibinfo
  {volume} {103}},\ \bibinfo {pages} {062133} (\bibinfo {year}
  {2021})}\BibitemShut {NoStop}%
\bibitem [{\citenamefont {Aravinda}\ \emph {et~al.}(2021)\citenamefont
  {Aravinda}, \citenamefont {Rather},\ and\ \citenamefont
  {Lakshminarayan}}]{Aravinda2021}%
  \BibitemOpen
  \bibfield  {author} {\bibinfo {author} {\bibfnamefont {S.}~\bibnamefont
  {Aravinda}}, \bibinfo {author} {\bibfnamefont {S.~A.}\ \bibnamefont
  {Rather}},\ and\ \bibinfo {author} {\bibfnamefont {A.}~\bibnamefont
  {Lakshminarayan}},\ }\href {https://doi.org/10.1103/physrevresearch.3.043034}
  {\bibfield  {journal} {\bibinfo  {journal} {Physical Review Research}\
  }\textbf {\bibinfo {volume} {3}},\ \bibinfo {pages} {043034} (\bibinfo {year}
  {2021})}\BibitemShut {NoStop}%
\bibitem [{\citenamefont {Stephen}\ \emph {et~al.}(2024)\citenamefont
  {Stephen}, \citenamefont {Ho}, \citenamefont {Wei}, \citenamefont
  {Raussendorf},\ and\ \citenamefont {Verresen}}]{Stephen2022}%
  \BibitemOpen
  \bibfield  {author} {\bibinfo {author} {\bibfnamefont {D.~T.}\ \bibnamefont
  {Stephen}}, \bibinfo {author} {\bibfnamefont {W.~W.}\ \bibnamefont {Ho}},
  \bibinfo {author} {\bibfnamefont {T.-C.}\ \bibnamefont {Wei}}, \bibinfo
  {author} {\bibfnamefont {R.}~\bibnamefont {Raussendorf}},\ and\ \bibinfo
  {author} {\bibfnamefont {R.}~\bibnamefont {Verresen}},\ }\href
  {https://doi.org/10.1103/physrevlett.132.250601} {\bibfield  {journal}
  {\bibinfo  {journal} {Physical Review Letters}\ }\textbf {\bibinfo {volume}
  {132}},\ \bibinfo {pages} {250601} (\bibinfo {year} {2024})}\BibitemShut
  {NoStop}%
\bibitem [{\citenamefont {Claeys}\ \emph
  {et~al.}(2022{\natexlab{a}})\citenamefont {Claeys}, \citenamefont {Henry},
  \citenamefont {Vicary},\ and\ \citenamefont {Lamacraft}}]{Claeys2022}%
  \BibitemOpen
  \bibfield  {author} {\bibinfo {author} {\bibfnamefont {P.~W.}\ \bibnamefont
  {Claeys}}, \bibinfo {author} {\bibfnamefont {M.}~\bibnamefont {Henry}},
  \bibinfo {author} {\bibfnamefont {J.}~\bibnamefont {Vicary}},\ and\ \bibinfo
  {author} {\bibfnamefont {A.}~\bibnamefont {Lamacraft}},\ }\href
  {https://doi.org/10.1103/physrevresearch.4.043212} {\bibfield  {journal}
  {\bibinfo  {journal} {Physical Review Research}\ }\textbf {\bibinfo {volume}
  {4}},\ \bibinfo {pages} {043212} (\bibinfo {year}
  {2022}{\natexlab{a}})}\BibitemShut {NoStop}%
\bibitem [{\citenamefont {Claeys}\ and\ \citenamefont
  {Lamacraft}(2022)}]{Claeys2022a}%
  \BibitemOpen
  \bibfield  {author} {\bibinfo {author} {\bibfnamefont {P.~W.}\ \bibnamefont
  {Claeys}}\ and\ \bibinfo {author} {\bibfnamefont {A.}~\bibnamefont
  {Lamacraft}},\ }\href {https://doi.org/10.22331/q-2022-06-15-738} {\bibfield
  {journal} {\bibinfo  {journal} {Quantum}\ }\textbf {\bibinfo {volume} {6}},\
  \bibinfo {pages} {738} (\bibinfo {year} {2022})}\BibitemShut {NoStop}%
\bibitem [{\citenamefont {Suzuki}\ \emph {et~al.}(2022)\citenamefont {Suzuki},
  \citenamefont {Mitarai},\ and\ \citenamefont {Fujii}}]{Suzuki2022}%
  \BibitemOpen
  \bibfield  {author} {\bibinfo {author} {\bibfnamefont {R.}~\bibnamefont
  {Suzuki}}, \bibinfo {author} {\bibfnamefont {K.}~\bibnamefont {Mitarai}},\
  and\ \bibinfo {author} {\bibfnamefont {K.}~\bibnamefont {Fujii}},\ }\href
  {https://doi.org/10.22331/q-2022-01-24-631} {\bibfield  {journal} {\bibinfo
  {journal} {Quantum}\ }\textbf {\bibinfo {volume} {6}},\ \bibinfo {pages}
  {631} (\bibinfo {year} {2022})}\BibitemShut {NoStop}%
\bibitem [{\citenamefont {Brahmachari}\ \emph {et~al.}(2024)\citenamefont
  {Brahmachari}, \citenamefont {Rajmohan}, \citenamefont {Rather},\ and\
  \citenamefont {Lakshminarayan}}]{Brahmachari2022}%
  \BibitemOpen
  \bibfield  {author} {\bibinfo {author} {\bibfnamefont {S.}~\bibnamefont
  {Brahmachari}}, \bibinfo {author} {\bibfnamefont {R.~N.}\ \bibnamefont
  {Rajmohan}}, \bibinfo {author} {\bibfnamefont {S.~A.}\ \bibnamefont
  {Rather}},\ and\ \bibinfo {author} {\bibfnamefont {A.}~\bibnamefont
  {Lakshminarayan}},\ }\href {https://doi.org/10.1103/physreva.109.022610}
  {\bibfield  {journal} {\bibinfo  {journal} {Physical Review A}\ }\textbf
  {\bibinfo {volume} {109}},\ \bibinfo {pages} {022610} (\bibinfo {year}
  {2024})}\BibitemShut {NoStop}%
\bibitem [{\citenamefont {Borsi}\ and\ \citenamefont
  {Pozsgay}(2022)}]{Borsi2022}%
  \BibitemOpen
  \bibfield  {author} {\bibinfo {author} {\bibfnamefont {M.}~\bibnamefont
  {Borsi}}\ and\ \bibinfo {author} {\bibfnamefont {B.}~\bibnamefont
  {Pozsgay}},\ }\href {https://doi.org/10.1103/physrevb.106.014302} {\bibfield
  {journal} {\bibinfo  {journal} {Physical Review B}\ }\textbf {\bibinfo
  {volume} {106}},\ \bibinfo {pages} {014302} (\bibinfo {year}
  {2022})}\BibitemShut {NoStop}%
\bibitem [{\citenamefont {Suchsland}\ \emph {et~al.}(2023)\citenamefont
  {Suchsland}, \citenamefont {Moessner},\ and\ \citenamefont
  {Claeys}}]{Suchsland2023}%
  \BibitemOpen
  \bibfield  {author} {\bibinfo {author} {\bibfnamefont {P.}~\bibnamefont
  {Suchsland}}, \bibinfo {author} {\bibfnamefont {R.}~\bibnamefont
  {Moessner}},\ and\ \bibinfo {author} {\bibfnamefont {P.~W.}\ \bibnamefont
  {Claeys}},\ }\href {https://doi.org/10.48550/arXiv.2308.03851} {\bibfield
  {journal} {\bibinfo  {journal} {arXiv:2308.03851}\ } (\bibinfo {year}
  {2023})}\BibitemShut {NoStop}%
\bibitem [{\citenamefont {Logarić}\ \emph {et~al.}(2024)\citenamefont
  {Logarić}, \citenamefont {Dooley}, \citenamefont {Pappalardi},\ and\
  \citenamefont {Goold}}]{Logaric2023}%
  \BibitemOpen
  \bibfield  {author} {\bibinfo {author} {\bibfnamefont {L.}~\bibnamefont
  {Logarić}}, \bibinfo {author} {\bibfnamefont {S.}~\bibnamefont {Dooley}},
  \bibinfo {author} {\bibfnamefont {S.}~\bibnamefont {Pappalardi}},\ and\
  \bibinfo {author} {\bibfnamefont {J.}~\bibnamefont {Goold}},\ }\href
  {https://doi.org/10.1103/physrevlett.132.010401} {\bibfield  {journal}
  {\bibinfo  {journal} {Physical Review Letters}\ }\textbf {\bibinfo {volume}
  {132}},\ \bibinfo {pages} {010401} (\bibinfo {year} {2024})}\BibitemShut
  {NoStop}%
\bibitem [{\citenamefont {Bertini}\ \emph
  {et~al.}(2020{\natexlab{a}})\citenamefont {Bertini}, \citenamefont {Kos},\
  and\ \citenamefont {Prosen}}]{Bertini2020}%
  \BibitemOpen
  \bibfield  {author} {\bibinfo {author} {\bibfnamefont {B.}~\bibnamefont
  {Bertini}}, \bibinfo {author} {\bibfnamefont {P.}~\bibnamefont {Kos}},\ and\
  \bibinfo {author} {\bibfnamefont {T.}~\bibnamefont {Prosen}},\ }\href
  {https://doi.org/10.21468/SciPostPhys.8.4.067} {\bibfield  {journal}
  {\bibinfo  {journal} {{SciPost} Physics}\ }\textbf {\bibinfo {volume} {8}}
  (\bibinfo {year} {2020}{\natexlab{a}})}\BibitemShut {NoStop}%
\bibitem [{\citenamefont {Claeys}\ and\ \citenamefont
  {Lamacraft}(2020)}]{Claeys2020}%
  \BibitemOpen
  \bibfield  {author} {\bibinfo {author} {\bibfnamefont {P.~W.}\ \bibnamefont
  {Claeys}}\ and\ \bibinfo {author} {\bibfnamefont {A.}~\bibnamefont
  {Lamacraft}},\ }\href {https://doi.org/10.1103/physrevresearch.2.033032}
  {\bibfield  {journal} {\bibinfo  {journal} {Physical Review Research}\
  }\textbf {\bibinfo {volume} {2}},\ \bibinfo {pages} {033032} (\bibinfo {year}
  {2020})}\BibitemShut {NoStop}%
\bibitem [{\citenamefont {Rampp}\ \emph {et~al.}(2023)\citenamefont {Rampp},
  \citenamefont {Moessner},\ and\ \citenamefont {Claeys}}]{Rampp2023}%
  \BibitemOpen
  \bibfield  {author} {\bibinfo {author} {\bibfnamefont {M.~A.}\ \bibnamefont
  {Rampp}}, \bibinfo {author} {\bibfnamefont {R.}~\bibnamefont {Moessner}},\
  and\ \bibinfo {author} {\bibfnamefont {P.~W.}\ \bibnamefont {Claeys}},\
  }\href {https://doi.org/10.1103/physrevlett.130.130402} {\bibfield  {journal}
  {\bibinfo  {journal} {Physical Review Letters}\ }\textbf {\bibinfo {volume}
  {130}},\ \bibinfo {pages} {130402} (\bibinfo {year} {2023})}\BibitemShut
  {NoStop}%
\bibitem [{\citenamefont {Huang}\ \emph {et~al.}(2023)\citenamefont {Huang},
  \citenamefont {Li}, \citenamefont {Huse},\ and\ \citenamefont
  {Chan}}]{Huang2023}%
  \BibitemOpen
  \bibfield  {author} {\bibinfo {author} {\bibfnamefont {K.}~\bibnamefont
  {Huang}}, \bibinfo {author} {\bibfnamefont {X.}~\bibnamefont {Li}}, \bibinfo
  {author} {\bibfnamefont {D.~A.}\ \bibnamefont {Huse}},\ and\ \bibinfo
  {author} {\bibfnamefont {A.}~\bibnamefont {Chan}},\ }\href
  {https://doi.org/10.48550/ARXIV.2308.16179} {\bibfield  {journal} {\bibinfo
  {journal} {arXiv:2308.16179}\ } (\bibinfo {year} {2023})}\BibitemShut
  {NoStop}%
\bibitem [{\citenamefont {Hunter-Jones}(2019)}]{HunterJones2019}%
  \BibitemOpen
  \bibfield  {author} {\bibinfo {author} {\bibfnamefont {N.}~\bibnamefont
  {Hunter-Jones}},\ }\href {https://doi.org/10.48550/arXiv.1905.12053}
  {\bibfield  {journal} {\bibinfo  {journal} {arXiv:1905.12053}\ } (\bibinfo
  {year} {2019})}\BibitemShut {NoStop}%
\bibitem [{\citenamefont {Kardar}(2007)}]{Kardar2007}%
  \BibitemOpen
  \bibfield  {author} {\bibinfo {author} {\bibfnamefont {M.}~\bibnamefont
  {Kardar}},\ }\href@noop {} {\emph {\bibinfo {title} {Statistical Physics of
  Fields}}}\ (\bibinfo  {publisher} {Cambridge University Press},\ \bibinfo
  {year} {2007})\ p.\ \bibinfo {pages} {370}\BibitemShut {NoStop}%
\bibitem [{\citenamefont {Zhou}\ and\ \citenamefont {Nahum}(2020)}]{Zhou2020}%
  \BibitemOpen
  \bibfield  {author} {\bibinfo {author} {\bibfnamefont {T.}~\bibnamefont
  {Zhou}}\ and\ \bibinfo {author} {\bibfnamefont {A.}~\bibnamefont {Nahum}},\
  }\href {https://doi.org/10.1103/physrevx.10.031066} {\bibfield  {journal}
  {\bibinfo  {journal} {Physical Review X}\ }\textbf {\bibinfo {volume} {10}},\
  \bibinfo {pages} {031066} (\bibinfo {year} {2020})}\BibitemShut {NoStop}%
\bibitem [{\citenamefont {Jonay}\ and\ \citenamefont {Zhou}(2023)}]{Jonay2023}%
  \BibitemOpen
  \bibfield  {author} {\bibinfo {author} {\bibfnamefont {C.}~\bibnamefont
  {Jonay}}\ and\ \bibinfo {author} {\bibfnamefont {T.}~\bibnamefont {Zhou}},\
  }\href {https://doi.org/10.48550/ARXIV.2310.04491} {\bibfield  {journal}
  {\bibinfo  {journal} {arXiv:2310.04491}\ } (\bibinfo {year}
  {2023})}\BibitemShut {NoStop}%
\bibitem [{\citenamefont {Lie}\ \emph {et~al.}(2022)\citenamefont {Lie},
  \citenamefont {Teo},\ and\ \citenamefont {Jeong}}]{Lie2022}%
  \BibitemOpen
  \bibfield  {author} {\bibinfo {author} {\bibfnamefont {S.~H.}\ \bibnamefont
  {Lie}}, \bibinfo {author} {\bibfnamefont {Y.~S.}\ \bibnamefont {Teo}},\ and\
  \bibinfo {author} {\bibfnamefont {H.}~\bibnamefont {Jeong}},\ }\href
  {https://doi.org/10.48550/arXiv.2204.00374} {\bibfield  {journal} {\bibinfo
  {journal} {arXiv:2204.00374}\ } (\bibinfo {year} {2022})}\BibitemShut
  {NoStop}%
\bibitem [{\citenamefont {Rather}\ \emph {et~al.}(2020)\citenamefont {Rather},
  \citenamefont {Aravinda},\ and\ \citenamefont {Lakshminarayan}}]{Rather2020}%
  \BibitemOpen
  \bibfield  {author} {\bibinfo {author} {\bibfnamefont {S.~A.}\ \bibnamefont
  {Rather}}, \bibinfo {author} {\bibfnamefont {S.}~\bibnamefont {Aravinda}},\
  and\ \bibinfo {author} {\bibfnamefont {A.}~\bibnamefont {Lakshminarayan}},\
  }\href {https://doi.org/10.1103/physrevlett.125.070501} {\bibfield  {journal}
  {\bibinfo  {journal} {Physical Review Letters}\ }\textbf {\bibinfo {volume}
  {125}},\ \bibinfo {pages} {070501} (\bibinfo {year} {2020})}\BibitemShut
  {NoStop}%
\bibitem [{\citenamefont {Caux}\ and\ \citenamefont {Mossel}(2011)}]{Caux2011}%
  \BibitemOpen
  \bibfield  {author} {\bibinfo {author} {\bibfnamefont {J.-S.}\ \bibnamefont
  {Caux}}\ and\ \bibinfo {author} {\bibfnamefont {J.}~\bibnamefont {Mossel}},\
  }\href {https://doi.org/10.1088/1742-5468/2011/02/p02023} {\bibfield
  {journal} {\bibinfo  {journal} {Journal of Statistical Mechanics: Theory and
  Experiment}\ }\textbf {\bibinfo {volume} {2011}},\ \bibinfo {pages} {P02023}
  (\bibinfo {year} {2011})}\BibitemShut {NoStop}%
\bibitem [{\citenamefont {Rigol}\ \emph {et~al.}(2007)\citenamefont {Rigol},
  \citenamefont {Dunjko}, \citenamefont {Yurovsky},\ and\ \citenamefont
  {Olshanii}}]{Rigol2007}%
  \BibitemOpen
  \bibfield  {author} {\bibinfo {author} {\bibfnamefont {M.}~\bibnamefont
  {Rigol}}, \bibinfo {author} {\bibfnamefont {V.}~\bibnamefont {Dunjko}},
  \bibinfo {author} {\bibfnamefont {V.}~\bibnamefont {Yurovsky}},\ and\
  \bibinfo {author} {\bibfnamefont {M.}~\bibnamefont {Olshanii}},\ }\href
  {https://doi.org/10.1103/physrevlett.98.050405} {\bibfield  {journal}
  {\bibinfo  {journal} {Physical Review Letters}\ }\textbf {\bibinfo {volume}
  {98}},\ \bibinfo {pages} {050405} (\bibinfo {year} {2007})}\BibitemShut
  {NoStop}%
\bibitem [{\citenamefont {Essler}\ and\ \citenamefont
  {Fagotti}(2016)}]{Essler2016}%
  \BibitemOpen
  \bibfield  {author} {\bibinfo {author} {\bibfnamefont {F.~H.~L.}\
  \bibnamefont {Essler}}\ and\ \bibinfo {author} {\bibfnamefont
  {M.}~\bibnamefont {Fagotti}},\ }\href
  {https://doi.org/10.1088/1742-5468/2016/06/064002} {\bibfield  {journal}
  {\bibinfo  {journal} {Journal of Statistical Mechanics: Theory and
  Experiment}\ }\textbf {\bibinfo {volume} {2016}},\ \bibinfo {pages} {064002}
  (\bibinfo {year} {2016})}\BibitemShut {NoStop}%
\bibitem [{\citenamefont {Castro-Alvaredo}\ \emph {et~al.}(2016)\citenamefont
  {Castro-Alvaredo}, \citenamefont {Doyon},\ and\ \citenamefont
  {Yoshimura}}]{CastroAlvaredo2016}%
  \BibitemOpen
  \bibfield  {author} {\bibinfo {author} {\bibfnamefont {O.~A.}\ \bibnamefont
  {Castro-Alvaredo}}, \bibinfo {author} {\bibfnamefont {B.}~\bibnamefont
  {Doyon}},\ and\ \bibinfo {author} {\bibfnamefont {T.}~\bibnamefont
  {Yoshimura}},\ }\href {https://doi.org/10.1103/physrevx.6.041065} {\bibfield
  {journal} {\bibinfo  {journal} {Physical Review X}\ }\textbf {\bibinfo
  {volume} {6}},\ \bibinfo {pages} {041065} (\bibinfo {year}
  {2016})}\BibitemShut {NoStop}%
\bibitem [{\citenamefont {Bertini}\ \emph {et~al.}(2021)\citenamefont
  {Bertini}, \citenamefont {Heidrich-Meisner}, \citenamefont {Karrasch},
  \citenamefont {Prosen}, \citenamefont {Steinigeweg},\ and\ \citenamefont
  {{\v{Z}}nidari{\v{c}}}}]{Bertini2021}%
  \BibitemOpen
  \bibfield  {author} {\bibinfo {author} {\bibfnamefont {B.}~\bibnamefont
  {Bertini}}, \bibinfo {author} {\bibfnamefont {F.}~\bibnamefont
  {Heidrich-Meisner}}, \bibinfo {author} {\bibfnamefont {C.}~\bibnamefont
  {Karrasch}}, \bibinfo {author} {\bibfnamefont {T.}~\bibnamefont {Prosen}},
  \bibinfo {author} {\bibfnamefont {R.}~\bibnamefont {Steinigeweg}},\ and\
  \bibinfo {author} {\bibfnamefont {M.}~\bibnamefont {{\v{Z}}nidari{\v{c}}}},\
  }\href {https://doi.org/10.1103/revmodphys.93.025003} {\bibfield  {journal}
  {\bibinfo  {journal} {Reviews of Modern Physics}\ }\textbf {\bibinfo {volume}
  {93}},\ \bibinfo {pages} {025003} (\bibinfo {year} {2021})}\BibitemShut
  {NoStop}%
\bibitem [{\citenamefont {Dubail}(2017)}]{Dubail2017}%
  \BibitemOpen
  \bibfield  {author} {\bibinfo {author} {\bibfnamefont {J.}~\bibnamefont
  {Dubail}},\ }\href {https://doi.org/10.1088/1751-8121/aa6f38} {\bibfield
  {journal} {\bibinfo  {journal} {Journal of Physics A: Mathematical and
  Theoretical}\ }\textbf {\bibinfo {volume} {50}},\ \bibinfo {pages} {234001}
  (\bibinfo {year} {2017})}\BibitemShut {NoStop}%
\bibitem [{\citenamefont {Alba}\ and\ \citenamefont
  {Calabrese}(2019)}]{Alba2019}%
  \BibitemOpen
  \bibfield  {author} {\bibinfo {author} {\bibfnamefont {V.}~\bibnamefont
  {Alba}}\ and\ \bibinfo {author} {\bibfnamefont {P.}~\bibnamefont
  {Calabrese}},\ }\href {https://doi.org/10.1103/physrevb.100.115150}
  {\bibfield  {journal} {\bibinfo  {journal} {Physical Review B}\ }\textbf
  {\bibinfo {volume} {100}},\ \bibinfo {pages} {115150} (\bibinfo {year}
  {2019})}\BibitemShut {NoStop}%
\bibitem [{\citenamefont {Bertini}\ \emph
  {et~al.}(2020{\natexlab{b}})\citenamefont {Bertini}, \citenamefont {Kos},\
  and\ \citenamefont {Prosen}}]{Bertini2020a}%
  \BibitemOpen
  \bibfield  {author} {\bibinfo {author} {\bibfnamefont {B.}~\bibnamefont
  {Bertini}}, \bibinfo {author} {\bibfnamefont {P.}~\bibnamefont {Kos}},\ and\
  \bibinfo {author} {\bibfnamefont {T.}~\bibnamefont {Prosen}},\ }\href
  {https://doi.org/10.21468/SciPostPhys.8.4.068} {\bibfield  {journal}
  {\bibinfo  {journal} {{SciPost} Physics}\ }\textbf {\bibinfo {volume} {8}}
  (\bibinfo {year} {2020}{\natexlab{b}})}\BibitemShut {NoStop}%
\bibitem [{\citenamefont {Lopez-Piqueres}\ \emph {et~al.}(2021)\citenamefont
  {Lopez-Piqueres}, \citenamefont {Ware}, \citenamefont {Gopalakrishnan},\ and\
  \citenamefont {Vasseur}}]{LopezPiqueres2021}%
  \BibitemOpen
  \bibfield  {author} {\bibinfo {author} {\bibfnamefont {J.}~\bibnamefont
  {Lopez-Piqueres}}, \bibinfo {author} {\bibfnamefont {B.}~\bibnamefont
  {Ware}}, \bibinfo {author} {\bibfnamefont {S.}~\bibnamefont
  {Gopalakrishnan}},\ and\ \bibinfo {author} {\bibfnamefont {R.}~\bibnamefont
  {Vasseur}},\ }\href {https://doi.org/10.1103/physrevb.104.104307} {\bibfield
  {journal} {\bibinfo  {journal} {Physical Review B}\ }\textbf {\bibinfo
  {volume} {104}},\ \bibinfo {pages} {104307} (\bibinfo {year}
  {2021})}\BibitemShut {NoStop}%
\bibitem [{\citenamefont {Medenjak}(2022)}]{Medenjak2022}%
  \BibitemOpen
  \bibfield  {author} {\bibinfo {author} {\bibfnamefont {M.}~\bibnamefont
  {Medenjak}},\ }\href {https://doi.org/10.1088/1751-8121/ac8fc4} {\bibfield
  {journal} {\bibinfo  {journal} {Journal of Physics A: Mathematical and
  Theoretical}\ }\textbf {\bibinfo {volume} {55}},\ \bibinfo {pages} {404002}
  (\bibinfo {year} {2022})}\BibitemShut {NoStop}%
\bibitem [{\citenamefont {Vanicat}\ \emph {et~al.}(2018)\citenamefont
  {Vanicat}, \citenamefont {Zadnik},\ and\ \citenamefont
  {Prosen}}]{Vanicat2018}%
  \BibitemOpen
  \bibfield  {author} {\bibinfo {author} {\bibfnamefont {M.}~\bibnamefont
  {Vanicat}}, \bibinfo {author} {\bibfnamefont {L.}~\bibnamefont {Zadnik}},\
  and\ \bibinfo {author} {\bibfnamefont {T.}~\bibnamefont {Prosen}},\ }\href
  {https://doi.org/10.1103/physrevlett.121.030606} {\bibfield  {journal}
  {\bibinfo  {journal} {Physical Review Letters}\ }\textbf {\bibinfo {volume}
  {121}},\ \bibinfo {pages} {030606} (\bibinfo {year} {2018})}\BibitemShut
  {NoStop}%
\bibitem [{\citenamefont {Ljubotina}\ \emph {et~al.}(2019)\citenamefont
  {Ljubotina}, \citenamefont {Zadnik},\ and\ \citenamefont
  {Prosen}}]{Ljubotina2019}%
  \BibitemOpen
  \bibfield  {author} {\bibinfo {author} {\bibfnamefont {M.}~\bibnamefont
  {Ljubotina}}, \bibinfo {author} {\bibfnamefont {L.}~\bibnamefont {Zadnik}},\
  and\ \bibinfo {author} {\bibfnamefont {T.}~\bibnamefont {Prosen}},\ }\href
  {https://doi.org/10.1103/physrevlett.122.150605} {\bibfield  {journal}
  {\bibinfo  {journal} {Physical Review Letters}\ }\textbf {\bibinfo {volume}
  {122}},\ \bibinfo {pages} {150605} (\bibinfo {year} {2019})}\BibitemShut
  {NoStop}%
\bibitem [{\citenamefont {Claeys}\ \emph
  {et~al.}(2022{\natexlab{b}})\citenamefont {Claeys}, \citenamefont
  {Herzog-Arbeitman},\ and\ \citenamefont {Lamacraft}}]{Claeys2022b}%
  \BibitemOpen
  \bibfield  {author} {\bibinfo {author} {\bibfnamefont {P.~W.}\ \bibnamefont
  {Claeys}}, \bibinfo {author} {\bibfnamefont {J.}~\bibnamefont
  {Herzog-Arbeitman}},\ and\ \bibinfo {author} {\bibfnamefont {A.}~\bibnamefont
  {Lamacraft}},\ }\href {https://doi.org/10.21468/SciPostPhys.12.1.007}
  {\bibfield  {journal} {\bibinfo  {journal} {{SciPost} Physics}\ }\textbf
  {\bibinfo {volume} {12}} (\bibinfo {year} {2022}{\natexlab{b}})}\BibitemShut
  {NoStop}%
\bibitem [{\citenamefont {Calabrese}(2020)}]{Calabrese2020}%
  \BibitemOpen
  \bibfield  {author} {\bibinfo {author} {\bibfnamefont {P.}~\bibnamefont
  {Calabrese}},\ }\href {https://doi.org/10.21468/SciPostPhysLectNotes.20}
  {\bibfield  {journal} {\bibinfo  {journal} {{SciPost} Physics Lecture Notes}\
  } (\bibinfo {year} {2020})}\BibitemShut {NoStop}%
\bibitem [{\citenamefont {Schumacher}\ and\ \citenamefont
  {Westmoreland}(2002)}]{Schumacher2002}%
  \BibitemOpen
  \bibfield  {author} {\bibinfo {author} {\bibfnamefont {B.}~\bibnamefont
  {Schumacher}}\ and\ \bibinfo {author} {\bibfnamefont {M.~D.}\ \bibnamefont
  {Westmoreland}},\ }\href {https://doi.org/10.1023/a:1019653202562} {\bibfield
   {journal} {\bibinfo  {journal} {Quantum Information Processing}\ }\textbf
  {\bibinfo {volume} {1}},\ \bibinfo {pages} {5} (\bibinfo {year}
  {2002})}\BibitemShut {NoStop}%
\bibitem [{\citenamefont {Bernard}\ and\ \citenamefont
  {Doyon}(2016)}]{Bernard2016}%
  \BibitemOpen
  \bibfield  {author} {\bibinfo {author} {\bibfnamefont {D.}~\bibnamefont
  {Bernard}}\ and\ \bibinfo {author} {\bibfnamefont {B.}~\bibnamefont
  {Doyon}},\ }\href {https://doi.org/10.1088/1742-5468/2016/06/064005}
  {\bibfield  {journal} {\bibinfo  {journal} {Journal of Statistical Mechanics:
  Theory and Experiment}\ }\textbf {\bibinfo {volume} {2016}},\ \bibinfo
  {pages} {064005} (\bibinfo {year} {2016})}\BibitemShut {NoStop}%
\bibitem [{\citenamefont {Nakata}\ \emph {et~al.}(2023)\citenamefont {Nakata},
  \citenamefont {Wakakuwa},\ and\ \citenamefont {Koashi}}]{Nakata2023a}%
  \BibitemOpen
  \bibfield  {author} {\bibinfo {author} {\bibfnamefont {Y.}~\bibnamefont
  {Nakata}}, \bibinfo {author} {\bibfnamefont {E.}~\bibnamefont {Wakakuwa}},\
  and\ \bibinfo {author} {\bibfnamefont {M.}~\bibnamefont {Koashi}},\ }\href
  {https://doi.org/10.22331/q-2023-02-21-928} {\bibfield  {journal} {\bibinfo
  {journal} {Quantum}\ }\textbf {\bibinfo {volume} {7}},\ \bibinfo {pages}
  {928} (\bibinfo {year} {2023})}\BibitemShut {NoStop}%
\end{thebibliography}%

\end{document}